\documentclass[12pt]{article}

\usepackage{amssymb,epsfig}

\renewcommand{\theequation}{\thesection.\arabic{equation}}

\newcommand{\be}{\begin{equation}}
\newcommand{\ee}{\end{equation}}
\newcommand{\ba}{\begin{eqnarray}}
\newcommand{\ea}{\end{eqnarray}}

\def\beq{\begin{equation}}
\def\eeq{\end{equation}}
\def\beqa{\begin{eqnarray}}
\def\eeqa{\end{eqnarray}}
\def\eq#1{Eq.~(\ref{#1})}

\def\IR{\hbox{\tiny IR}}

\def\UV{\hbox{\tiny UV}}

\newcommand{\zsl}{\mbox{$z$\hspace{-0.5em}\raisebox{0.1ex}{$/$}}}
\newcommand{\Deltasl}{\mbox{$\Delta$\hspace{-0.5em}\raisebox{0.1ex}{$/$}}}

\begin{document}

\begin{titlepage}

\begin{flushright}
\begin{tabular}{l}
Cavendish-HEP-04/06\\
 hep-ph/0401158
\end{tabular}
\end{flushright}
\vspace{1.5cm}

\begin{center}
{\LARGE \bf
Renormalon approach to higher--twist distribution amplitudes
and the convergence of the conformal expansion}

\vspace{1cm}

{\sc Vladimir M.~Braun}${}^1$,
{\sc Einan Gardi}${}^{1,2}$ and
{\sc Stefan Gottwald}${}^{1}$
\\[0.5cm]
\vspace*{0.1cm} ${}^1${\it
   Institut f\"ur Theoretische Physik, Universit\"at
   Regensburg, \\ D-93040 Regensburg, Germany
                       } \\[0.2cm]
\vspace*{0.1cm} ${}^2$ {\it
 Cavendish Laboratory, University of Cambridge\\
Madingley Road, Cambridge, CB3 OHE, United Kingdom\footnote{Address after January 1st 2004.}
                       } \\[1.0cm]

\vskip1.2cm
{\bf Abstract:\\[10pt]} \parbox[t]{\textwidth}{
Power corrections to exclusive processes are usually
calculated using models for twist--four distribution amplitudes
(DA) which are based on the leading--order terms in the conformal
expansion. In this work we develop a different approach which does
not rely on conformal symmetry but is based instead on 
renormalon analysis. This way we obtain an upper bound for the 
contributions of higher conformal spin operators, which 
translates into a bound on the end--point behavior of DA. The existence of 
such a bound is important for proving factorization theorems.  
For the two--particle twist--four DA we find in the renormalon model 
that the conformal expansion converges but it does not converge 
uniformly near the end points. This means that power corrections 
to observables which are particularly sensitive to the region 
where one valence quark is soft, may be underestimated when using 
the first few terms in the conformal expansion. 
The renormalon models of twist--four DA of the pion and the $\rho$ 
meson are constructed and can be used as a viable alternative to 
existing models. 
}
  \vskip1cm  
{\em Submitted to Nuclear Physics B }\\[1cm] 
\end{center}

\begin{center}
\end{center}

\end{titlepage}

{\tableofcontents}

\newpage

\section{Introduction\label{int}}
\setcounter{equation}{0}

The relevant non-perturbative degrees of freedom in hard exclusive
processes are described by hadron distribution amplitudes (DA)
\cite{ER,BL} that detail the momentum--fraction distributions
of partons in the infinite momentum frame by integrating out the
transverse momentum dependence. The leading--twist parton
distributions appear in the QCD description of hard exclusive
reactions to the leading power accuracy and refer to parton
configurations with the minimal number of constituents. The
higher--twist distributions, in turn, are more numerous and are
used to take into account a variety of effects due to parton
virtuality, transverse momentum, and contributions of higher Fock
states that are relevant for the description of power--suppressed
corrections in the hard momentum. It should be mentioned that
application of QCD factorization to exclusive processes beyond the
leading--twist approximation presents a serious challenge because
of end--point divergences related to the contributions of soft
partons. Up to now such applications have been mostly in
connection with the so--called light--cone sum rules (LCSR)
\cite{LCSR} in which case the end--point divergences are removed by
construction. More generally, end--point contributions have to be
added and there is a hope that the precise separation of hard
(twist--expandable) and soft (end--point) contributions can be
achieved within the soft--collinear effective theory, see e.g.
\cite{SCET}. Having in mind the existing and potential
applications it is worthwhile to have a fresh look at the
theoretical description of higher--twist  DA and the related
uncertainties. The present work presents a step in this direction.

The existing theoretical framework for the description of  DA is
based on the conformal symmetry of the QCD Lagrangian, see
\cite{conformal} for a detailed review. The symmetry can be used
to separate the dependence of the hadron wave function on the
longitudinal momentum fractions and the transverse coordinates
(that are later traded for the renormalization scale) in very much
the same way as the rotational symmetry of the potential in
quantum mechanics allows to separate the angular and the radial
dependence. The orthogonal polynomials appearing in the expansion
of distribution amplitudes \cite{ER,BL} are nothing but the
irreducible representations of the collinear conformal group and
play the same role as spherical harmonics in quantum mechanics,
with the orbital angular momentum replaced by conformal spin. One
motivation for using the conformal expansion is that contributions
with different conformal spin do not mix with each other under QCD
evolution to the leading logarithmic accuracy \cite{ER,BL}.
Another rationale \cite{BF90} is that QCD equations of motion (EOM) only
relate terms with the same conformal spin so that any exact
EOM--based relation between  DA can be
satisfied order by order in the conformal expansion. It follows
that any parametrization of  DA based on a truncated conformal
expansion is consistent with the EOM and is preserved by
evolution.

Assuming that  DA are dominated by first few terms having the
lowest conformal spin, the conformal expansion provides a
practical framework for constructing models for DA
\cite{CZ81,CZ84,COZ87,CZreport,BF90,BBKT98,BB98,Ball99,BFMS00,BBK02}
which are both consistent with the QCD constraints and involve
just a few non-perturbative parameters. This approach has been
used extensively for phenomenology.
Its main limitation is that just the first one or two terms in the
expansion are included; the increasing number of parameters at
higher conformal spins makes this program impractical otherwise,
so the assumption that the conformal expansion converges fast is
absolutely essential.

From the theoretical point of view, convergence of the conformal
expansion is expected because anomalous dimensions of conformal
operators are rising logarithmically with the spin\footnote{This
property is proven for twist--two and twist--three operators
and is probably true for all twists.}
and, therefore, higher--spin contributions are suppressed at
asymptotically large scales. However, this suppression is
numerically weak and not sufficient to guarantee convergence
at the scales of practical interest. {}For leading twist this
question can eventually be decided by experiment and indeed
the qualitative success of quark counting rules and, more importantly,
the CLEO data on the $\pi(\eta)\gamma\gamma^*$ transition form
factor \cite{CLEO} strongly suggest that the meson DA are not very
far from their asymptotic shape. For higher twist, using
experimental data to constrain DA seems totally unrealistic and up
to now there had been no arguments whatsoever whether truncation
of the conformal expansion is legitimate in this case or not.

The aim of this work is to construct alternative models for 
twist--four meson DA that do not rely on the conformal expansion 
and present plausible bounds for the higher--spin contributions 
to these DA.  To this end, we approach the problem from a
different angle, using the concept of
renormalons~\cite{ren_review}. Owing to the breaking of scale
invariance in the quantum theory through the running of the
coupling, operators of different twist mix with each other
under renormalization. Independence of a physical observable on
the factorization scale implies intricate cancellations between
different twists --- the so--called cancellation of renormalon
ambiguities --- and the existence of these ambiguities can be used
to estimate power--suppressed corrections in a similar way as the
logarithmic scale dependence is used to estimate the accuracy of
fixed--order perturbative calculations. 

Most applications of
renormalons have so far been in inclusive or semi--inclusive cross
sections \cite{ren_review,ren_structure_functions}, 
although some work has already been done in the context
of exclusive processes. This includes analysis of the
large--order behavior of the Brodsky--Lepage evolution
kernel~\cite{Mikhailov98}, and studies of infrared renormalons in
specific processes, e.g. the $\gamma^*\gamma\pi_0$ form
factor~\cite{GK97}, the pion electromagnetic form
factor~\cite{Agaev98}, and deeply--virtual Compton
scattering~\cite{BS98}. Here we do not consider a specific process
but instead develop a general framework for estimating
higher--twist contributions to exclusive processes involving
pseudoscalar and vector mesons by constructing a renormalon model
for DA, continuing the work by J.~Andersen~\cite{Jeppe}. We trace
the cancellation of renormalons and in this way establish an
explicit connection between previous renormalon calculations and
the operator product expansion (OPE) for exclusive processes, for
cases that the latter exists. We then use the renormalon model to
study the convergence of the conformal expansion.

In order to explain how the concept of renormalons becomes useful
for constructing a consistent model for higher--twist DA let us
discuss a concrete example. Consider the following matrix element:
\begin{eqnarray}
\label{NL_int}
&&\left\langle 0\left \vert {\rm T}\{ \bar{d}(x_2)\Deltasl
\gamma_{5} u(x_1)\} \right\vert \pi^+(p)
\right\rangle_{\mu^2\simeq 1/|\Delta^2|} \,=\,  \\ && \hspace*{+1pt}
{}= i(p\Delta) \,f_\pi\!\int_0^1\!\!du \,{\rm e}^{-iupx_1-i\bar{u}px_2}  \,
\left[ \int_0^1\!\!dv\,C(v,u,\Delta^2,\mu_F^2) \phi_{\pi}(v;\mu_F^2)
+\Delta^2 g_1(u;\mu_F^2)+\cdots  \right]\!, \nonumber
\end{eqnarray}
where $\bar{u}\equiv 1-u$, $f_\pi$ is the pion decay constant and
the fields are taken at a small space-like separation:
$\Delta\equiv x_1-x_2$ with $\Delta^2< 0$, $|\Delta^2| \ll
1/\Lambda_{\rm QCD}^2$. Also, $\mu$ stands for the ultraviolet
renormalization scale and for this discussion we assume that
$\mu^2\simeq 1/|\Delta^2|$ to avoid large logarithms.
 The expansion in the square brackets on the right--hand side
of \eq{NL_int} is nothing but the OPE: $C(v,u,\Delta^2,\mu_F^2)=
\delta(u-v) +{\cal O}(\alpha_s)$ is the twist--two coefficient function,
$\phi_{\pi}(v;\mu_F^2)$ is the standard,  twist--two pion DA, and
$\Delta^2 g_1(u;\mu_F^2)$ represents the twist--four contribution;
$\mu_F^2$ is a factorization scale. The function $g_1(u;\mu_F^2)$
can be interpreted as the distribution of the transverse momentum
(squared) of the quark in the pion \cite{CZreport,BF90} and is one
example of the higher--twist DA that we will be interested in.

The conformal expansion of $g_1(u;\mu_F^2)$ has been constructed
in~\cite{BF90} using the expansion of three--particle DA involving
a quark, an antiquark and a gluon, and  EOM. In
Sect.~\ref{expansion} we shall explain how this is done. Here we
just quote the result:
\begin{eqnarray}
\label{ce_1_int}
g_1(u;\mu_F^2)&=&\bigg\{
g^{(J=3)}(\mu_F^2)\bigg[ \,u^{2}\,\bar{u}^{2}\bigg]
+g^{(J=4)}(\mu_F^2)\bigg[ u\bar{u}\,(13\,u\bar{u} + 2)
\\ \nonumber
&+&  2\,(6\,u^{2} + 3\,u + 1)\,\bar{u}^{3}\,\mathrm{ln}(\bar{u})
+  2 \,( 6\,\bar{u}^2 + 3\,\bar{u} + 1)\,\,u^{3} \mathrm{ln}(u) \bigg]
+\cdots\bigg\},
\end{eqnarray}
where the coefficients $g^{(J)}(\mu_F^2)$ renormalize
multiplicatively, and $J=3,4,\ldots$ is the conformal spin. The
corresponding logarithmic scale dependence cancels against the
scale dependence of the twist--four coefficient function which is
not shown in \eq{NL_int} for brevity. To the leading logarithmic
accuracy  and with the usual choice $\mu_F=\mu \simeq 1/|\Delta|$
this function reduces to a constant which can be included in the
definition of $g_1(u;\mu_F^2)$. The dots in \eq{ce_1_int} stand
for the contribution of conformal spins $J\geq 5$, which were
usually neglected. In this paper we want to study their
significance.

To understand the r{\^o}le of renormalons one needs to carefully
examine the separation made in \eq{NL_int} between twist two and
twist four. For a qualitative discussion it is convenient to use a
hard cutoff $1/|\Delta^2| \sim \mu^2 \gg \mu_F^2\gg \Lambda^2$ for
factorization: loop momenta $k^2>\mu_F^2$ contribute to the
coefficient function $C(v,u,\Delta^2,\mu_F^2)$ while momenta
$k^2<\mu_F^2$ contribute to the DA. Upon expanding the coefficient
function near the light--cone \hbox{$|\Delta^2|\mu_F^2\ll1$}, one
obtains:
\begin{equation}
C(v,u,\Delta^2,\mu_F^2)=\bigg(1+c_1\alpha_s
+c_2\alpha_s^2+\cdots\bigg)
-d(v,u)\mu_F^2\Delta^2\cdots,
\label{IRR_int}
\end{equation}
where $\alpha_s=\alpha_s(\mu^2 =1/|\Delta^2|)$. It is the usual
perturbative series with coefficients
 $c_i=c_i\left(v,u;\ln\left(\mu_F^2/\mu^2\right)\right)$
depending logarithmically on the scales, and the $d(v,u)$ term
represents the leading power correction that arises because the
low--momentum regions are cut off. Since the left--hand side of
\eq{NL_int} does not depend on $\mu_F^2$, any such dependence in
$C(v,u,\Delta^2,\mu_F^2)$  should cancel within the square
brackets on the right--hand side. In particular, the logarithmic
dependence of $C(v,u,\Delta^2,\mu_F^2)$ on $\mu_F^2$ is canceled
by that of the twist--two distribution amplitude
$\phi_\pi(v;\mu_F^2)$. The cancellation of power dependence must
involve the twist--four term $\Delta^2 g_1(u;\mu_F^2)$, so it is
expected that in this factorization prescription
\begin{equation}
g_1(u;\mu_F^2)\longrightarrow \mu_F^2\,\int_0^1dv\,d(v,u)  \phi_{\pi}(v;\mu_F^2)\,
+\,\bar{g}_1(u;\mu_F^2)\,,
\label{UV_div_int}
\end{equation}
where the second term depends on $\mu_F^2$ at most
logarithmically. Indeed, upon renormalization the relevant
twist--four operators show not only logarithmic ultraviolet
divergence which is related to their anomalous dimension, but also
quadratic ultraviolet divergence. Using a hard cutoff, the
dependence of twist--four on $\mu_F^2$ is indeed that of
\eq{UV_div_int}: the twist--four operators mix with the
leading--twist such that the dependence on $\mu_F^2$ in
\eq{NL_int} cancels out. Clearly the function $d(v,u)$ can be
computed either by considering the dependence of the twist--two
coefficient function on $\mu_F^2$, i.e. its sensitivity to small
loop momenta, or by considering the dependence of twist--four
operators on $\mu_F^2$ i.e. their sensitivity to large loop
momenta. Cancellation of $\mu_F^2$ to power accuracy requires that
these two regularizations will be done using the same
prescription.

In practice, a hard cutoff is difficult to implement. Usually,
dimensional regularization is used instead. In this case power
terms in the coefficient function (\eq{IRR_int}) do not appear.
However, the coefficients $c_i$ computed in a MS--like scheme
diverge factorially with the order $i$. The factorial divergence
implies that the sum of the perturbative series is only defined to a
power accuracy and this ambiguity (renormalon ambiguity) must be
compensated by adding a non-perturbative higher--twist correction.
A detailed analysis shows \cite{ren_review} that the asymptotic
large--order behavior of the coefficients (the renormalons) is in
one--to--one correspondence with the sensitivity to extreme (small
or large) loop momenta and that infrared renormalons in 
the leading--twist coefficient function are compensated by ultraviolet
renormalons in the matrix elements of twist--four operators. At
the end the picture described above re-appears:
only the details depend on the factorization method.

Returning to \eq{UV_div_int} we observe that the quadratic term in
$\mu_F$ is spurious since its sole purpose is to cancel the
similar contribution to the coefficient function. It does not
contribute, therefore, to any physical observable. The idea of the
renormalon model~\cite{ren_review,ren_structure_functions}  
is that, with a replacement of $\mu_F$ by a
suitable non-perturbative scale, this contribution should be of
the same order and have roughly the same functional form as the
physical second contribution on the right--hand side of
\eq{UV_div_int}, which is the only one of interest. Assuming this
``ultraviolet dominance'' \cite{Braun95,BBM97,ren_review} we get
the following  model:
\begin{equation}
g_1(u;\mu_F^2)\simeq k \Lambda_{\rm QCD}^2\,\int_0^1dv\,d(v,u)  \phi_{\pi}(v;\mu_F^2),
\label{UV_Lambda_int}
\end{equation}
where by explicit calculation (see  Sect.~\ref{Pion_DA}) one finds
for  $d(v,u)$ at leading order
\begin{eqnarray}
\label{dvu}
d(v,u)&=&
\frac{1}{v^2}\,\bigg[u+(v-u)\ln\left(1-\frac{u}{v}\right)\bigg] \theta(v>u)\nonumber
  \\ &&\hspace*{40pt} +\,
 \frac{1}{(1-v)^2}\,\bigg[1- u+(u-v)
\ln\left(1-\frac{1- u}{1- v}\right)\bigg] \theta(v<u).
\end{eqnarray}
The overall coefficient in this model $k={\cal O}(1)$ can be fixed by the
normalization integral $\int_0^1 du\, g_1(u;\mu^2)$ corresponding to 
the matrix element of a local operator which can be estimated by QCD
sum rules \cite{CZreport,BF90} or calculated on the lattice. To
the accuracy of \eq{dvu} the logarithmic $\mu_F$ dependence on the
left--hand side and on the right--hand side of \eq{UV_Lambda_int}
do not match: one expects this model to be relevant at low
scales of the order of a few times~$\Lambda_{\rm QCD}$.

The ``ultraviolet dominance'' assumption used to derive
\eq{UV_Lambda_int}, is in fact sufficient to derive the full set
of two-- and three--particle DA of twist--four in terms of the
leading--twist DA. It is important that this approximation is
fully consistent with the OPE and respects all constraints imposed by
EOM. On the other hand, it does not assume any
hierarchy of contributions of the increasing conformal spin and
indeed the expressions in \eq{ce_1_int} and \eq{UV_Lambda_int} are
very different for any reasonable choice of the leading--twist DA.
Since twist--four anomalous dimensions are not taken into account,
one should expect that this model overestimates contributions of
conformal operators with high spin. It is therefore complementary
to the usual models based on using a first few terms with the
lowest spin and provides an upper--bound estimate of the neglected
contributions.

The presentation is organized as follows. In Sect.~2 we formulate
our program in more precise terms, introducing the relevant
techniques (Borel transform) and the systematic approximation
(large--$N_f$ expansion) that will be used throughout the rest of
the work. The expression in \eq{dvu} is derived and the
cancellation of the renormalon ambiguity for the matrix element of
the type in \eq{NL_int} is demonstrated by an explicit
calculation. The renormalon model of the pion DA of twist--four is
presented in Sect.~3 and in Sect.~4 we discuss its conformal
expansion. The generalization of these results for the case of
vector mesons is considered in Sect.~5 while Sect.~6 is reserved
for the conclusions.

\section{Cancellation of renormalon ambiguities in the OPE\label{Pion_DA}}
\setcounter{equation}{0}

\subsection{Definitions\label{definit}}

In this section we compute the renormalon ambiguity of the leading--twist
coefficient function in the simplest exclusive amplitude involving a
pseudoscalar meson, and demonstrate how the unique result is restored  by
adding the twist--four contributions in the OPE.
To begin with, let us set up the necessary definitions.

We choose to consider the gauge--invariant time--ordered product of
two quark ``currents'' at a small (but non-vanishing) light--cone
separation, which can be parametrized in terms of two
Lorentz--invariant amplitudes $G_1(u,\Delta^2;\mu^2)$ and
$G_2(u,\Delta^2;\mu^2)$ defined as
\begin{eqnarray}
\label{NL}
&&\left\langle 0\left \vert {\rm T}\{ \bar{d}(x_2) \gamma_{\nu}
\gamma_{5} [x_2,x_1] u(x_1)\} \right\vert \pi^+(p)
\right\rangle_{\mu^2} \,=\,  \\ && \hspace*{+40pt}
= i \,f_\pi\, \int_0^1\,du \,{\rm e}^{-iupx_1-i\bar{u}px_2}  \,
\left[ G_1(u,\Delta^2;\mu^2) p_{\nu} + G_2(u,\Delta^2;\mu^2)
 \left( \frac{p\cdot\Delta }{\Delta^2} \Delta_{\nu}
-p_{\nu}\right) \right]. \nonumber
\end{eqnarray}
Here $\Delta\equiv x_1-x_2$ with $|\Delta^2|\ll 1/\Lambda_{\rm QCD}^2$, $\Delta^2< 0$ 
playing the r\^ole of the hard
scale, $\bar{u}=1-u$ and $\mu^2$ is
the ultraviolet renormalization scale.
We use the notation $[x_2,x_1]$ for the
Wilson line connecting the points $x_2$ and $x_1$,
\begin{equation}
  \label{def_wilson_line}
  [x_2,x_1]=P \exp\left[-ig\int_0^1\!dt\,\Delta_\mu A^\mu(x_2+t\Delta)\right].
\end{equation}
Note that the $\mu^2$ dependence comes entirely from the 
wave--function renormalization of the quark and the antiquark fields in
\eq{NL} and can be removed by adding the corresponding
$Z$--factors. Up to this additional renormalization, the functions
$G_{1,2}(u,\Delta^2;\mu^2)$ can be viewed as physical amplitudes.
Their advantage is that they are simpler than exclusive amplitudes which
are relevant for phenomenology (such as the two--photon one) 
and at the same time retain all the features that are important in the present context.

The asymptotic behavior of $G_{i}(u,\Delta^2;\mu^2)$, $i=1,2$ in
the light--cone limit $\Delta^2\longrightarrow 0$ with
$\Delta\cdot p$ fixed can be studied by the OPE, schematically
\begin{equation}
\label{OPE} G_i(u,\Delta^2)=  C_i^{(2)} \otimes \phi^{(2)} \,+\,
\Delta^2 \sum C_i^{(4)} \otimes \phi_{i}^{(4)} \, +\,{\cal
O}(\Delta^4),
\end{equation}
where $C_i^{(t)}$ are the coefficient functions and $\phi^{(t)}$
are the pion DA given by vacuum--to--pion matrix elements of
renormalized non-local operators, $t$ refers to twist and the
summation goes over all independent contributions of given twist.
The convolution is defined as
\begin{equation}
C_i \otimes \phi = \int_0^1 dv \,C_i(v,u,\Delta^2,\mu^2,\mu_F^2)
\,\phi(v;\mu_F^2),
\end{equation}
where $u,v$ have the meaning of the light--cone momentum fractions and we have
indicated the dependence on the factorization scale $\mu_F^2$ and other variables.
The Lorentz structures in \eq{NL} are chosen such that the coefficient functions
$C_i$ depend on $\Delta^2$ logarithmically as follows from power counting.

The leading--twist pion DA $\phi_\pi(u) = \phi^{(2)}(u)$ is defined as usual by
\begin{eqnarray}
\label{phi_pi}
\left\langle 0\left \vert \bar{d}(-z) \not\!z  \gamma_{5} [-z,z] u(z) \right\vert
 \pi^+(p)
\right\rangle_{\mu_F^2} &=&
i f_\pi (pz)\, \int_0^1du \,{\rm e}^{-ipz(u-\bar{u})} \phi_{\pi}(u;\mu_F^2)\,,
\end{eqnarray}
where the normalization condition is
\begin{equation}
\int_0^1 du\, \phi_{\pi}(u;\mu_F^2)=1.
\label{phi_pi_norm}
\end{equation}
Here and below $z_\mu$ is the light--like vector, $z^2=0$.
 Although we will usually be using covariant
notation, it is sometimes convenient to refer to light--cone
coordinates  where $p =(p_+,0,0)$ and  $x=(0,x_{-},x_{\perp})$.
The light--cone limit corresponds to $x_\perp\longrightarrow 0$
with $x_-$ fixed such that $z=(0,x_-,0)$.

In the present paper we will consider the leading--twist
coefficient function to all orders in $\alpha_s(\mu^2)$ but
restrict ourselves to the leading order in higher--twist
contributions: $C^{(4)}(u,v) = \delta(u-v)+ {\cal O}(\alpha_s)$.
To this accuracy there is a single twist--four DA contributing to
each of the functions $G_{1,2}$:
\begin{eqnarray}
\label{PT_exp}
G_1(u,\Delta^2)&=&\bigg[1+c_1\alpha_s+c_2\alpha_s^2+\cdots\bigg]\otimes
\phi_{\pi} +\Delta^2\phi_1^{(4)}(u;\mu_F^2)+\cdots\nonumber
\\
G_2(u,\Delta^2)&=&\bigg[\widetilde{c}_1\alpha_s+\widetilde{c}_2\alpha_s^2+\cdots\bigg]\otimes
\phi_{\pi}+\Delta^2\phi_2^{(4)}(u;\mu_F^2)+\cdots.
\end{eqnarray}
In physical terms $\phi^{(4)}_1(u;\mu_F^2)$ and
$\phi^{(4)}_2(u;\mu_F^2)$ correspond to contributions
of valence--quark transverse momenta and of the ``wrong'' components of
the quark spinors, respectively. In the notations  of Ref.~\cite{BF90}
\[
g_1(u)=\phi_1^{(4)}(u) \qquad\qquad{\rm and} \qquad\qquad g_2(u)=\frac{d}{du}\phi_2^{(4)}(u).
\]
For an operator definition, it is convenient to use the concept of
non-local operator {\it off} the light cone \cite{BB89} which is
defined as the generating function (formal Taylor expansion) for
renormalized local operators. For example,
\begin{eqnarray}
\label{exam1}
   \Big[\bar{d}(-x) \not\!x \gamma_{5} [-x,x] u(x)\Big]_{\mu_F^2} &=&
   \sum_k\frac{2^{k+1}}{k!}x_{\nu}x_{\nu_1}\ldots x_{\nu_k}
  \bigg\{
   \Big[
   d \gamma_\nu \stackrel{\leftrightarrow}{D}_{\nu_1}\ldots 
        \stackrel{\leftrightarrow}{D}_{\nu_1} u - \mbox{\rm Traces}
   \Big]_{\mu_F^2}^{\rm twist-2}
\nonumber\\
 &&+ \big[\mbox{\rm Traces}\big]^{\rm twist-4}_{\mu_F^2}
+ \big[\mbox{\rm Traces}\big]^{\rm twist-6}_{\mu_F^2}
+\ldots \bigg\},
\end{eqnarray}
where $\stackrel{\leftrightarrow}{D}_{\nu_1} =
\stackrel{\rightarrow}{D}_{\nu_1}-\stackrel{\leftarrow}{D}_{\nu_1}$
is the covariant derivative, see \cite{BB89,BB91} for details.
Note that the twist expansion of a non-local operator corresponds
to its expansion over irreducible representations of the Lorentz
group. This involves rearrangement of traces and symmetrization
over groups of indices if necessary, {\it cf.} \cite{Lazar}. This
compact notation is widely used in
\cite{BF90,CZreport,BBKT98,BB89,Ball99,BBK02}. In particular, we
have \cite{BF90}
\begin{eqnarray}
\label{off}
&&\langle 0 \vert \Big[\bar{d}(x_2) \gamma_{\nu}
\gamma_{5} [x_2,x_1] u(x_1)\}\Big]_{\mu_F^2}\vert \pi^+(p)
 \rangle \,=\, \nonumber \\ &&\hspace*{+30pt}
= i \,f_\pi p_\nu \int_0^1du \,{\rm e}^{-iupx_1-i\bar{u}px_2}
\left[\phi^{(2)}(u;\mu_F^2)+\Delta^2\phi^{(4)}_1(u;\mu_F^2) +{\cal O}(\Delta^4)\right]
\nonumber\\ &&\hspace*{+40pt}
 + \, i\,f_\pi\Big(\Delta_\nu(p\Delta)-p_{\nu}\Delta^2\Big)
\int_0^1du \,{\rm e}^{-iupx_1-i\bar{u}px_2}
\left[\phi^{(4)}_2(u;\mu_F^2) +{\cal O}(\Delta^2)\right].
\end{eqnarray}
Despite the similar appearance, beyond the tree level \eq{NL} and \eq{off} describe
different objects. As follows from the Taylor expansion \eq{exam1}
the  non--local operator on the left--hand side of \eq{off} is an
analytic function of $\Delta^2$ at $\Delta^2\longrightarrow 0$. It
corresponds to the analytic part of the amplitude in \eq{NL}
\cite{BB91} while the coefficient functions (by definition) take
into account singular contributions. The most striking difference
is that the non-local operator in \eq{off} does not include the
whole twist--two part of  $G_2(u,\Delta^2)$ in~\eq{NL}, which is $\sim
1/\Delta^2$. This contribution was overlooked in \cite{Jeppe}. The
non-local operator also does not include any of the coefficient
functions appearing in~\eq{NL}, which absorb all logarithms $\sim
\ln (-\Delta^2\mu_F^2)$.

Using the EOM the two--particle pion DA
$\phi^{(4)}_1(u;\mu_F^2)$ and $\phi^{(4)}_2(u;\mu_F^2)$ can be
expressed in terms of the Fock components involving an extra gluon
field. Following~\cite{BF90} we define the three--particle pion DA
of twist four
\begin{eqnarray}
\label{Phi}
&&\left\langle 0\left \vert \bar{d}(-z) [-z,vz] \gamma_{\nu} \gamma_{5}
gG_{\mu\rho} (vz) [vz,z] u(z) \right\vert
 \pi^+(p)
\right\rangle \,=\,
  \, \nonumber \\ && \hspace*{+35pt} =\,f_\pi
\int \!{\cal D}\alpha_i\,\,
{\rm e}^{-ipz(\alpha_1-\alpha_2+\alpha_3v)}\bigg\{
\frac{p_{\nu}}{pz}\,(p_{\mu}z_{\rho}-p_{\rho}z_{\mu})\Phi_{\parallel}(\alpha_1,\alpha_2,\alpha_3)
 \nonumber\\ && \hspace*{+60pt}+\left[\,p_{\rho}\left(g_{\mu\nu}-\frac{z_{\mu}p_{\nu}}{pz}\right)
 -p_{\mu}\left(g_{\rho\nu}-\frac{z_{\rho}p_{\nu}}{pz}\right)\right]
\Phi_{\perp}(\alpha_1,\alpha_2,\alpha_3)    \bigg\},
\end{eqnarray}
where the longitudinal momentum fraction of the gluon is
$\alpha_3$ and the integration measure is defined as
\begin{equation}
\label{measure}
   \int\! {\cal D}\alpha_i = \int_0^1d\alpha_1d\alpha_2d\alpha_3\,\delta(1-\alpha_1-\alpha_2-\alpha_3)\,.
\end{equation}
 One obtains \cite{BF90} (see also
Appendix A in \cite{BB98}):
\begin{eqnarray}
\label{using_Phi}
\phi_2^{(4)}(u)&=&\int_0^u dv\int_0^{v}d\alpha_1\int_0^{1-v}d\alpha_2\,
\frac{1}{\alpha_3}
\left[2\Phi_\perp-\Phi_\parallel\right](\alpha_1,\alpha_2,\alpha_3) \nonumber\\      \\  \nonumber
\phi_1^{(4)}(u)+\phi_2^{(4)}(u)&=&\frac12\int_0^ud\alpha_1\int_0^{1-u}d\alpha_2\,
\frac{\bar u \alpha_1-u\alpha_2}{\alpha_3^2}
\left[2\Phi_\perp-\Phi_\parallel\right](\alpha_1,\alpha_2,\alpha_3),
\end{eqnarray}
where $\alpha_3=1-\alpha_1-\alpha_2$.
The derivation of these relations relies on exact operator identities \cite{BB89} which relate
integrals over $v$ of the quark--gluon--antiquark operator in \eq{Phi} to derivatives of the
quark--antiquark operator appearing in \eq{off}.

For completeness, we present here the definitions of the other three--particle
 twist--four pion DA. These distributions do not contribute to \eq{NL} to leading order because
of our specific choice of the (simple) correlation function, but
are important for other applications. We will use these
definitions in Sect.~3 where we construct the renormalon
model.

First of all, there exist another pair of DA that correspond to
the substitution\footnote{We are using the sign conventions for the
$\epsilon^{\mu\rho\alpha\beta}$ tensor and the $\gamma_5$ matrix
from Bjorken and Drell \cite{Bjorken}. In particular, ${\rm
Tr}(\gamma^{\mu}\gamma^{\rho}\gamma^{\alpha}\gamma^{\beta}\gamma_5)
=
 4i \epsilon^{\mu\rho\alpha\beta}$ and $\widetilde\sigma_{\alpha\beta} = i\sigma_{\alpha\beta}\gamma_5$.}
 of $\gamma_{5}G_{\mu\rho}$ by $i\widetilde{G}^{\mu\rho}\equiv \frac{i}2 \epsilon^{\mu \rho
\alpha \beta} G_{\alpha\beta}$ in \eq{Phi}.
To the twist--four accuracy
\begin{eqnarray}
\label{Psi}
&&\langle 0\vert \bar{d}(-z) [-z,vz] \gamma_{\nu}
ig\widetilde{G}_{\mu\rho} (vz) [vz,z] u(z) \vert
 \pi^+(p)
\rangle \,=\,
  \, \nonumber \\ && \hspace*{+35pt} =\,f_\pi
\int \!{\cal D}\alpha_i\,\,
{\rm e}^{-ipz(\alpha_1-\alpha_2+\alpha_3v)}\bigg\{
\frac{p_{\nu}}{pz}\,(p_{\mu}z_{\rho}-p_{\rho}z_{\mu})\Psi_{\parallel}(\alpha_1,\alpha_2,\alpha_3)
 \nonumber\\ && \hspace*{+60pt}+\left[\,p_{\rho}\left(g_{\mu\nu}-\frac{z_{\mu}p_{\nu}}{pz}\right)
 -p_{\mu}\left(g_{\rho\nu}-\frac{z_{\rho}p_{\nu}}{pz}\right)\right]
\Psi_{\perp}(\alpha_1,\alpha_2,\alpha_3)    \bigg\}.
\end{eqnarray}
The functions $\Psi_{\perp,\parallel}$ and  $\Phi_{\perp,\parallel}$
are not independent. In particular $\Psi_{\perp}$ and  $\Phi_{\perp}$ can be obtained as the 
symmetric and the antisymmetric part, respectively,
of a more general DA \cite{Ball99}
\begin{eqnarray}
\label{H}
&&\langle 0 \vert \bar{d}(-z) [-z,vz] \gamma_-\gamma_{\nu}\gamma_+
ig\widetilde{G}_{\mu\rho} (vz) [vz,z] u(z) \vert
 \pi^+(p)
 \rangle \,=\,
  \, \\ && \hspace*{0pt} =\,f_\pi \left[\,p_{\rho}\left(g_{\mu\nu}-\frac{z_{\mu}p_{\nu}}{pz}\right)
 -p_{\mu}\left(g_{\rho\nu}-\frac{z_{\rho}p_{\nu}}{pz}\right)\right]
\int \!{\cal D}\alpha_i\,\,
{\rm e}^{-ipz(\alpha_1-\alpha_2+\alpha_3v)}
H^{\downarrow\uparrow}(\alpha_1,\alpha_2,\alpha_3)\,.
\nonumber
\end{eqnarray}
One obtains \cite{Ball99}
\begin{eqnarray}
    \Psi_\perp(\alpha_1,\alpha_2,\alpha_3)
    = -\frac12\left[H^{\downarrow\uparrow}(\alpha_1,\alpha_2,\alpha_3)
     +H^{\downarrow\uparrow}(\alpha_2,\alpha_1,\alpha_3)\right],
\nonumber\\
    \Phi_\perp(\alpha_1,\alpha_2,\alpha_3)
    = -\frac12\left[H^{\downarrow\uparrow}(\alpha_1,\alpha_2,\alpha_3)
    -H^{\downarrow\uparrow}(\alpha_2,\alpha_1,\alpha_3)\right],
\end{eqnarray}
see also Appendix A in \cite{Belyaev95}.

In addition, we introduce a new three--particle DA $\Xi_{\pi}(\alpha_i)$ as
\begin{eqnarray}
\label{Xi} &&\left\langle 0\left \vert \bar{d}(-z)
\gamma_{\mu}  \gamma_{5} gD^{\alpha}G_{\alpha\beta}(vz)[vz,z]u(z)
\right\vert \pi^+(p)\right\rangle=
\nonumber \\&&\hspace*{60pt}=\,
if_{\pi}p_{\mu}p_{\beta}
\int \!{\cal D}\alpha_i\,\,
{\rm e}^{-ipz(\alpha_1-\alpha_2+v\alpha_3)}\,\Xi_{\pi}(\alpha_1,\alpha_2,\alpha_3).
\end{eqnarray}
The Lorentz structure $p_{\mu}p_{\beta}$ is the only one relevant
at twist four. Thanks to the equation of motion
$D^{\alpha}G^A_{\alpha\beta} = -g\sum_q \bar q t^A \gamma_\beta q$
where the summation goes over all light flavors, $\Xi_{\pi}(\alpha_i)$
can be viewed as describing  either a quark--antiquark--gluon or a
specific four--quark component of the pion: with the
quark--antiquark pair in a color--octet state and at the same
space--time point. This DA was not considered previously because
its conformal expansion  starts with a higher spin $J=5$ (see
below) whereas in \cite{BF90,Ball99} only the terms with $J=3$ and
$J=4$ were included.

There exist further twist--four four--quark operators where all
quark light--cone coordinates are separated, and also twist--four
operators which include two gluons in addition to the quark--antiquark
pair. They give rise to four--particle DA and will not be
considered in this work. Although such operators contribute to
exclusive amplitudes at twist four, they do not appear to leading
order in the flavor expansion ({\it cf.}~\cite{GKRT}) and can be
systematically neglected to our accuracy.

\subsection{Infrared renormalons  in the twist--two coefficient functions\label{ren_calc}}

The scale separation made in \eq{OPE} is arbitrary in two respects. To a
given {\em logarithmic accuracy}\, the separation between coefficient
functions and operator matrix elements is ambiguous. However, the
logarithmic dependence on the
factorization scale $\mu_F^2$ cancels between the coefficient
function $C^{(t=2)}$ and the corresponding DA
$\phi^{(t=2)}$ leaving their product invariant.
This can be phrased through perturbative evolution equations.
To {\em power accuracy}\, the separation between terms of different
twist is ambiguous and  the scale--independence of ``structure functions'' $G$
is only restored in the sum of all twists.

As discussed in the introduction the intuitive way to see this is
to imagine using a cutoff scale $\mu_F^2$ to implement
factorization: this scale serves as an infrared cutoff for the
coefficient functions while it acts as an ultraviolet cutoff for
the DA. The twist--two contribution would then depend on the scale
as $\sim \mu_F^2\Delta^2\phi^{(2)}$. To keep $G$ invariant this
dependence should cancel against a term of the form $\mu_F^2
\phi^{(2)}$ in $\phi_i^{(4)}$. In other words, the twist--four
operator has a quadratic ultraviolet divergence through which it
mixes with the twist--two operator. In dimensional regularization the
power--like cutoff dependence does not occur, but the ambiguity
still persists because the perturbative series develops a factorial
large--order behaviour. The perturbation theory, therefore, diverges
(renormalon divergence)  and its sum is only defined to power--like
accuracy. This is the renormalon ambiguity which we are going to address
now.

\subsubsection{An all--order  calculation\label{IR_calc}}

In order to evaluate the infrared--renormalon ambiguity
in the twist--two coefficient functions
\begin{eqnarray}
C_1^{(2)}(v,u,\Delta^2,\mu^2,\mu_F^2)&=&\delta(u-v)+c_1
\alpha_s(\mu^2)+c_2\alpha_s^2(\mu^2)+\cdots\nonumber \\
C_2^{(2)}(v,u,\Delta^2,\mu^2,\mu_F^2)&=&\widetilde{c}_1
\alpha_s(\mu^2)+\widetilde{c}_2\alpha_s^2(\mu^2)+
\cdots
\end{eqnarray}
we need to calculate them to all orders in the coupling. Of
course, a full all--order calculation cannot be performed.
Instead, as in other applications~\cite{ren_review}, we restrict
ourselves to the perturbative series generated by the
running--coupling effects in one--loop diagrams, i.e. using QCD
coupling at the scale of the gluon virtuality\footnote{This
approximation is sometimes referred to as the large--$\beta_0$
limit and can formally be defined considering first the large
$N_f$ limit with  $\alpha_s N_f$ is fixed, and replacing $N_f$ by
$-6\beta_0$ to recover the non-Abelian contributions.}. A
convenient tool to perform this calculation is the Borel
representation of the running coupling,
\begin{equation}
  \frac{\beta_0\alpha_s(-k^2)}{\pi}=\int_0^\infty dw \,{\rm e}^{\frac53 w}
  \,\left(\frac{\Lambda^2}{-k^2}\right)^{w},
  \label{Borel_def}
\end{equation}
where
$\beta_0$ is the leading--order coefficient
of the $\beta$ function,
\begin{equation}
\frac{d(\alpha_s/\pi)}{d\ln\mu^2}=-\beta_0(\alpha_s/\pi)^2+\ldots
\hspace*{60pt}\beta_0=\frac{11}{12}C_A-\frac16 N_f\,,
\label{beta_0}
\end{equation}
$\Lambda =\Lambda_{\rm QCD}^{\overline {\rm{MS}}}$ and the
exponential factor ${\rm e}^{\frac53 w}$ originates from the
renormalization of the fermion loop in the $\overline {\rm{MS}}$
scheme. This way the calculation reduces to one loop with a
modified gluon propagator
\begin{equation}
\label{propa}
\frac{1}{-k^2}\longrightarrow \frac{(\Lambda^2)^w}{(-k^2)^{1+w}}.
\end{equation}
and the result for the coefficient functions takes the form of a Borel integral
\begin{eqnarray}
C_1(v,u,\Delta^2,\mu^2,\mu_F^2)&=&\delta(u-v)\,+\, \int_0^\infty dw
\,{\rm e}^{\frac53 w} \,B_1(w;v,u,\Delta^2\mu^2,\mu_F^2/\mu^2)
\,\left(-\Delta^2\Lambda^2\right)^{w},\nonumber \\
C_2(v,u,\Delta^2,\mu^2,\mu_F^2)&=&\int_0^\infty dw \,{\rm e}^{\frac53 w}
\,B_2(w;v,u,\Delta^2\mu^2,\mu_F^2/\mu^2)
\,\left(-\Delta^2\Lambda^2\right)^{w}.\label{B}
\end{eqnarray}
The perturbative expansion of $C_i$ can be recovered order by order from the expansion of
their Borel transforms $B_i$ near $w=0$ observing  that for one--loop coupling
\[
\int_0^\infty \,dw \,w^n\left(\Lambda^2/\mu^2\right)^w=\frac{n!}{\left(\ln \mu^2/\Lambda^2\right)^{n+1}}=
n!\left(\alpha_s\beta_0/\pi\right)^{n+1}.
\]
The fixed--sign factorial behavior of the coefficients in this
expansion in high orders manifests itself in that the functions
$B_i$ develop singularities at finite values of $w$ on the
positive real axis (renormalon singularities) rending the
integrals in \eq{B} ill--defined. The imaginary part that
arises when the integration contour is moved above (or below) of
the nearest singularity (to the origin) can be taken as a measure
of the ambiguity of the summation of the perturbative series, see
\cite{ren_review} for details.

\begin{figure}[t]
  \centerline{\epsfig{file=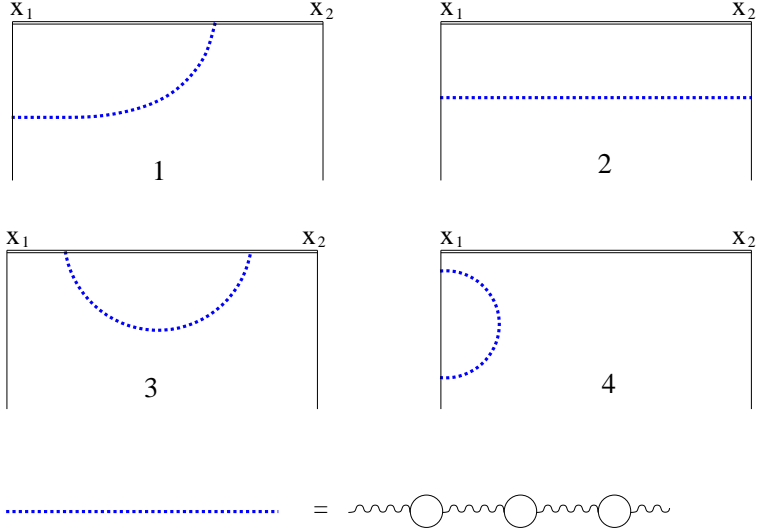,angle=0,width=10cm}}
  \vspace*{20pt}
  \caption{\small One--loop diagrams appearing in the non-local matrix element of
    \eq{NL}.
    The upper double line represents a path--ordered exponential (a Wilson
    line) connecting $x_1$ and $x_2$, while the dashed line represents a
    dressed gluon (dressing is equivalent to summing over any number of
    fermion--loops insertions in the large--$N_f$ limit). }
  \label{diagrams}
\end{figure}

The relevant diagrams are shown in Fig. \ref{diagrams}.
The calculation is presented in detail in Appendix~\ref{IRR}.
We consider quarks on--shell and set the space-time dimension to $d=4$ from the beginning
since $w\not=0$ acts as a regulator. The result for the (not yet renormalized)
coefficient functions reads
\begin{eqnarray}
\label{B1}
   B_1(w;u,v)&=& \frac{C_F}{2\beta_0}\frac{\Gamma(-w)}{\Gamma(1+w)}4^{-w}
  \bigg\{\theta(u>v)\frac{1}{\bar v}\left[(1+w)f_+^{(w)}\left(1-\frac{\bar u}{\bar v}\right)-wf^{(w)}\left(1-\frac{\bar u}{\bar v}\right)\right]
\nonumber\\&&{}\hspace*{75pt}+\,
\theta(u<v)\frac{1}{v}\left[(1+w)f_+^{(w)}\left(1-\frac{u}{v}\right)-wf^{(w)}\left(1-\frac{u}{v}\right)\right]
\nonumber\\&&{}\hspace*{75pt}+\,
\frac{1-w}{1+w}  \left[\theta(u>v)\left(\frac{\bar u}{\bar v}\right)^{w+1}+
        \theta(u<v)\left(\frac{ u}{ v}\right)^{w+1}\right]
\nonumber\\&&\hspace*{75pt}{}
-\,\frac{1}{1-2w}\delta(u-v)\bigg\}
\end{eqnarray}
and
\begin{eqnarray}
\label{B2}
   B_2(w;u,v)&=& \frac{C_F}{\beta_0}\frac{\Gamma(1-w)}{\Gamma(2+w)}4^{-w}
  \left[\theta(u>v)\left(\frac{\bar u}{\bar v}\right)^{w+1}+
        \theta(v>u)\left(\frac{ u}{ v}\right)^{w+1}\right],
\end{eqnarray}
where we used a short--hand notation
\begin{eqnarray}
  \label{fw}
  f^{(w)}(\beta)
  &=&\frac{\beta^{2w-1} (1-\beta)^{1+w}}{1+w} \,_2\!
  F_1(2w,w+1,2+w;1-\beta)\,.
\end{eqnarray}
In particular
\begin{eqnarray}
\label{fvalues}
 f^{(w=0)}(\beta) &=& \frac{1-\beta}{\beta},
\nonumber\\
 f^{(w=1)}(\beta) &=&  1-\beta+\beta\ln\beta\,.
\end{eqnarray}
The ``$+$" prescription is defined as usual by:
\begin{equation}
  \label{def_plus}
  f^{(w)}_+(\beta)=f^{(w)}(\beta)-\delta(\beta)\int_0^1 d\tilde{\beta}
  f^{(w)}(\tilde{\beta}).
\end{equation}
The terms $\sim f^{(w)}$ in \eq{B1} correspond, in Feynman gauge, to the contribution
of the vertex correction (Diagram 1) in Fig.~\ref{diagrams} and its symmetric
counterpart, the contribution in the third line
in \eq{B1} and the entire \eq{B2} originate from the box diagram
(Diagram 2) and the remaining term $\sim 1/(1-2w)$ in the last line of \eq{B1}
stands for the self--energy insertion in the Wilson line (Diagram 3).

\subsubsection{Singularities of the Borel transform\label{Borel_sing}}

The answer for $B_1(w;u,v)$ in \eq{B1} has a simple pole at $w=0$. This singularity
is expected \cite{ren_review} and has to be removed by the subtraction of ultraviolet
(UV) logarithmic divergences due to the wave--function renormalization of the quark
fields, and of infrared (IR) logarithmic divergences that correspond to the
renormalization of the leading--twist DA. To this end, note that vanishing of
the Diagram 4 in Fig.~\ref{diagrams} is a result of an exact cancellation between
IR and UV divergent integrals. Upon introducing a scale which regulates one singularity,
the other one will appear as a $1/w$ pole. Schematically
\begin{equation}
\label{UV-IR}
  B_1(w;u,v)^{\rm Diagram-4} = \frac{C_F}{4\beta_0}
        \left( \left[\frac{1}{w}\right]_{\UV} -\left[\frac{1}{w}\right]_{\IR}\right).
\end{equation}
In addition to renormalizing Diagram 4, ultraviolet subtraction removes the $1/w$
pole of Diagram 3 in Fig.~\ref{diagrams}.

Having performed the UV renormalization the remaining $1/w$ singularities from Diagrams 1, 2
and 4 are removed upon performing IR factorization. The counter-term
which is to be subtracted from the DA $\phi^{(2)}$ and added to
the coefficient function $B_1(w;u,v)$ is
\begin{equation}
\label{B1reg}
  B_1(w;u,v)^{\rm c.t.} = \frac{C_F}{2\beta_0} \frac{K(u,v) + {\cal O}(w)}{w}
\end{equation}
where $K(u,v)$ is obtained as the limit at $w\to0$ of the expression in the curly brackets in \eq{B1}
and adding the IR--pole contribution in \eq{UV-IR}:
\begin{equation}
K(u,v)\equiv \left[\frac{v}{u}\left(1+\frac1{u-v}\right) \theta(u>v)+
\frac{1-v}{1-u}\left(1+\frac1{v-u}\right) \theta(u<v)\right]_+.
\end{equation}
$K(u,v)$ can easily be identified as the leading--order
Brodsky--Lepage kernel~\cite{ER,BL} 
controlling the $\mu_F^2$--evolution of the pion DA~$\phi_{\pi}(u;\mu_F^2)$:
\begin{equation}
\frac{d\phi_{\pi}(v;\mu_F^2)}{d\ln\mu_F^2}=\frac{C_F\alpha_s}{2\pi}\int_0^1 du \, K(u,v)\,
\phi_{\pi}(u;\mu_F^2)\,.
\end{equation}
The terms ${\cal O}(w)$ in \eq{B1reg} correspond to
scheme--dependent higher--order contributions to the
Brodsky--Lepage evolution kernel. In MS--like
schemes the kernel has an expansion in $\alpha_s$ with a finite
radius of convergence, see \cite{Mikhailov98} for explicit
expressions in the large--$\beta_0$ limit. In other words in such
schemes the infrared counter-term is free of Borel singularities.

To summarize, the subtraction of logarithmic UV and IR divergences
removes the $1/w$ singularity in the Borel transformed coefficient
functions in much the same way as poles $1/(d-4)$ are subtracted
in renormalized amplitudes when using  dimensional regularization.
For the subsequent discussion it is important that in MS-like
schemes the subtracted terms are analytic functions of the Borel
variable $w$ and do not influence the structure of singularities
of $B_i$ at $w>0$ which we are going to address now. For this
reason we can work with non-subtracted amplitudes in what follows.

{}First,  we note the presence of a Borel singularity at $w=1/2$
in the last term in \eq{B1} which comes from the self-energy
insertion in the Wilson line and reflects a linear divergence in
the UV region (UV renormalon). Such singularities are well--known
in the context of the heavy--quark effective theory \cite{BB94} in
which case they reflect ambiguities in the non-perturbative
definition of the heavy--quark mass \cite{BSUV94,BB94}. In our
case, this singularity is an artifact of choosing an
oversimplified ``exclusive process'' in \eq{NL} where a dynamical
quark propagating between the points $x_1$ and $x_2$ is replaced
by a  path--ordered exponential. It has nothing to do with the
twist expansion and will not appear in realistic  physics
applications. Therefore we will not consider this singularity
further in this paper.

The remaining singularities at positive integer $w_0=1,2,\ldots$
have IR origin and are called IR renormalons. They obstruct the
Borel integration in \eq{B} and render the sum of perturbation
theory ambiguous to power accuracy $\sim
(\Lambda^2\Delta^2)^{w_0}$. Hence we concentrate on the IR
renormalon with $w_0=1$ which is the closest one to the origin and
the only one relevant for the calculation to the twist--four
accuracy $\sim \Delta^2$. For definiteness, we choose the residue
at the $w=1$ singularity times $\pi=3.14\ldots$ as a measure of
the ambiguity of the Borel integral (\ref{B}), i.e.
\begin{equation}
\label{ambi}
 \delta_{\IR}\left\{\frac{f(w)}{1-w}\right\} = - \pi f(w=1)\,.
\end{equation}
Using \eq{B1} and \eq{B2} we obtain the ambiguity of the twist--two approximation for the
amplitudes $G_1$ and $G_2$
\begin{eqnarray}
\label{ambi2}
 \delta_{\IR}\left\{C_1\otimes\phi_\pi\right\} &=&
-c\Lambda^2\Delta^2\!\!\int_0^1\!\!dv\,\phi_{\pi}(v)
\left\{\frac{1}{v}\,f^{(1)}\left(1\!-\!\frac{u}{v}\right) \theta(v>u)\right.
\left.+\,\frac{1}{\bar v}\,f^{(1)}\left(1\!-\!\frac{\bar u}{\bar v}\right) \theta(v<u)\right\}  \nonumber \\
 &=&-c\Lambda^2\Delta^2\!\!
\int_0^1\!\!dv\,\phi_{\pi}(v)\bigg\{
 \frac{1}{v^2}\,\bigg[u+(v-u)\ln\left(1-\frac{u}{v}\right)\bigg] \theta(v>u)\nonumber
  \\ &&\hspace*{90pt} +\,
 \frac{1}{\bar v^2}\,\bigg[\bar u+(u-v)\ln\left(1-\frac{\bar u}{\bar v}\right)\bigg] \theta(v<u)\bigg\},
 \nonumber \\ 
\delta_{\IR}\left\{C_2\otimes\phi_\pi\right\} &=&c\Lambda^2\Delta^2
\int_0^1dv\,\phi_{\pi}(v)\left\{
\left(\frac{u}{v}\right)^2 \theta(v>u)+
\left(\frac{\bar u}{\bar v}\right)^2 \theta(v<u)\right\},
\end{eqnarray}
respectively, where we used \eq{fvalues} and where the overall normalization
\begin{equation}
c\equiv \frac{\pi C_F}{8\beta_0}{\rm e}^{\frac53}\simeq 0.7
\label{c}
\end{equation}
corresponds to the convention in \eq{ambi}. The given number is for $N_F=3$.
We are going to demonstrate that this ambiguity is exactly canceled by the UV renormalon
ambiguity in the twist--four DA, which is reminiscent of quadratic UV divergence of the
contributing operators.

\subsection{UV renormalons in higher--twist operators and cancellation of ambiguities\label{tw4_3particle}}

\begin{figure}[t]
  \centerline{\epsfig{file=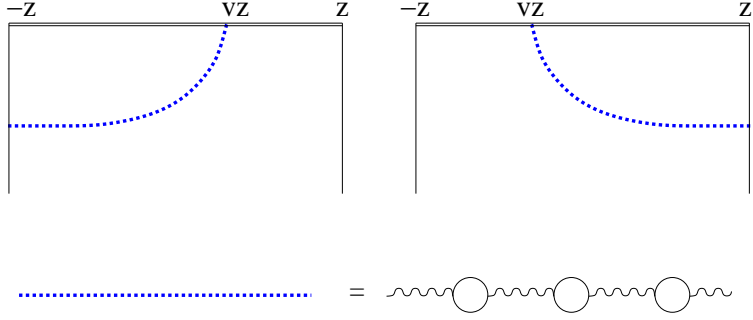,angle=0,width=10cm}}
  \vspace*{20pt}
  \caption{\small One--loop diagrams appearing in the non-local matrix element
    of quark--antiquark--gluon light--cone operators.
    The upper double line represents a path--ordered exponential (a Wilson
    line) along the light cone
    connecting between $-z$ and $vz$ and between 
    $vz$ and $z$ while the dashed line represents a
    dressed gluon (dressing is equivalent to summing over any number of
    fermion--loops insertions in the large--$N_f$ limit). }
  \label{diagrams_tw4} 
\end{figure}

In order to reveal the UV--renormalon divergence in the twist--four
DA we consider the perturbative series generated by
running--coupling effects in the matrix element of a generic
quark--antiquark--gluon operator
\begin{equation}
\bar{d}(-z) [-z,vz] \Gamma
gG_{\alpha\beta} (vz) [vz,z] u(z),
\label{qGq_op}
\end{equation}
sandwiched between quarks states with momenta $q_1$ and $q_2$ with
$q_i^2 \not= 0$. Here $\Gamma$ stands for an arbitrary Dirac
structure and $z^2=0$. The relevant diagrams are shown in
Fig.~\ref{diagrams_tw4}.

One obtains
\footnote{In the following for brevity we do not show the Wilson lines in the non-local operators.}
\begin{eqnarray}
\label{UV_div_Gamma}
\left\langle q_2\left\vert \bar{d}(-z) \Gamma gG_{\alpha\beta} (vz) u(z)\right\vert q_1 \right\rangle
 &=&
 {\rm e}^{-i(q_1+q_2)z}\,\left(g_{\lambda\alpha}g_{\beta}^{\mu}
 -g_{\lambda\beta}g_{\alpha}^{\mu} \right)\frac{4\pi^2 C_F}{\beta_0}\,
 \,\int_0^{\infty}\!\! dw\,{\rm e}^{\frac53 w} (-\Lambda^2)^{w}\,
\nonumber  \\ &&{}\hspace*{-3cm}\times
 \Big[\bar{d}_{q_2} \gamma_{\mu}\gamma_{\rho}\Gamma u_{q_1}\,
 I^{\lambda\rho}(q_2,(1+v)z)
 -\bar{d}_{q_2}\Gamma\gamma_{\rho}\gamma_{\mu} u_{q_1}\,
 I^{\lambda\rho}(q_1,(1-v)z)\Big],
\end{eqnarray}
where $\bar{d}_{q_2}$ and $u_{q_1}$ are quark spinors and
$I^{\lambda\rho}(q,z)$ stands for the momentum integral
\begin{equation}
I^{\lambda\rho}(q,z)=\int\frac{d^4k}{(2\pi)^4}\frac{k^{\lambda}
(q+k)^{\rho}}{\left(k^2 \right)^{(1+w)}(q+k)^2}\,{\rm
e}^{-ikz}
\end{equation}
which is UV divergent at $w=1$.
Performing the integral and extracting the $w=1$ pole, we obtain:
\begin{eqnarray}
\left. I^{\lambda\rho}(q,z)\right\vert_{w=1}&=&\frac{-i}{32\pi^2(1-w)}\,
\int_0^1\!da\,a\,{\rm e}^{iqz(1-a)}
\nonumber\\ & &{}\times
\bigg[iq^{\lambda}z^{\rho}(1-a)-iq^{\rho}z^{\lambda}a
+g^{\lambda\rho}+\frac12 a(1-a)q^2z^{\lambda}z^{\rho}\bigg].
\end{eqnarray}
As a consequence  of the $w=1$ singularity the Borel integral in \eq{UV_div_Gamma} is ill--defined. Using \eq{ambi}
to quantify the ambiguity and specifying for the relevant Dirac and Lorentz structures the final result can be brought
to a form of operator relations ({\it cf.}\ \cite{ren_review,Jeppe})
\begin{eqnarray}
\label{UV_div}
\hspace*{-5pt} \delta_{\UV}\left\{
z^{\mu}\,\bar{d}(-z) \gamma^{\nu} \gamma_{5} gG_{\mu\nu} (vz) u(z) \right\}
 &=& 2i\,c\Lambda^2 \int_0^1\! da\, (1-a)\Big[\bar d(-y)\zsl\gamma_5 u(z)-
  \bar d(-z)\zsl\gamma_5 u(\widetilde{y})  \Big],
         \nonumber       \\
\hspace*{-5pt}  \delta_{\UV}\left\{z^{\mu}\,\bar{d}(-z) \zsl \gamma_{5}
gG_{\mu\rho} (vz) u(z) \right\}
 &=& -2i\,c\Lambda^2\, z_{\rho}\int_0^1 da \,a\,\Big[\bar d(-y)\zsl\gamma_5 u(z)-
  \bar d(-z)\zsl\gamma_5 u(\widetilde{y})
 \Big],\nonumber \\
\hspace*{-5pt} \delta_{\UV}\left\{z^{\mu}\bar{d}(-z) \gamma^{\nu}
ig\widetilde{G}_{\mu\nu} (vz) u(z) \right\}
 &=& 2i\,c\Lambda^2 \int_0^1\! da\, (1+a)\,\Big[\bar d(-y)\zsl\gamma_5 u(z)+
  \bar d(-z)\zsl\gamma_5 u(\widetilde{y})
 \Big],            \nonumber       \\
\hspace*{-5pt} \delta_{\UV}\left\{z^{\mu}\bar{d}(-z) \zsl
ig\widetilde{G}_{\mu\rho} (vz) u(z) \right\}
 &=&2i\,c\Lambda^2 z_{\rho}\int_0^1 da \,a\,\Big[\bar d(-y)\zsl\gamma_5 u(z)+
  \bar d(-z)\zsl\gamma_5 u(\widetilde{y})
 \Big],
\nonumber\\
\end{eqnarray}
where $y\equiv z(1-(1+v)(1-a))$  and $\widetilde{y}\equiv
z(1-(1-v)(1-a))$, and $c$ is the constant defined in \eq{c}.
In order to arrive at these expressions  we performed integration by parts over $a$ in order to remove
factors of $q\cdot z$ and then converted the results into operator
relations in configuration space.
The relations in \eq{UV_div} can be viewed as the
mixing under renormalization between the twist--four
quark--gluon--antiquark operators and the twist--two
quark--antiquark operator and the first two of them  were derived in \cite{Jeppe}.
In a similar manner --- see Appendix \ref{DG_eq_of_motion} --- we obtain one more operator relation
\begin{eqnarray}
\label{DG}
\lefteqn{\hspace*{-50pt}\delta_{\UV}\left\{\bar d(-z)z^{\beta}
gD^{\alpha}G_{\alpha\beta}(vz)\zsl\gamma_5 u(z)
\right\}=}
\nonumber\\
&=& -4ic\Lambda^2\int_0^1da\,
a\frac{d}{da}\left[\frac{1}{1+v}\,\bar d(-y)\zsl\gamma_5
u(z)\,+\,\frac{1}{1-v}\bar d(-z)\zsl\gamma_5 u(\widetilde{y}) \right].
\end{eqnarray}
{}Taking matrix elements of Eqs.~(\ref{UV_div}) and (\ref{DG}) between the vacuum and the pion state
we derive UV ambiguities of three--particle pion DA defined in Sect.~2.1 in terms of the leading--twist pion DA:
\begin{eqnarray}
\label{three_part_dist_amb}
\delta_{\UV}\left\{\Phi_{\perp}(\alpha_1,\alpha_2,\alpha_3)\right\}&=&
c\Lambda^2\left[
\frac{\phi_{\pi}(\alpha_1)}{1-\alpha_1}-\frac{\phi_{\pi}(\alpha_2)}{1-\alpha_2}
\right],
\nonumber \\
\delta_{\UV}\left\{\Phi_{\parallel}(\alpha_1,\alpha_2,\alpha_3)\right\}&=&2c\Lambda^2\left[
\frac{\alpha_2\phi_{\pi}(\alpha_1)}{(1-\alpha_1)^2}
-\frac{\alpha_1\phi_{\pi}(\alpha_2)}{(1-\alpha_2)^2}
\right],
\nonumber\\
\delta_{\UV}\left\{\Psi_{\perp}(\alpha_1,\alpha_2,\alpha_3)\right\}&=&
c\Lambda^2\left[
\frac{\phi_{\pi}(\alpha_1)}{1-\alpha_1}+\frac{\phi_{\pi}(\alpha_2)}{1-\alpha_2}
\right],
\nonumber \\
\delta_{\UV}\left\{\Psi_{\parallel}(\alpha_1,\alpha_2,\alpha_3)\right\}&=&-2c\Lambda^2\left[
\frac{\alpha_2\phi_{\pi}(\alpha_1)}{(1-\alpha_1)^2}
+\frac{\alpha_1\phi_{\pi}(\alpha_2)}{(1-\alpha_2)^2}
\right],
\nonumber\\
\delta_{\UV}\left\{\Xi_{\pi}(\alpha_1,\alpha_2,\alpha_3)\right\}&=&
-4c\Lambda^2
\left[\frac{\alpha_2\,\phi_{\pi}(\alpha_1)}{1-\alpha_1}
-\frac{\alpha_1\,\phi_{\pi}(\alpha_2)}{1-\alpha_2}\right],
\end{eqnarray}
where we used the symmetry property $\phi_\pi(\alpha) = \phi_\pi(1-\alpha)$ to arrive at
the given expressions.

Last but not least, we use EOM in \eq{using_Phi} to calculate the
UV--renormalon ambiguity in the two--particle twist--four pion DA
$\phi_{1,2}^{(4)}$ and find that it coincides identically with the
IR renormalon ambiguity of the twist--two result in \eq{ambi2} but
has opposite sign. It follows that the ``structure functions''
$G_i$ are unambiguous to the twist--four accuracy
\begin{eqnarray}
\label{cancel}
 \delta\left\{G_1\right\} &=& \delta_{\IR}\left\{C_1\otimes\phi_\pi\right\}
                              + \Delta^2  \delta_{\UV}\left\{\phi_1^{(4)}\right\} =0\,,
\nonumber\\
  \delta\left\{G_2\right\} &=& \delta_{\IR}\left\{C_2\otimes\phi_\pi\right\}
                              + \Delta^2  \delta_{\UV}\left\{\phi_2^{(4)}\right\} =0\,,
\end{eqnarray}
as expected. For the cancellation to hold, it is important that both the leading--twist
coefficient functions and the matrix elements of higher--twist operators are calculated
using the same regularization and renormalization prescription.
This can be seen as a consistency check for the OPE analysis.

\section{Renormalon model for twist--four DA of the pion\label{pion_model}}
\setcounter{equation}{0}

The UV--renormalon ambiguities in the twist--four DA should be viewed as indicative
of the size and the momentum--fraction dependence of ``genuine'' non-perturbative effects.
We define the renormalon model for twist--four DA of the pion by taking the
functional form of the corresponding UV--renormalon ambiguities, replacing the overall normalization
constant $c\Lambda^2$ by a suitable non-perturbative parameter. The crucial observation is
that although the absolute normalization of the renormalon ambiguity in \eq{ambi} is essentially {\it ad hoc},\
the relative normalization for the different DA in \eq{three_part_dist_amb} is meaningful since
the running--coupling  calculation satisfies all
constraints imposed  by Lorentz symmetry and the EOM. 
Therefore, the renormalon model for the entire set of twist--four DA of the pion 
has just one free parameter: the overall normalization. 
This parameter can be related to the matrix element of the local operator
\begin{equation}
\label{delta}
   \langle 0 \vert \bar{d} \gamma_{\nu}
ig\widetilde{G}_{\mu\rho}u \vert \pi^+(p) \rangle \,=\,\frac13 f_\pi \delta^2
   [p_\rho g_{\mu\nu}-p_\mu g_{\rho\nu}]\,,\qquad \delta^2 \simeq 0.2~\mbox{GeV}^2,
\end{equation}
where the number comes from QCD sum rules \cite{UseAndMisuse}. The UV--renormalon ambiguity
of the same matrix element is, on the other hand
\begin{equation}
\label{delta1}
  \delta_{\UV}\left\{\langle 0\vert \bar{d} \gamma_{\nu}
ig\widetilde{G}_{\mu\rho}u \vert \pi^+(p) \rangle\right\} \,=\,2f_\pi c\Lambda^2
   [p_\rho g_{\mu\nu}-p_\mu g_{\rho\nu}]\,,
\end{equation}
so in \eq{three_part_dist_amb} we replace
\begin{equation}
  c\Lambda^2 \to \frac16 \delta^2 \simeq (180~\mbox{\rm MeV})^2
\end{equation}
and end up with the set of three--particle DA
\begin{eqnarray}
\label{RMthree}
\Phi_{\perp}(\alpha_1,\alpha_2,\alpha_3)&=&
\frac{\delta^2}{6}\left[
\frac{\phi_{\pi}(\alpha_1)}{1-\alpha_1}-\frac{\phi_{\pi}(\alpha_2)}{1-\alpha_2}
\right],
\nonumber \\
\Phi_{\parallel}(\alpha_1,\alpha_2,\alpha_3)&=&\frac{\delta^2}{3}\left[
\frac{\alpha_2\phi_{\pi}(\alpha_1)}{(1-\alpha_1)^2}
-\frac{\alpha_1\phi_{\pi}(\alpha_2)}{(1-\alpha_2)^2}
\right],
\nonumber\\
\Psi_{\perp}(\alpha_1,\alpha_2,\alpha_3)&=&
\frac{\delta^2}{6}\left[
\frac{\phi_{\pi}(\alpha_1)}{1-\alpha_1}+\frac{\phi_{\pi}(\alpha_2)}{1-\alpha_2}
\right],
\nonumber \\
\Psi_{\parallel}(\alpha_1,\alpha_2,\alpha_3)&=&-\frac{\delta^2}{3}\left[
\frac{\alpha_2\phi_{\pi}(\alpha_1)}{(1-\alpha_1)^2}
+\frac{\alpha_1\phi_{\pi}(\alpha_2)}{(1-\alpha_2)^2}
\right],
\nonumber\\
\Xi_{\pi}(\alpha_1,\alpha_2,\alpha_3)&=&
-\frac{2\delta^2}{3}
\left[\frac{\alpha_2\,\phi_{\pi}(\alpha_1)}{1-\alpha_1}
-\frac{\alpha_1\,\phi_{\pi}(\alpha_2)}{1-\alpha_2}\right].
\end{eqnarray}
Note that we have made the same replacement in all five DA. It
must be so because these DA are all related --- see Appendix
\ref{DG_eq_of_motion}. Using the EOM,
\eq{RMthree}, ({\it cf.}\ \eq{ambi2}),
\begin{eqnarray}
\label{RMtwo}
 \phi_1^{(4)}(u)
 &=&\frac{\delta^2}{6}\!\!\int_0^1\!\!dv\,\phi_{\pi}(v)\bigg\{
 \frac{1}{v^2}\,\bigg[u+(v-u)\ln\left(1-\frac{u}{v}\right)\bigg] \theta(v>u)\nonumber
  \\ &&\hspace*{90pt} +\,
 \frac{1}{\bar v^2}\,\bigg[\bar u+(u-v)\ln\left(1-\frac{\bar u}{\bar v}\right)\bigg] \theta(v<u)\bigg\},
 \nonumber \\ 
 \phi_2^{(4)}(u)
&=&-\frac{\delta^2}{6}\int_0^1dv\,\phi_{\pi}(v)\left\{
\left(\frac{u}{v}\right)^2 \theta(v>u)+
\left(\frac{\bar u}{\bar v}\right)^2 \theta(v<u)\right\}
\end{eqnarray}
which completes the calculation.

The given expressions are valid for an arbitrary leading--twist pion DA $\phi_\pi(u)$. For practical applications
it may be worthwhile to choose the asymptotic expression
\begin{equation}
\phi_{\pi}(u)=6u(1-u)
\label{asympt}
\end{equation}
which is known to provide one with a reasonable accuracy \cite{CLEO}. With this choice
\begin{eqnarray}
\label{RMthreeAs}
\Phi_{\perp}(\alpha_1,\alpha_2,\alpha_3)&=&
 \delta^2[\alpha_1-\alpha_2]\,,
\nonumber \\
\Phi_{\parallel}(\alpha_1,\alpha_2,\alpha_3)&=& 2\delta^2
\alpha_1\alpha_2\left[\frac{1}{1-\alpha_1}-\frac{1}{1-\alpha_2}\right]\,,
\nonumber\\
\Psi_{\perp}(\alpha_1,\alpha_2,\alpha_3)&=&
 \delta^2 \left[\alpha_1+\alpha_2\right]\,,
\nonumber \\
\Psi_{\parallel}(\alpha_1,\alpha_2,\alpha_3)&=&- 2 \delta^2
\alpha_1\alpha_2\left[\frac{1}{1-\alpha_1}+\frac{1}{1-\alpha_2}\right]\,,
\nonumber\\
\Xi_{\pi}(\alpha_1,\alpha_2,\alpha_3)&=& 0\,.
\end{eqnarray}
Note that $\Xi_{\pi}(\alpha_1,\alpha_2,\alpha_3)$ is vanishing in this approximation, but in general it is 
not zero and has to be taken into account if corrections to the asymptotic pion DA are included.
 
{}For the two particle DA one gets \cite{Jeppe}
\begin{eqnarray}
\label{2_particle_model}
\phi_1^{(4)}(u)&=&\delta^2\bigg\{\bar u \bigg[\ln(\bar u) - \mbox{\rm Li}_2 (\bar u)\bigg]
+u\bigg[\ln (u)- \mbox{\rm Li}_2( u )\bigg]
-u\bar u+\frac{\pi^2}6\bigg\}, \nonumber \\
\phi_2^{(4)}(u)&=&\delta^2\,\bigg\{u^2\ln(u) +\bar u^2\ln(\bar u)+u\bar u\bigg\},
\end{eqnarray}
where ${\rm Li}_a(x)\equiv \sum_{n=1}^{\infty}x^n/n^a$.
In the next section we shall compare these expressions with the model of \cite{BF90,Ball99} based on
conformal expansion.

\section{Conformal expansion\label{expansion}}
\setcounter{equation}{0}

\subsection{General formalism}
{}For fields ``living'' on the light--cone the conformal transformations reduce to the three--parameter group
$SL(2,\mathbb{R})$ with the algebra of hyperbolic rotations, see \cite{conformal} for a review.
The conformal transformations for quantum fields are governed by their conformal spin which is defined
as
\begin{equation}
\label{cspin}
  j =\frac12(s+\ell)\,,
\end{equation}
where $\ell$ is the scaling dimension, $\ell =3/2$ for quarks and $\ell=1$ for gluons,  and $s$ is the spin
projection on the light cone. For Dirac spinors (quarks) the different spin components can be separated
with the help of projection operators $q_{\pm}=\Pi_{\pm}q$ where
\begin{equation}
\Pi_{+}=\frac12 \gamma{-}\gamma{+},\qquad\qquad \Pi_{-}=\frac12 \gamma{+}\gamma{-},\qquad\qquad
\Pi_{+}+\Pi_{-} =1\,.
\end{equation}
The $q_{+}$ and $q_{-}$ fields have spin projections $s=1/2$ and $s=-1/2$, and hence different conformal spins
$j=1$ and $j=1/2$, respectively.  Similarly, the gluon field strength has to be decomposed
in different spin components: $G_{+\perp}$ has spin projection $s=1$ and conformal spin $j=3/2$;
$G_{\perp\perp}$ and $G_{+-}$ both have $s=0$ and $j=1$; finally
$G_{-\perp}$ has $s=-1$ and $j=1/2$.

Conformal expansion of DA presents an example of the classical
problem of spin summation in quantum mechanics, with the only
difference being that the total conformal spin $J$ of the
multi-parton system is always larger or equal to the sum of spins
of constituents:
\begin{equation}
\label{Jspin}
  J = J_{\rm min} +N\,, \qquad J_{\rm min} =  j_1 +\ldots + j_k\,.
\end{equation}
The integer number $N$ can be identified with the total number of covariant derivatives $D_+$
in the corresponding local operator. Adding a derivative increases the conformal spin by one unit
and does not change the twist of the operator, defined as dimension minus spin.
The generic conformal expansion for two--particle DA has
the form \cite{Makeenko81}
\begin{eqnarray}
\label{two-expand}
\phi(u)=
\frac{\Gamma(2j_1+2j_2)}{\Gamma(2j_1)\Gamma(2j_2)}\,
 u^{2j_1-1}\,(1-u)^{2j_2-1}\,
\sum_{N=0}^{\infty} \,c_{J}\,P^{(2j_1-1,2j_2-1)}_{N}(2u-1)\,,
\end{eqnarray}
where $P_N^{(\alpha,\beta)}[x]$  are Jacobi polynomials \cite{BE}, $u$ and $1-u$ stand for the momentum
fractions of the parton with spin $j_1$ and $j_2$, respectively,  and the coefficients $c_J$
correspond to the contribution of the total conformal spin $J=j_1+j_2+N$.
The factor in front of the sum is called the asymptotic distribution amplitude.

{}For three partons, a generic DA can be written as a double sum
\begin{eqnarray}
\label{three-expand}
\Phi(\alpha_i)=
\frac{\Gamma(2j_1+2j_2+2j_3)}{\Gamma(2j_1)\Gamma(2j_2)\Gamma(2j_3)}\,
\alpha_1^{2j_1-1}\,\alpha_2^{2j_2-1}\,\alpha_3^{2j_3-1}\,
\sum_{N=0}^{\infty} \, \sum_{n=0}^{N}\,C_{Jj}\,
Y_{Jj}^{(12)3}(\alpha_i)\,,
\end{eqnarray}
where \cite{BDKM99,BKM01}
\begin{eqnarray}
\label{conf_basis}
\!\!\!Y_{Jj}^{(12)3}(\alpha_i)=(1-\alpha_3)^{j-j_1-j_2}\,
P_{J-j-j_3}^{(2j_3-1,2j-1)}[1-2\alpha_3]\,
P_{j-j_1-j_2}^{(2j_1-1,2j_2-1)}\left[\frac{\alpha_2-\alpha_1}{1-\alpha_3}\right]
\end{eqnarray}
are the basis functions \cite{conformal} corresponding to the total conformal spin $J=j_1+j_2+j_3+N$ and
the fixed conformal spin of the $(1,2)$--parton pair $j=j_1+j_2+n$.
The functions
$Y_{Jj}^{(12)3}(\alpha_i)$ form a complete basis and  are mutually orthogonal
with respect to the conformal
scalar product:
\begin{equation}
\label{scalprod}
\int_0^1 {\cal D}\alpha_i \,\,
\alpha_1^{2j_1-1} \alpha_2^{2j_2-1} \alpha_3^{2j_3-1}\,
Y_{Jj}^{(12)3}(\alpha_i) Y_{J'j'}^{(12)3}(\alpha_i) = {\cal
N}_{Jj}\,\delta_{JJ'}\,\delta_{jj'}\,,
\end{equation}
where
\begin{eqnarray}
\label{normY}
{\cal N}_{Jj}
&=&
\frac{\Gamma(j\!+\!j_{1}\!-\!j_{2})\Gamma(j\!-\!j_{1}\!+\!j_{2})}
{\Gamma(j\!-\!j_1\!-\!j_2\!+\!1)\Gamma(j\!+\!j_1\!+\!j_2\!-\!1) (2j\!-\!1)}\,
\frac{\Gamma(J\!-\!j\!+\!j_3)\Gamma(J\!+\!j\!-\!j_3)}
{\Gamma(J\!-\!j\!-\!j_3\!+\!1)\Gamma(J\!+\!j\!+\!j_3\!-\!1)(2J\!-\!1)}\,.
\nonumber\\[-5pt]
\end{eqnarray}
Using this orthogonality property the coefficients in the expansion of any DA can be obtained by projection:
\begin{eqnarray}
\label{C_Jj}
C_{Jj}&=&
\frac{\Gamma(2j_1)\Gamma(2j_2)\Gamma(2j_3)}{\Gamma(2j_1+2j_2+2j_3)} \,\frac{1}{{\cal N}_{Jj}}
  \,  \int_0^1 {\cal D}\alpha_i\, \Phi(\alpha_i)
\,Y_{Jj}^{(12)3}(\alpha_i).
\end{eqnarray}
The use of conformal symmetry is that, for a given twist,
only the coefficients $C_{J,j}$ with $j_1+j_2 \le j \le J-j_3$ in \eq{three-expand} and $c_J$ in \eq{two-expand}
for the same value of the {\it total spin}\  $J$
can be related by EOM and/or renormalization group evolution to the
leading logarithmic accuracy. It follows that any parametrization of DA based on a truncated
conformal expansion is consistent with the EOM and is preserved by evolution.

\subsection{The two lowest orders}
The relevant light--cone projections of the twist--four light--cone operators corresponding to the
three--particle pion  DA are
\begin{eqnarray}
\label{conf_op}
{\cal H}^{\downarrow\uparrow}_\perp&=&
\bar{d}_{-}(-z)\gamma^{\perp}\gamma_5 g G_{\perp +}(vz) u_{+}(z)\,,
\nonumber \\
{\cal O}_{\parallel}&=&\bar{d}_{+}(-z)\gamma_{+}\gamma_5 g G_{+-}(vz) u_{+}(z)\,,
\nonumber \\
\widetilde {\cal O}_{\parallel}&=&\bar{d}_{+}(-z)\gamma_{+} i g \widetilde G_{+-}(vz) u_{+}(z)\,,
\nonumber \\
{\cal D} &=& \bar{d}_{+}(-z)\gamma_{+}\gamma_5 g D^\mu G_{\mu+}(vz) u_{+}(z)\,.
\end{eqnarray}
In the first operator $j_d=\frac12$, $j_u=1$ and $j_g=3/2$ while in the
second and the third operators
$j_d=1$, $j_u=1$ and $j_g=1$. In all these cases the sum is the same, $J=j_d+j_u+j_g = 3$.
This is the minimum conformal spin corresponding to the asymptotic DA. In the fourth operator
$j_d=j_u=1$ but owing to the derivative $j_g=2$, making the minimal conformal spin $4$.
In addition, the $G$--parity demands that DA $\Phi_{\parallel}$ and $\Xi_{\pi}$ are antisymmetric 
under the interchange of  
$\alpha_1\leftrightarrow \alpha_2$. This implies that the matrix elements  of the operators with minimal conformal spin
vanish identically in these two cases and the conformal expansion for  $\Phi_{\parallel}$ and $\Xi_{\pi}$ 
starts one unit of spin higher at $J=4$ and $J=5$, respectively.

The model developed in \cite{BF90,Ball99} corresponds to taking into account contributions
of the lowest two conformal spins $J=3$ and $J=4$. To this accuracy
\begin{eqnarray}
\label{modelBF}
\Phi_{\perp}(\alpha_1,\alpha_2,\alpha_3)&=&
10\,\delta^2\,(\alpha_1-\alpha_2)\alpha_3^2\left[1+6\,
\epsilon\,(1-2\alpha_3)\right]\,,
\nonumber \\
\Phi_{\parallel}(\alpha_1,\alpha_2,\alpha_3)&=& 120 \epsilon \delta^2
\alpha_1\alpha_2\alpha_3(\alpha_1-\alpha_2)\,,
\nonumber\\
\Psi_{\perp}(\alpha_1,\alpha_2,\alpha_3)&=&
 10 \delta^2 \alpha_3^2 (1-\alpha_3)\left[1+6\epsilon(1-2\alpha_3)\right]\,,
\nonumber \\
\Psi_{\parallel}(\alpha_1,\alpha_2,\alpha_3)&=&- 40 \delta^2
\alpha_1\alpha_2\alpha_3 \left[1+3\epsilon(1-3\alpha_3)\right]\,,
\nonumber\\
\Xi_{\pi}(\alpha_1,\alpha_2,\alpha_3)&=& 0\,.
\end{eqnarray}
\begin{figure}[t]
\centerline{
\epsfig{file=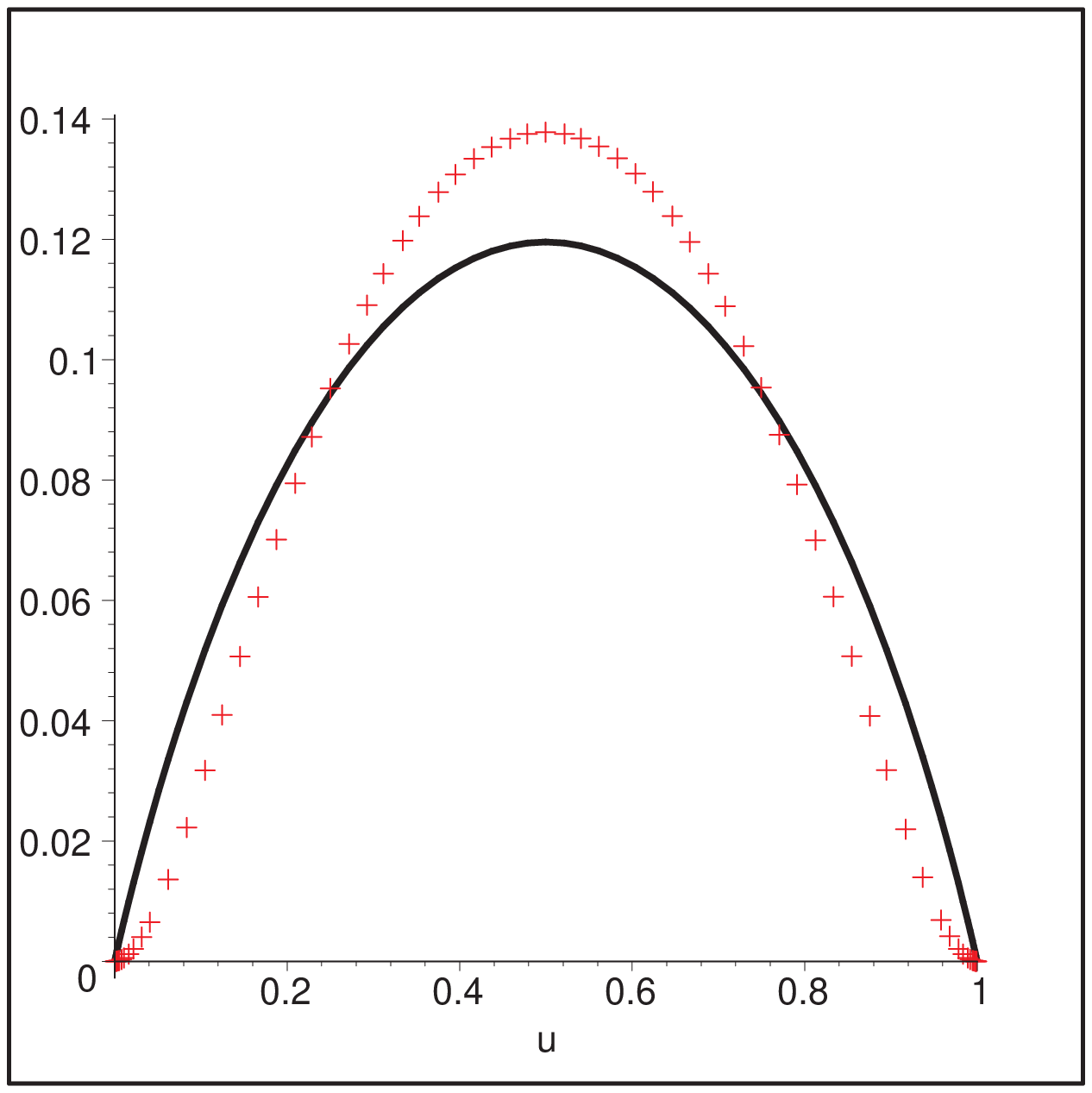,angle=-90,width=6.4cm}\hspace*{20pt}
\epsfig{file=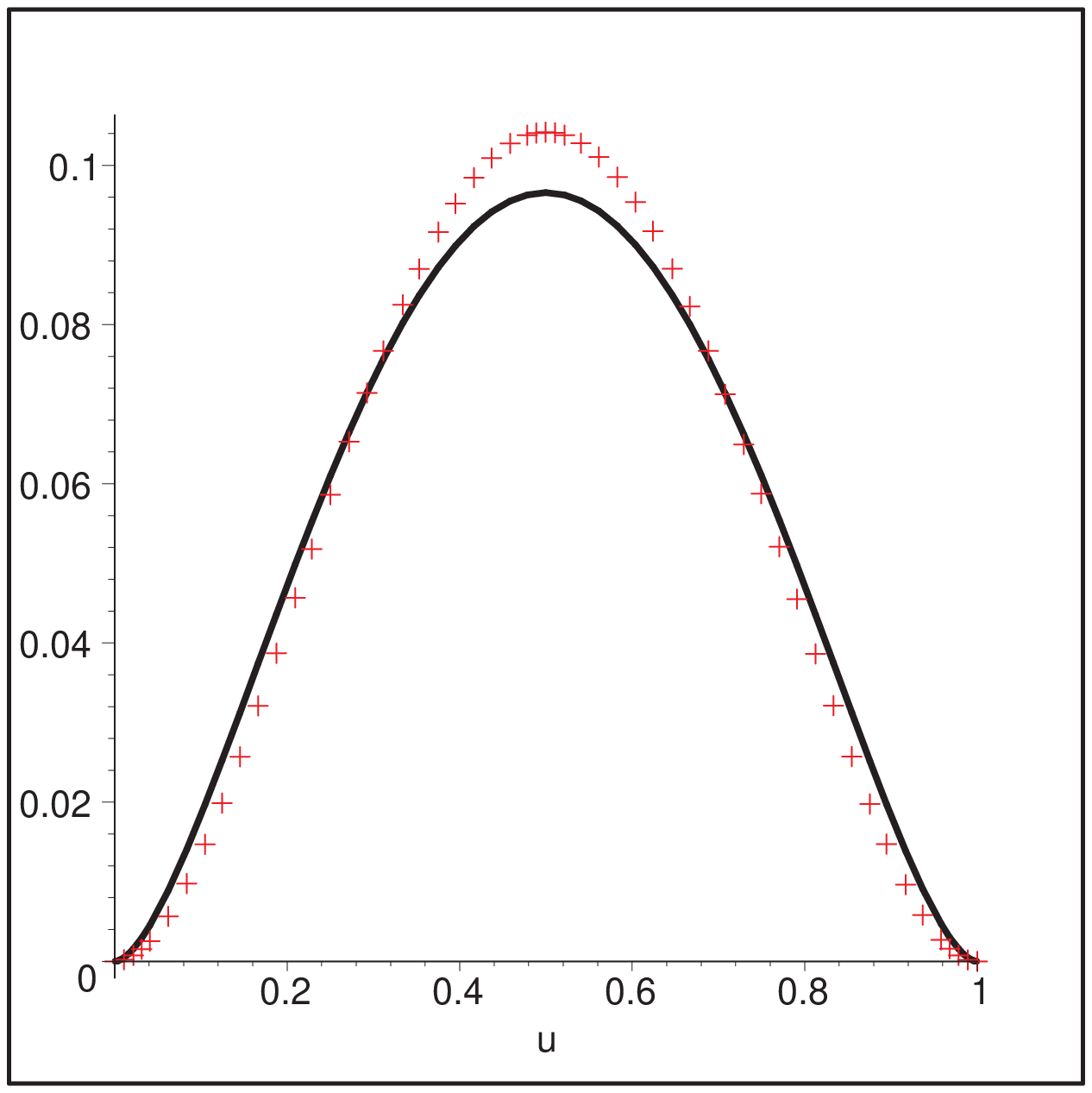,angle=-90,width=6.4cm}
}
\vspace*{20pt}
\caption{\small The two--particle twist--four DA $\phi_1^{(4)}$ (left)
and $-\phi_2^{(4)}$ (right) based on the renormalon model (full line)
and model of Ref. \cite{BF90} (crosses) which uses the first two orders in the conformal expansion
with the sum--rule estimate $\epsilon=0.5$. The normalization is chosen to be the same $\delta=1$.}
\label{phi1_and2_vs_BF} 
\end{figure}
The terms  $\sim \delta^2$ are $J=3$ and define the asymptotic DA
while contributions $\sim \epsilon \delta^2$ correspond to $J=4$.
For each spin $J=3$ and $J=4$ there exists only one independent non-perturbative parameter and from conformal
symmetry it follows that they have to have an autonomous scale dependence. This can be checked by a direct
computation. 

We have defined an alternative,  renormalon model of higher--twist DA in Sect.~3 by requiring that
it reproduces the correct normalization in \eq{delta}. It is now a matter of a simple algebra
to expand the renormalon model in \eq{RMthree} in contributions of increasing conformal spin (\ref{three-expand})
and compare with \eq{modelBF}. We find that the structure in \eq{modelBF} is reproduced
and the parameter $\epsilon$ proves to be independent on the choice of the leading--twist pion DA. One obtains
\begin{equation}
      \left.\epsilon\right\vert_{\rm Ren}= {7}/{12} \simeq 0.58
\label{eps_ren}
\end{equation}
which is comfortably close to the QCD sum--rule estimate
\begin{equation}
      \left.\epsilon\right\vert_{\rm SR } \simeq  0.5\pm 0.1 \quad \mbox{\cite{BF90}} \,.
\label{eps_SR}
\end{equation}
It is important  that the conformal expansion of all four DA in \eq{three-expand} yields the same value of $\epsilon$,
which illustrates that the renormalon model is consistent with the EOM.
Note that $\Xi_{\pi}(\alpha_1,\alpha_2,\alpha_3)=0$ to this accuracy; as mentioned above, 
its conformal expansion starts with $J=5$.

The two--particle twist--four DA of \eq{OPE} are related to the three--particle ones through the EOM, 
\eq{using_Phi}. Thus, when truncated to order $J=4$ in the conformal expansion, the renormalon model essentially 
coincides  with the model of \cite{BF90}, with the sole difference being the replacement of the sum--rule estimate 
for $\epsilon$ in \eq{eps_SR}, by that in \eq{eps_ren}.
However, in the renormalon model there is a priori no need to truncate the expansion at $J=4$:
~\eq{2_particle_model} represents the sum of all conformal spins.
{} Figure \ref{phi1_and2_vs_BF} compares the two--particle DA between the model of~\cite{BF90} and that 
of~\eq{2_particle_model}.
In general the two models are close, especially for $\phi_2^{(4)}$. 
We note, however, a qualitative difference in the end--point behavior. This is the subject of the next section.

\subsection{Higher orders and the end--point behavior}

\subsubsection{Three--particle distributions}

Let us start with the three--particle functions. The difference
between the models in \eq{RMthree} and \eq{modelBF} corresponds to
the contribution of higher conformal spins $J\ge 5$. Most striking
is that the end--point behavior of the renormalon model
expressions for small gluon momenta $\alpha_3 \equiv \alpha_g \to
0$ is $\Phi \sim {\it const}$ for all five three--particle
distributions in question whereas for each order in the conformal
expansion $\Phi_\parallel, \Psi_\parallel \sim \alpha_3^1$,
$\Phi_\perp, \Psi_\perp \sim \alpha_3^2$ and $\Xi_{\pi} \sim
\alpha_3^3$.

The difference indicates that the conformal expansion
is not converging.  Indeed,
assuming the asymptotic leading--twist pion DA in \eq{asympt} we
derive\footnote{To project the different DA onto the basis one uses the
coefficients of \eq{C_Jj}. The integration over $\alpha_1$ and $\alpha_2$
separates into a product of two independent integrals upon
changing variables to
$\tau=2(\alpha_1+\alpha_2)-1$ and $\sigma=(\alpha_2-\alpha_1)/(\alpha_1+\alpha_2)$ with both
integrals ranging from $-1$ to $1$.}
the following formal expansion of the DA in \eq{RMthreeAs} in contributions of the conformal spin $J$:
\begin{eqnarray}
\label{three_conf_exp}
\left\{\begin{array}{c}\Psi_\perp(\alpha_i) \\ \Phi_{\perp}(\alpha_i)\end{array}\right\}&=&
\delta^2\,
\left\{\begin{array}{c} \alpha_1 +\alpha_2\\ \alpha_1 - \alpha_2\end{array}\right\}
\,\alpha_3^2\,\sum_{J=3}^{\infty}
\frac{(2J-1)(J+1)}{J-1}\,P_{J-3}^{(2,2)}[1-2\alpha_3]\,, \nonumber\\
\left\{\begin{array}{c}\Psi_\parallel(\alpha_i) \\ \Phi_{\parallel}(\alpha_i)\end{array}\right\}
&=&-2\delta^2\,\alpha_1\alpha_2\alpha_3\, \sum_{J=3}^{\infty}
\sum_{j=2}^{J-1}
\left\{\begin{array}{c} 1 + (-1)^j\\1 - (-1)^j\end{array}\right\}
\frac{(2J-1)(2j-1)(J+j-1)}{(J-1)J(j-1)}
\nonumber \\&&\hspace*{10pt}
\times (1-\alpha_3)^{j-2}\,P_{J-j-1}^{(1,2j-1)}[1-2\alpha_3]\,
P_{j-2}^{(1,1)}\left[\frac{\alpha_2-\alpha_1}{1-\alpha_3}\right].
\end{eqnarray}
Taken literally the expansions
in \eq{three_conf_exp} 
are badly divergent for any fixed value of the gluon momentum fraction $\alpha_3$
and have to be understood as
distributions in mathematical sense, i.e. they have to be
convoluted with a suitable test function.
Note that for the ``transverse'' distributions the double sum
disappears since the only non-vanishing contributions come from
$j=3/2$ ($n=0$) terms, {\it cf.}\  \eq{three-expand}. Owing to
G--parity the DA $\Psi_\perp, \Psi_\parallel$ are symmetric and
$\Phi_\perp, \Phi_\parallel$ are antisymmetric under replacement
$\alpha_1\leftrightarrow \alpha_2$. This symmetry is realized
differently for the transverse and the longitudinal components:
for $\Phi_{\perp}, \Psi_{\perp}$ it is associated with the
explicit overall factor $\alpha_1\pm \alpha_2$ while the expansion
itself is symmetric, depending only on
$\alpha_1+\alpha_2=1-\alpha_3$. {}For $\Phi_{\parallel},
\Psi_\parallel$, on the other hand, the proper symmetry is
obtained by the selection of odd/even values of $j$ and taking
into account that
$P_{j-2}^{(1,1)}\left[x\right]=(-1)^j\,P_{j-2}^{(1,1)}\left[-x\right]$.

Absence of convergence may be an artifact of the single--dressed--gluon
approximation. As well known \cite{ren_review}, this accuracy is
sufficient to identify the position of the singularity $w=1$ but
does not distinguish between singularities of different strength
so that all renormalons appear as simple poles\footnote{This
corresponds to picking up the leading quadratic
   divergence of the twist--four operators and
   ignoring possible logarithmic enhancement/suppression. Also note that
   we do not show in \eq{Anom_singularity} the term $\sim \beta_1/\beta_0^2$
   which usually appears in a
   sum with ${\gamma_0^{(J,h)}}/{\beta_0}$ because we tacitly imply using the
   scheme--invariant Borel transform where $u$ is conjugate to
   $\ln(\mu^2/\Lambda^2)$ and not to $1/(\alpha_s(\mu)\beta_0/\pi)$. }.
In the full theory they will be converted to branch points and
instead of a pole at $w=1$ for a given $J$ there will be a sum of
terms with different singular behavior
\begin{equation}
\frac{1}{1-w} \longrightarrow
\sum_{h=0}^N\frac{r_{J,h}}{(1-w)^{1-{\gamma_0^{(J,h)}}/{\beta_0}}},
\label{Anom_singularity}
\end{equation}
where $\beta_0$ is defined in \eq{beta_0} and $\gamma_0^{(J,h)}$
are the eigenvalues of the leading--order anomalous dimension
matrix ${\gamma_0^{(J)}}$ for operators ${\cal O}_J(\mu^2)$ with
conformal spin $J$:
\begin{equation}
\mu^2\frac{{\cal O}_J(\mu^2)}{d\mu^2}=-\gamma^{(J)} \,{\cal
O}_J(\mu^2), \qquad\qquad
\gamma^{(J)}={\gamma_0^{(J)}}\alpha_s/\pi+\cdots.
\end{equation}
For large spins the anomalous dimensions are dominated by soft--gluon 
emission and for quark--antiquark--gluon operators one
expects \cite{BKM01}
\begin{equation}
      C_F\, \ln J \le \gamma_0^{(J,h)} \le N_c\, \ln J\,,
\end{equation}
where the prefactors are nothing but color charges corresponding
to the possible classical geometries of color flow. The
logarithmic rise of anomalous dimensions translates to the
suppression of contributions of higher conformal spin operators at
large scales $\mu_F^2$
\[ \left(\frac{\alpha_s(\mu_F^2)}{\alpha_s(\mu_0^2)}\right)^{\mbox{\scriptsize\rm const}\cdot \ln J}
     \sim J^{- \mbox{\scriptsize\rm const}\cdot\ln\ln \mu_F^2}\]
in the same way as higher--spin contributions get 
suppressed for the leading--twist DA \cite{ER,BL}.
This suppression will improve the expansion in \eq{three_conf_exp} and make it 
convergent at very large scales. 
In this sense, the renormalon model in \eq{RMthree}, 
\eq{RMthreeAs} can be regarded as representing a
worst--case scenario for the convergence of the conformal expansion.

\subsubsection{Two--particle distributions}

The large higher--spin contributions to three--particle pion DA 
in the soft--gluon region $\alpha_3\to 0$ do not necessarily yield large corrections to physical 
observables because the gluon momentum fraction is always integrated over. 
Two--particle DA are more directly relevant. We find
that  $\phi_1^{(4)}(u)$ and  $\phi_2^{(4)}(u)$ in \eq{2_particle_model} in the renormalon model 
have the asymptotic behavior in the end--point regions $\sim u(1-u)$ and $\sim u^2(1-u)^2\ln u(1-u)$, respectively,
which is to be compared with the  $\sim u^2(1-u)^2$ behavior of the leading conformal spin 
contributions (asymptotic DA) in both cases \cite{BF90}.
\begin{figure}[t]
\centerline{
\epsfig{file=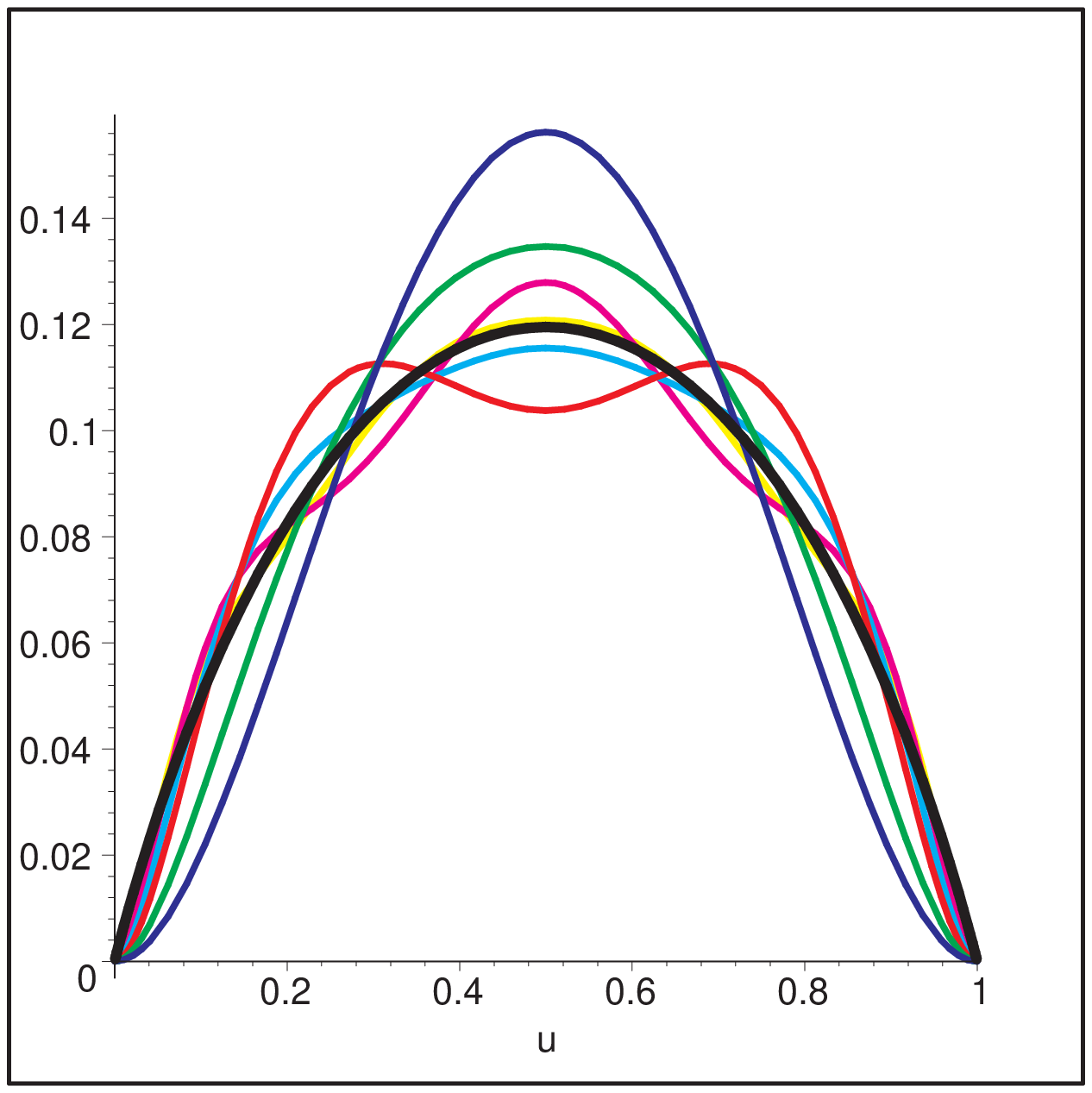,angle=-90,width=4.9cm}
\epsfig{file=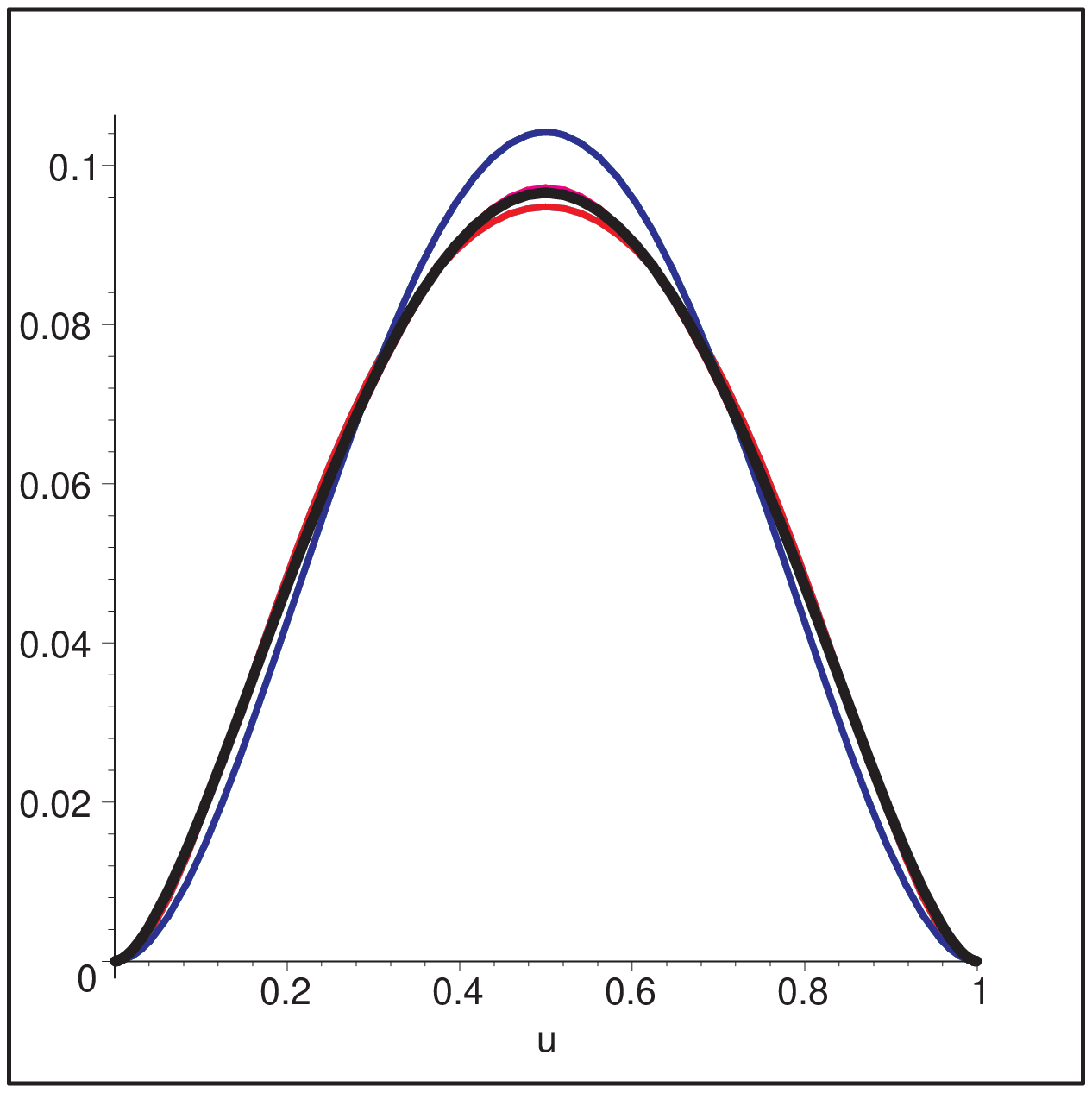,angle=-90,width=4.9cm}
\epsfig{file=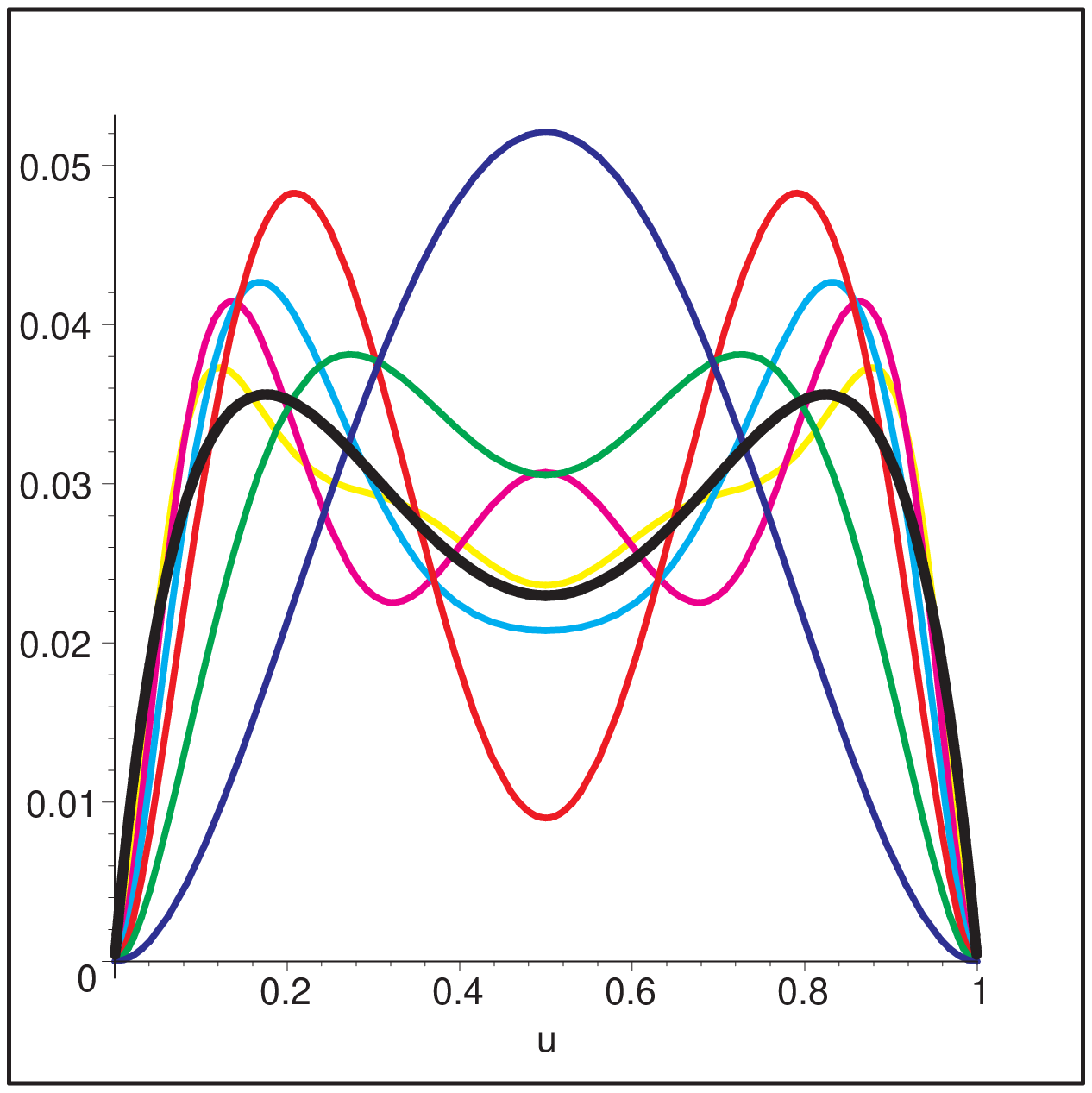,angle=-90,width=4.9cm}} \vspace*{20pt}
\caption{\small The first few orders (partial sums) in the conformal expansion of the  renormalon model for 
the two--particle twist--four DA $\phi_1^{(4)}$ (left), $-\phi_2^{(4)}$ (middle) 
and $\phi_1^{(4)}+\phi_2^{(4)}$ (right).
The thick black line represents~\eq{2_particle_model}.} 
\label{phi1_and2} 
\end{figure}

The expansions we obtained for the three--particle DA can readily be inserted into \eq{using_Phi} to 
yield the conformal expansions of the two--particle twist--four
amplitudes. In the case of $\phi_2^{(4)}$ the integration can easily be performed 
with the result 
\begin{equation}
\phi_2^{(4)}(u)=-4\delta^2\,u^2(1-u)^2\sum_{J=3,5,7,\ldots}^{\infty}\frac{2J-1}{J(J-1)^2(J-2)}\,
P_{J-3}^{(2,2)}(2u-1).
\label{ce_2}
\end{equation}
Away from the end--points one can use the asymptotic expansion \cite{BE}
\begin{equation}
\label{large_n}
 P_n^{(\alpha,\beta)}(\cos \theta) = \frac{\cos\{[n+(\alpha+\beta+1)/2]\theta-(2\alpha+1)\pi/4\}}
                              {\sqrt{\pi n}\,[\sin(\theta/2)]^{\alpha+1/2}[\cos(\theta/2)]^{\beta+1/2}}
 +{\cal O}(n^{-3/2})
\end{equation}
to see that \eq{ce_2} is convergent. It is not converging uniformly at the end points, however, which explains
the logarithmic enhancement compared to the asymptotic DA (and the model of \cite{BF90}). 
As noted in \cite{BF90}, the second derivative
\footnote{In notations of \cite{BF90} $(d^2/du^2)\phi_2^{(4)}(u) = (d/du) g_2(u)$.}
 $(d^2/du^2)\phi_2^{(4)}(u)$
 corresponds to the non-local light--cone operator $\bar{d}(-z)\gamma^{-}\gamma_5\, u(z) = 
\bar{d}_{-}(-z)\gamma^{-}\gamma_5\,u_{-}(z)$ with both quarks having spin projection 
$s=-1/2$ alias $j_{\bar d} =j_u =1/2$.  
It follows that the conformal expansion of $(d^2/du^2)\phi_2^{(4)}(u)$ goes over
Legendre polynomials, or  $P^{(0,0)}_{J-1}(2u-1)$. This result is consistent with \eq{ce_2} since
\begin{equation}
  \frac{d^2}{du^2}\,\left[ u^2(1-u)^2 P_{J-3}^{(2,2)}(2u-1)\right] = (J-2)(J-1)\, P_{J-1}^{(0,0)}(2u-1)\,.
\end{equation}

For $\phi_1^{(4)}$ a similar, closed--form all--order expression
is not available. However, it is straightforward to compute the conformal expansion
order by order in $J$, {\it cf.}\ \eq{ce_1_int}:
\begin{eqnarray}
\label{ce_1}
\phi_1^{(4)}(u)&=&\delta^2\,
\bigg\{\frac {5}{2}\bigg[ \,u^{2}\,\bar{u}^{2}\bigg]_{J=3}
+\frac {7}{24}\bigg[ u\bar{u}\,(13\,u\bar{u} + 2)\nonumber
 \\ \nonumber  &+&  2\,(6\,u^{2} + 3\,u + 1)\,\bar{u}^{3}\,\mathrm{ln}(\bar{u})
+  2 \,( 6\,\bar{u}^2 + 3\,\bar{u} + 1)\,\,u^{3} \mathrm{ln}(u) \bigg]_{J=4}
  \\  \nonumber
&+& \bigg[ {\frac{1}{40} {u\bar u \,({140\,u^{2}\bar{u}^2 +243\,u\bar{u}+42})}}
 + \frac{21}{20} \,(6\,u^{2} + 3\,u + 1)\,\bar{u}^{3}\,\mathrm{ln}(\bar{u})
  \\ 
&&  {}+{ \frac {21}{
20}} \,( 6\,\bar{u}^2 + 3\,\bar{u} + 1)\,u^{3}\,\mathrm{ln}(u)\bigg]_{J=5}
+\cdots\bigg\}.
\end{eqnarray}
\begin{figure}[t]
\centerline{\epsfig{file=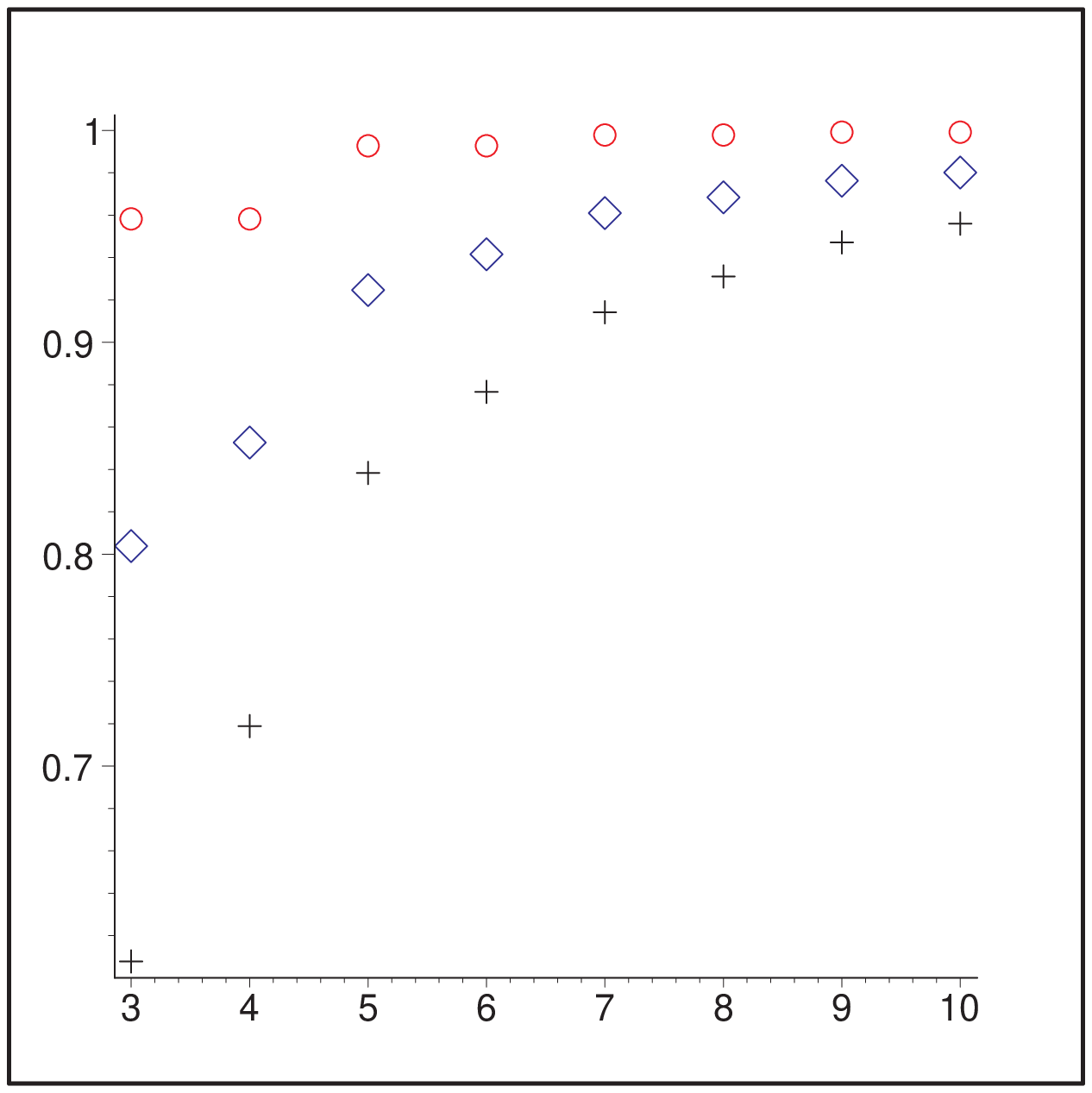,angle=-90,width=6.4cm}
\hspace*{10pt}\epsfig{file=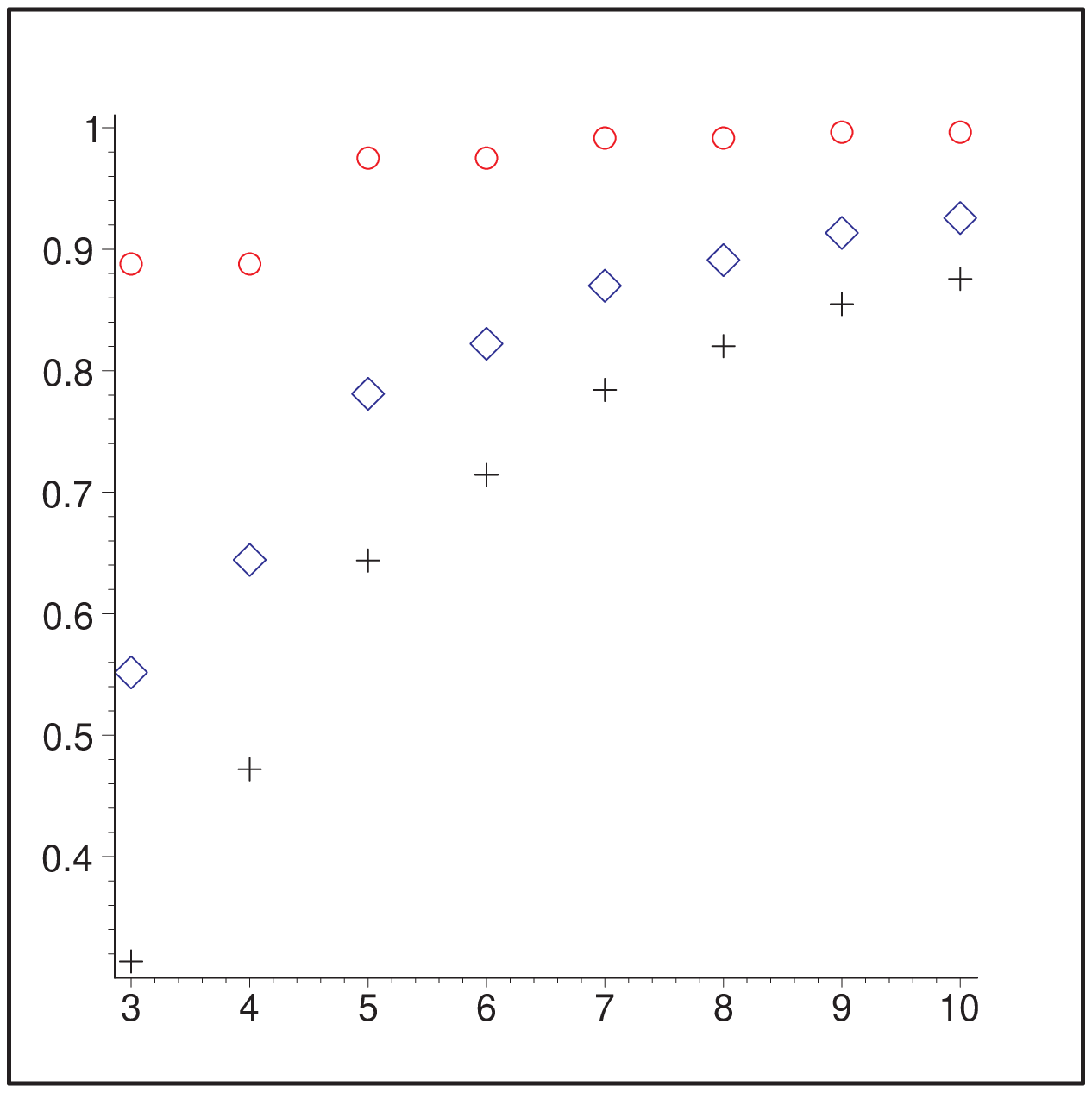,angle=-90,width=6.4cm}}
\vspace*{20pt} \caption{\small Subsequent approximations by the conformal expansion 
truncated at spin $J_{\rm max}=3,4,\ldots$
for the integrals $I_0$ (left panel) and $I_1$ (right panel) defined in \eq{typ} for 
$\phi_1^{(4)}(u)$ (diamonds),
$\phi_2^{(4)}(u)$ (circles) and $\phi_1^{(4)}(u)+\phi_2^{(4)}(u)$
(crosses). The results are all normalized to the exact value obtained using
 \eq{2_particle_model}. 
Note the different scales on the vertical axis.}
\label{conv} 
\end{figure}

Note that each term in the conformal expansion is of the order of  $u^2(1-u)^2$ near the 
end--points but the expansion is not converging uniformly so that
the $\sim u(1-u)$ behavior emerges in the sum of all spins.

The first few orders (partial sums) in the conformal expansion of the  renormalon model are
compared to the full result in Fig.~\ref{phi1_and2} for the three functions:
 $\phi_1^{(4)}$, $\phi_2^{(4)}$ and $\phi_1^{(4)}+\phi_2^{(4)}$. The convergence is worst for
the latter because of partial cancellation of the leading terms.   
Absence of uniform convergence at the end points means that the limit $u\to 0,1$ and the summation 
over conformal spins cannot be interchanged; the partial sums $\sum_{J=3}^{J_{\rm max}}$ diverge  
as $\phi_1^{(4)}(u) \sim (\delta^2/2) u^2(1-u)^2 J_{\rm max}^2$ and 
$\phi_2^{(4)}(u) \sim -2\delta^2 u^2(1-u)^2 \ln J_{\rm max}$, respectively.    

Last but not least, we show in Fig.~\ref{conv} the subsequent approximations by the conformal expansion 
truncated at spin $J_{\rm max}=3,4,\ldots$ for the typical integrals that one encounters in the description
of hard exclusive processes in QCD:
\begin{eqnarray}
\label{typ}
 I_0 &=& \int_0^1 \frac{du}{u}\, \phi^{(4)}_i(u)\,,
\nonumber\\
 I_1 &=& \int_0^1 \frac{du}{u} \ln(1/u)\,\phi^{(4)}_i(u)\,.
\end{eqnarray}
The difference between the renormalon model and the model of \cite{BF90} ($J_{\rm max}=4$)
is of order 10-30\% for the first integral and somewhat larger for the second one.

\section{Renormalon model for twist--four DA of the rho\label{tw4_vector}}
\setcounter{equation}{0}

\setcounter{footnote}{1}

A useful feature of the renormalon approach is its universality: with minor modifications
it can be applied to DA of vector mesons as well. For definiteness we consider here 
$\rho^+$ meson. 
One difference to the pion case is that because of spin  the number of DA proliferates significantly.  
We will conform to the definitions and notations of Ref.~\cite{BB98} and in particular distinguish between 
chiral--even and chiral--odd DA corresponding to the operators with odd/even number of $\gamma$--matrices,
respectively, 
between the quark fields. A second difference is that because of a sizable $\rho$--meson mass the 
twist--four $\rho$--meson DA receive the Wandzura--Wilczek--type contributions of 
the operators with geometric twist--two given in terms of the leading--twist $\rho$--meson DA with the same 
(longitudinal or transverse) polarization, and the operators of geometric twist--three that are expressed 
in terms of twist--three DA with the opposite polarization
\footnote{The mismatch is due to different 
twist definition: ``dimension minus spin projection'' (collinear twist) 
for DA {\it vs.}\ ``dimension minus spin'' (geometric twist) for operators, 
see \cite{BBKT98,conformal} for details.}.  
The Wandzura--Wilczek contributions to the vector--meson 
DA have been calculated in Refs.~\cite{BBKT98,BB98}. 
They have to be added to the ``genuine'' twist--four contributions considered here.
In this section we collect the necessary definitions and summarize the results; see Appendix C 
for more details.

In what follows $P_\mu$ stands for the $\rho$--meson momentum, $P^2=m_\rho^2$,  and $e_\mu^{(\lambda)}$ 
is the polarization vector $e\cdot P =0$. We use the notation
\[
p_{\mu}=P_\mu-\frac{m_\rho^2}{2p\cdot z}z_\mu, \qquad p^2=0,
\]
and define the transverse polarization vector by 
\[
e^\perp_{\mu}=e_\mu-
\frac{e\cdot z}{p\cdot z}
\left(p_\mu-\frac{m_\rho^2}{2p\cdot z}z_\mu \right).
\] 
We will also use the projector onto the directions orthogonal to $p$ and $z$:

\[
  g_{\mu\nu}^\perp = g_{\mu\nu} - \frac{1}{pz}(p_\mu z_\nu + p_\nu z_\mu)\,.
\]

\subsection{Chiral--even distribution amplitudes}

\subsubsection{Definitions}

We start by quoting the necessary definitions from Ref.~\cite{BB98} and in this section
consider matrix elements involving an odd number of $\gamma$ matrices, which we refer to as chiral-even in what follows.  
For the vector operator the light--cone expansion to twist--four accuracy reads:
\begin{eqnarray}
\langle 0|\Big[\bar d(-x) \gamma_\mu u(x)\Big]_{\mu_F^2}|\rho^+(P,\lambda)\rangle 
 &=& f_\rho m_\rho \Bigg\{
\frac{e^{(\lambda)}x}{Px}\, P_\mu \int_0^1 du \,e^{-i\xi Px}
\Big[\phi_\parallel(u)
+\frac{m^2_\rho x^2}{4}\,  {\mathbb A}(u)\Big]
\nonumber\\
&&{}+\left(e^{(\lambda)}_\mu-P_\mu\frac{e^{(\lambda)}x}{Px}\right)
\int_0^1 du\, e^{-i\xi Px} \,{\mathbb B}(u)
\nonumber\\&&{}
-\frac{1}{2}x_\mu \frac{e^{(\lambda)}x}{(Px)^2} m^2_\rho \int_0^1 du 
\, e^{-i\xi Px}\, {\mathbb C} (u)
\Bigg\}.
\label{eq:OPEvector}
\end{eqnarray}
We do not consider the axial--vector operator because its light--cone expansion 
only includes contributions of twist three, five, etc., that are not relevant in the 
present context. 
For brevity, in this section we do not show gauge factors between the 
quark and the antiquark fields; we also use the short-hand notation
\[\xi = 2u-1.\] 
The vector decay constant 
$f_\rho$ is defined, as usual, as
\begin{equation}
\langle 0|\bar d(0) \gamma_{\mu} u(0)|\rho^+(P,\lambda)\rangle =  f_{\rho}m_{\rho}
e^{(\lambda)}_{\mu}\,.
\label{eq:fr}
\end{equation}
The expansion in (\ref{eq:OPEvector})
involves three Lorentz invariant amplitudes which we have 
to interpret in terms of meson DA.
Definitions of the latter involve non-local operators at 
strictly light--like separations and can most conveniently be 
written for longitudinal and transverse meson polarizations separately.
Following \cite{BBKT98,BB98}, we define chiral--even 
two--particle DA of the $\rho^+$ meson as
\begin{eqnarray}
\lefteqn{\langle 0|\Big[\bar d(-z) \gamma_{\mu} u(z)\Big]_{\mu_F^2}|\rho^+(P,\lambda)\rangle 
 = f_{\rho} m_{\rho} \left[ p_{\mu}
\frac{e^{(\lambda)} z}{p  z}
\int_{0}^{1} \!du\, e^{-i \xi p  z} \phi_{\parallel}(u, \mu_F^{2}) \right. 
}\hspace*{0.6cm}\nonumber \\
&+&\left. e^{(\lambda)}_{\perp \mu}
\int_{0}^{1} \!du\, e^{-i \xi p  z} g_{\perp}^{(v)}(u, \mu_F^{2}) 
- \frac{1}{2}z_{\mu}
\frac{e^{(\lambda)} z }{(p  z)^{2}} m_{\rho}^{2}
\int_{0}^{1} \!du\, e^{-i \xi p  z} g_{3}(u, \mu_F^{2})
\right].\hspace*{0.4cm}
\label{eq:vda}
\end{eqnarray}
The distribution amplitude $\phi_\parallel$ is of twist two,
$g_\perp^{(v)}$ of twist three and $g_3$ of twist four. 
All three functions $\phi=\{\phi_\parallel,
g_\perp^{(v)},g_3\}$ are normalized as
\begin{equation}
\int_0^1\!du\, \phi(u) =1.
\label{eq:norm}
\end{equation}
This can be checked by comparing the two sides
of the  defining equations in the limit $z_\mu\to 0$ and using the EOM.

Comparing (\ref{eq:vda})  with the light--cone expansion
in (\ref{eq:OPEvector}) one finds \cite{BB98}
\begin{eqnarray}
\label{BC_relations}
{\mathbb B}(u) &=& g^{(v)}_\perp(u),    
\nonumber\\
{\mathbb C}(u) &=& g_3(u)+\phi_\parallel(u) -2 g^{(v)}_\perp(u)\,.    
\end{eqnarray}
The remaining invariant amplitude ${\mathbb A}(u)$ accounts for the 
transverse momentum distribution in the valence component of the 
wave function. We end up with two two--particle twist--four 
DA of the longitudinally polarized $\rho$--meson, ${\mathbb A}(u)$ and $g_3(u)$ which are counterparts 
of the pion DA $\phi_1$ and $\phi_2$ (a precise correspondence will be given below).

Three--particle chiral--even distributions are rather numerous and can be 
defined by the following matrix elements:
\begin{eqnarray}
\langle 0|\bar d(-z) g\widetilde G_{\mu\nu}(vz)\gamma_\alpha\gamma_5 
  u(z)|\rho^+(P,\lambda)\rangle & = &
  f_\rho m_\rho p_\alpha[p_\nu e^{(\lambda)}_{\perp\mu}
 -p_\mu e^{(\lambda)}_{\perp\nu}]{\cal A}(v,pz)
\nonumber\\ &
 & {}+f_\rho m_\rho^3\frac{e^{(\lambda)} z}{pz}
[p_\mu g^\perp_{\alpha\nu}-p_\nu g^\perp_{\alpha\mu}] \widetilde\Phi(v,pz)
\nonumber\\& & {}+
 f_\rho m_\rho^3\frac{e^{(\lambda)} z}{(pz)^2}
p_\alpha [p_\mu z_\nu - p_\nu z_\mu] \widetilde\Psi(v,pz),\label{eq:even1}\\
\langle 0|\bar d(-z) g G_{\mu\nu}(vz)i\gamma_\alpha 
  u(z)|\rho^+(P,\lambda)\rangle &=&
  f_\rho m_\rho p_\alpha[p_\nu e^{(\lambda)}_{\perp\mu} 
  - p_\mu e^{(\lambda)}_{\perp\nu}]{\cal V}(v,pz)
\nonumber\\&&
{}+ f_\rho m_\rho^3\frac{e^{(\lambda)} z}{pz}
[p_\mu g^\perp_{\alpha\nu} - p_\nu g^\perp_{\alpha\mu}] \Phi(v,pz)
\nonumber\\& & {}+f_\rho m_\rho^3\frac{e^{(\lambda)} z}{(pz)^2}
p_\alpha [p_\mu z_\nu - p_\nu z_\mu] \Psi(v,pz),\label{eq:even2}
\end{eqnarray}
where 
\begin{equation}
   {\cal A}(v,pz) =\int {\cal D}{\alpha_i} 
e^{-ipz(\alpha_u-\alpha_d+v\alpha_g)}{\cal A}(\alpha_i),
\end{equation}
etc., and ${\alpha_i}$ is the set of three momentum fractions
$\{\alpha_1,\alpha_2,\alpha_3\}=\{\alpha_u,\alpha_d,\alpha_g\}$.
The integration measure is defined in \eq{measure}.
 
Similarly to the pion case, all higher--twist two--particle 
DA of the $\rho$--meson  do not present genuine independent degrees
of freedom but can be expressed in terms of three--particle DA. 
The corresponding relations~\cite{BB98} are given in \eq{equ_motion_even} below. 

In addition, we introduce a new twist--four DA
\begin{equation}
\label{Xi_rho}
\left\langle 0\left \vert \bar{d}(-z) \gamma_{\mu}
 gD^{\alpha}G_{\alpha\beta}(vz) u(z) \right\vert \rho^+(P,\lambda)\right\rangle
= f_{\rho}m_{\rho}^3\,p_{\mu}p_{\beta}\,\frac{e^{(\lambda)} z}{p z} \Xi_\rho(v,pz)\,,
\end{equation}
which can be viewed, equivalently, either as a three--particle quark--antiquark--gluon DA,
or as a special case of a four-quark DA with the quark-antiquark pair in a color--octet state 
and at the same space point, {\it cf.}\ \eq{Xi}.

\subsubsection{Renormalon model and comparison with Ref. \cite{BB98} \label{ren_model_even}}

The renormalon model for twist--four DA of the longitudinally polarized $\rho$--meson can most 
simply be derived from the UV--renormalon ambiguity in twist--four operators.
The calculation is very similar to that in the pion case. One obtains in the single--dressed--gluon 
approximation
\begin{eqnarray}
\label{UV_div_rho_even}
\hspace*{-5pt} \delta_{\UV}\Big\{z^{\mu}\,\bar{d}(-z) \gamma^{\nu} \gamma_{5}
g\widetilde{G}_{\mu\nu} (vz) u(z) \Big\}
&=&
2\,c\Lambda^2\, \int_0^1 da (1+a)\,\bigg[\bar d(-y)\zsl u(z)+ \bar d(-z)\zsl u(\widetilde{y}) \bigg],            
\nonumber       \\
\hspace*{-5pt}  \delta_{\UV}\Big\{z^{\mu}\,\bar{d}(-z) \zsl \gamma_{5}
g\widetilde{G}_{\mu\rho} (vz) u(z) \Big\}
&=&
2\,c\Lambda^2\, z_{\rho}\,\int_0^1 da \,a\,\bigg[\bar d(-y)\zsl u(z)+ \bar d(-z)\zsl u(\widetilde{y})
 \bigg], 
\nonumber      \\
\hspace*{-5pt} \delta_{\UV}\Big\{z^{\mu}\,\bar{d}(-z) \gamma^{\nu}
igG_{\mu\nu} (vz) u(z) \Big\}
&=&
-2\,c\Lambda^2\, \int_0^1 da (1-a)\,\bigg[\bar d(-y)\zsl u(z)- \bar d(-z)\zsl u(\widetilde{y})
 \bigg],
\nonumber       \\
\hspace*{-5pt}  \delta_{\UV}\Big\{z^{\mu}\,\bar{d}(-z) \zsl
igG_{\mu\rho} (vz) u(z) \Big\}
&=&
2\,c\Lambda^2\, z_{\rho}\,\int_0^1 da \,a\,\bigg[\bar d(-y)\zsl u(z)- \bar d(-z)\zsl u(\widetilde{y})
 \bigg],
\nonumber\\
\hspace*{-5pt} \delta_{\UV}\Big\{ z^{\beta}\bar d(-z)
gD^{\alpha}G_{\alpha\beta}(vz)\zsl u(z)\Big\}
&=&
 -4ic\Lambda^2\int_0^1da\,a\frac{d}{da}\bigg[\frac{1}{1+v}\,\bar d(-y)\zsl
u(z)
\nonumber\\&&\hspace*{83pt}+\,\frac{1}{1-v}\bar d(-z)\zsl u(\widetilde{y}) \bigg],
\end{eqnarray}
where $c$ is given in \eq{c} and the variables $y$ and $\widetilde{y}$ are defined below \eq{UV_div}.
To obtain the renormalon model, we take the matrix elements of the operators in
\eq{UV_div_rho_even} between the vacuum and a  $\rho$--meson state, 
project onto the relevant Lorentz structures and fix the normalization by the local matrix 
element 
\begin{equation}
  \label{def_parameter_zeta_4}
  \langle 0|\bar{d}(0)g\widetilde{G}_{\mu\nu}(0)\gamma^\nu\gamma_5
  u(0)|\rho^+(P,\lambda)\rangle = f_\rho m_\rho^3e^{(\lambda)}_\mu \zeta_4\,, ~~~ \zeta_4 = 0.15\pm 0.10~\mbox{\cite{BK85}} 
\end{equation}
making the substitution
\begin{equation}
     c\Lambda^2 \to \frac16\zeta_4\,m^2_\rho\,.
\label{zeta_4}
\end{equation}
The resulting twist--four DA, expressed in terms of the twist--two one, read:
\begin{eqnarray}
\label{three_rho_even}
\widetilde{\Psi}(\alpha_i)&=&
\frac13 \zeta_4\left[
\frac{\alpha_2\phi_{\parallel}(\alpha_1)}{(1-\alpha_1)^2}
+\frac{\alpha_1\phi_{\parallel}(\alpha_2)}{(1-\alpha_2)^2}
\right],
\nonumber\\
\Psi(\alpha_i)\Big\}&=&
\frac13 \zeta_4\left[
\frac{\alpha_2\phi_{\parallel}(\alpha_1)}{(1-\alpha_1)^2}
-\frac{\alpha_1\phi_{\parallel}(\alpha_2)}{(1-\alpha_2)^2}
\right],
\nonumber\\
 \widetilde{\Phi}(\alpha_i)&=&
\frac16 \zeta_4
\left[
\frac{\phi_{\parallel}(\alpha_1)}{1-\alpha_1}+\frac{\phi_{\parallel}
(\alpha_2)}{1-\alpha_2}
\right],
\nonumber\\ 
\Phi(\alpha_i)&=&
-\frac16 \zeta_4
\left[
\frac{\phi_{\parallel}(\alpha_1)}{1-\alpha_1}-\frac{\phi_{\parallel}
(\alpha_2)}{1-\alpha_2}
\right], 
\nonumber\\      
\Xi_{\rho}(\alpha_i) &=& -
\frac23 \zeta_4 
\left[\frac{\alpha_2}{1-\alpha_1}\,\phi_{\parallel}(\alpha_1)
-\frac{\alpha_1}{1-\alpha_2}\phi_{\parallel}(\alpha_2)
\right],
\end{eqnarray}
where we used the symmetry $\phi_{\parallel}(u)=\phi_{\parallel}(1-u)$.
Note the similarity of this result to the pion case, \eq{RMthree}: upon replacing in the latter 
$\delta^2$ by $\zeta_4$ one recovers the former with 
$- \Phi_{\perp}^{\pi} \longrightarrow \Phi^{\rho},\,
\Phi_{\parallel}^{\pi} \longrightarrow \Psi^{\rho},\,
\Psi_{\perp}^{\pi} \longrightarrow \widetilde{\Phi}^{\rho},\,
- \Psi_{\parallel}^{\pi} \longrightarrow\widetilde{\Psi}^{\rho}$ and 
$\Xi_{\pi} \longrightarrow \Xi_{\rho}$.

The two--particle twist--four DA can now be restored using EOM \cite{BB98}. 
The calculation (see Appendix \ref{EOM_UV}) gives:
\begin{eqnarray}
\label{two_rho_even}
  \mathbb{A}(u)
&=&
\frac83 \zeta_4
\int_0^1 dv\,
\phi_\parallel(v) \bigg\{\theta(u>v)\;\frac{1}{\bar v ^2}\left[
{\bar{u}+\bar{u}^2+(u-v)  \ln\frac{u-v}{\bar v}}\right]\nonumber\\
  &&\hspace*{93pt}+\theta(u<v)\;\frac{1}{v^2}\left[{u+u^2+(v-u)\ln\frac{v-u}{v}}\right]\bigg\},
\nonumber\\
   g_3(u)
&=&
-\frac23 \zeta_4
\left\{\int_0^1\!dv\,
\phi_\parallel(v)\left[\theta(u>v)\;\frac{1}{\bar{v}^2}+\theta(u<v)\;
\frac{1}{v^2}\right]-\frac{\phi_\parallel(u)}{u\bar u}\right\}.
\end{eqnarray}
Continuing the comparison with the pion case $\mathbb{A}(u)$
is similar to $16(\phi_1-\phi_2)$ and $g_3$ to $2{d^2\phi_2}/{du^2}$.

Eqs.~(\ref{three_rho_even}) and (\ref{two_rho_even}) are valid for an arbitrary leading--twist DA. 
Choosing the asymptotic expression  
$\phi_\parallel(u) = 6u(1-u)$ yields a simple model  
\begin{eqnarray}
\label{three_rho_even_as}
 \widetilde{\Psi}(\alpha_i)&=&2 \zeta_4
\alpha_1\alpha_2\left[\frac{1}{1-\alpha_1}+\frac{1}{1-\alpha_2}\right],
\nonumber\\ 
           {\Psi}(\alpha_i)&=&2 \zeta_4
\alpha_1\alpha_2\left[\frac{1}{1-\alpha_1}+\frac{1}{1-\alpha_2}\right]
,
\nonumber\\
\widetilde{\Phi}(\alpha_i)&=&\zeta_4 \left[\alpha_1+\alpha_2\right],
\nonumber\\      
\Phi(\alpha_i)&=&\zeta_4 [\alpha_2-\alpha_1],
\nonumber\\
\Xi_{\rho}(\alpha_i) &=& 0 
\end{eqnarray}
and
\begin{eqnarray}
\label{two_rho_even_as}
  \mathbb{A}(u)
&=&16\zeta_4 \bigg\{\bar u \bigg[u\ln(\bar u) - \mbox{\rm Li}_2 (\bar u)\bigg]
+u\bigg[\bar{u}\ln (u)- \mbox{\rm Li}_2( u )\bigg]
-2u\bar u+\frac{\pi^2}6\bigg\},
\nonumber\\
   g_3(u)
&=&4\zeta_4 \bigg\{\ln(u) +\ln(\bar u)+2\bigg\}.
\end{eqnarray}
This model should be compared with that of Ball and Braun (BB) in Ref.~\cite{BB98} based on the two first orders in the 
conformal expansion:
\begin{eqnarray}
\label{BB_three_part}
\widetilde{\Psi}^{\rm BB}(\alpha_i) &=& 40\,\zeta_4\, \alpha_1 \alpha_2 \alpha_3
\Big[1+\frac{63}{4} \omega_4^A(3\alpha_3-1)\Big],
\nonumber\\
\Psi^{\rm BB} (\alpha_i) &=& 630\,\zeta_4\, \omega_4^A \, \alpha_1 \alpha_2 \alpha_3
(\alpha_2-\alpha_1)\,,
\nonumber\\
\widetilde{\Phi}^{\rm BB}(\alpha_i) &=& 10\,\zeta_4\, \alpha_3^2 (1-\alpha_3)
\Big[1+\frac{63}{2}\omega_4^A (2\alpha_3-1)\Big],
\nonumber\\
\Phi^{\rm BB} (\alpha_i) &=& 10\, \zeta_4\, \alpha_3^2 (\alpha_2-\alpha_1)
\Big[1+\frac{63}{2}\omega_4^A(2\alpha_3-1)\Big],
\nonumber\\
\Xi_{\rho}^{\rm BB}(\alpha_i) &=& 0\,,
\end{eqnarray}
where the new parameter $\omega_4^A$ is defined by the matrix element of the $J=4$ local operator 
in Eq.~(4.21) in Ref.~\cite{BB98}. From QCD sum rules \cite{BB98} one obtains a crude estimate:
\begin{equation}
    \left. \omega_4^A \right\vert_{\rm SR}= 0.8\pm 0.8\,.
\label{omega_SR}
\end{equation}
To the same $J=4$ accuracy 
\begin{eqnarray}
\label{two_rho_even_BB}
  \mathbb{A}^{\rm BB}(u) 
&=&
 \frac{200}{3} \zeta_4 u^2\bar u^2  -42 \zeta_4\omega_4^A
 \Big\{    u\bar u(2+13u\bar u)+   2 u^3(1+3\bar u+6 \bar u^2)\ln  u 
\nonumber\\&&{}+
   2\bar u^3(1+3 u+6 u^2)\ln \bar u \Big\},
\nonumber\\
   g_3^{\rm BB}(u) &=& - \frac{20}{3}\zeta_4 (1-6 u\bar u)\,.
\end{eqnarray}
To avoid misunderstanding, note that for this comparison we 
suppressed the Wandzura--Wilczek contributions of 
twist--two and twist--three operators to the coefficients 
in the conformal expansion in \cite{BB98} and only retained the genuine twist--four contributions.

Performing the conformal expansion of the renormalon model, \eq{three_rho_even_as}, up to $J=4$ we 
recover the structure of \eq{BB_three_part} predicting:
\begin{equation}
   \left.\omega_4^A \right\vert_{\rm Ren}= -\frac19,  
\label{omega_ren}
\end{equation}
which can be contrasted with the sum--rule result in \eq{omega_SR}. Similar to the pion case, this
number is not sensitive to the shape of the leading--twist DA.

\begin{figure}[t]
\centerline{
\epsfig{file=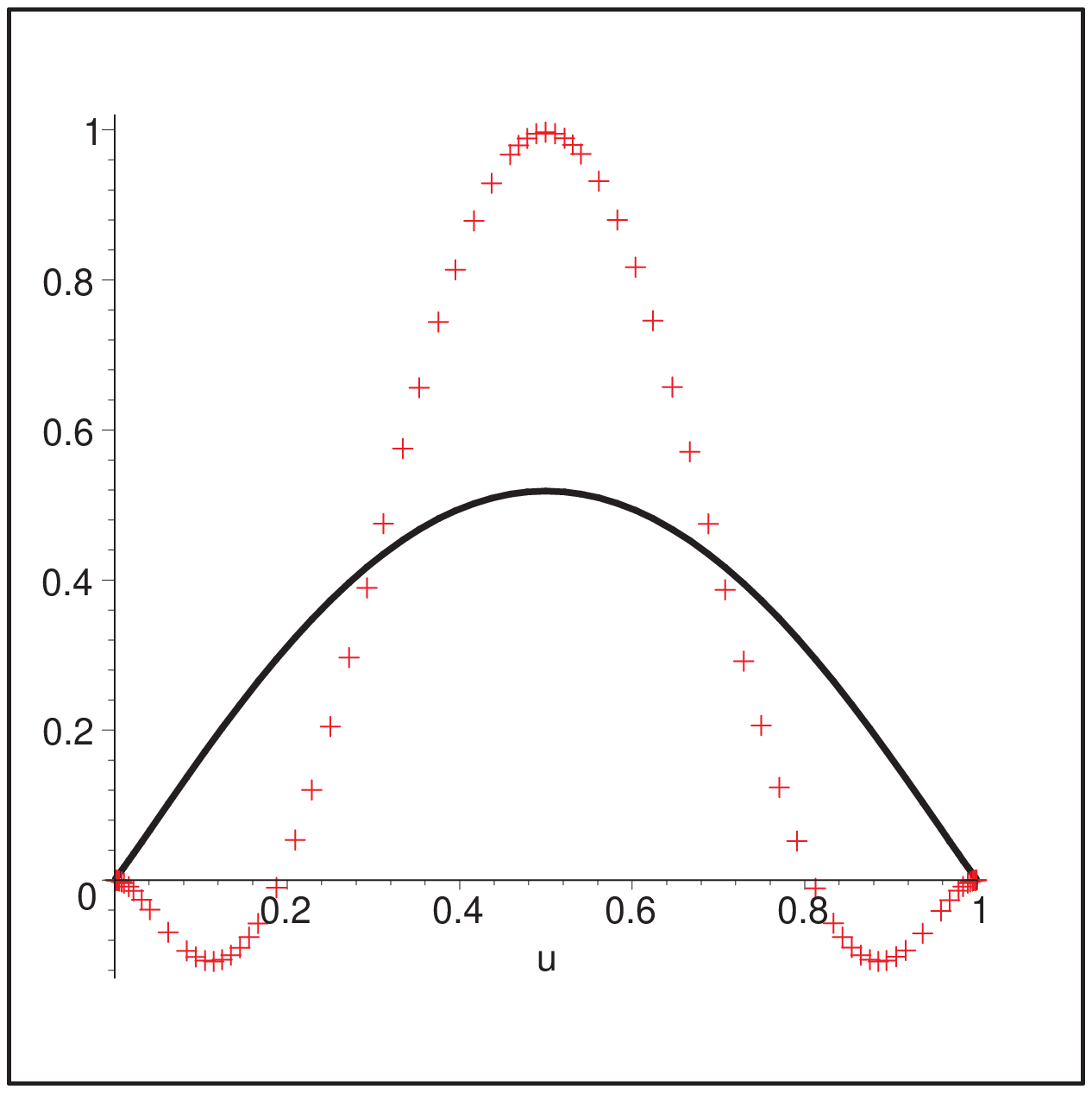,angle=-90,width=6.4cm}\hspace*{20pt}
\epsfig{file=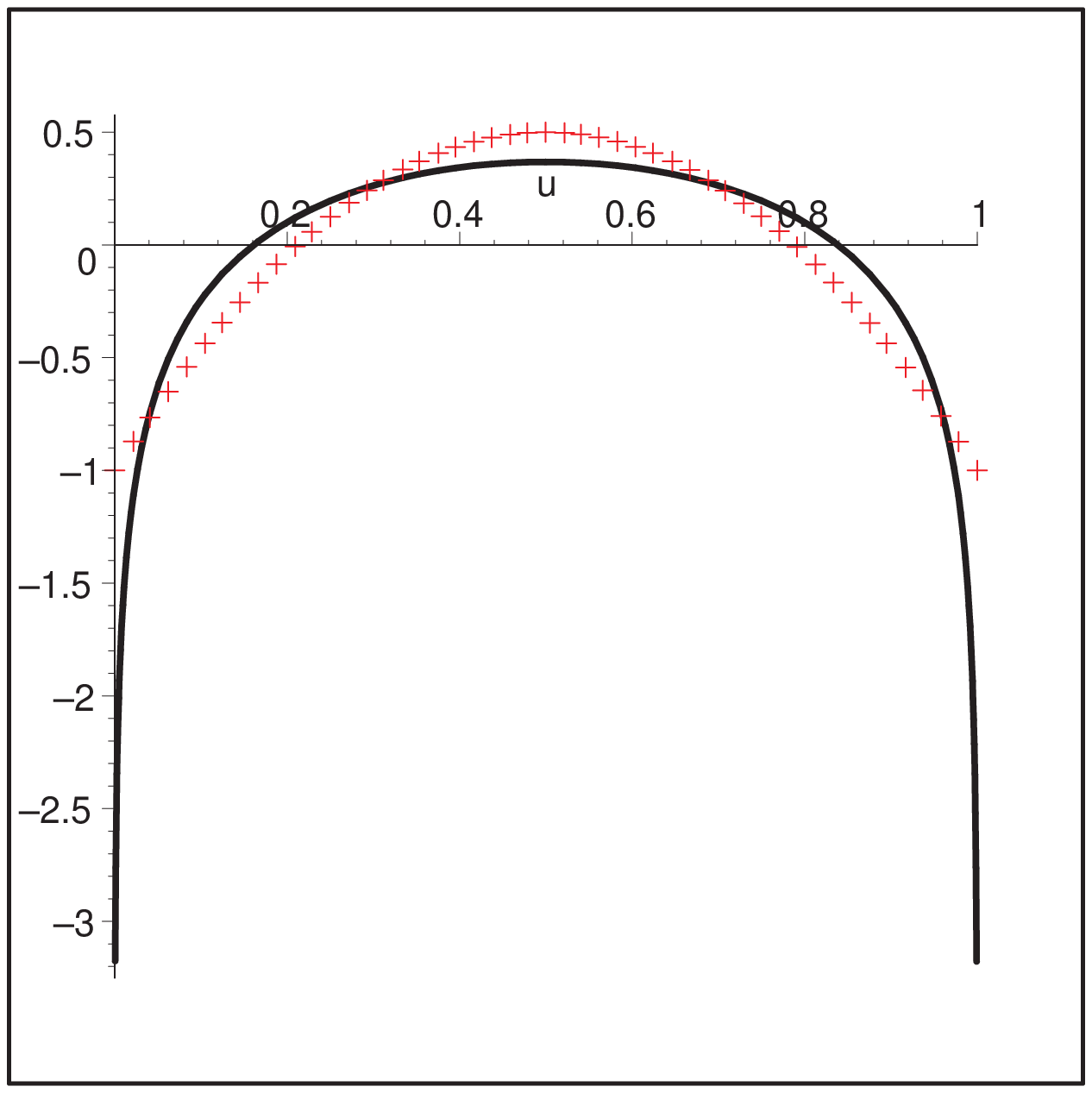,angle=-90,width=6.4cm}}
 \vspace*{20pt}
\caption{\small The two--particle twist--four DA of the $\rho$ in the chiral--even sector, 
$\mathbb{A}(u)$ (left) and $g_3(u)$ (right) based on the renormalon model of \eq{two_rho_even_as} 
(full line)
and model of Ref. \cite{BB98}, given by \eq{two_rho_even_BB} (crosses), 
which uses the first two orders in the conformal expansion
with the sum--rule estimate $\omega_4^A=0.8$. The normalization is chosen to be the same $\zeta_4=0.15$.
\label{A_and_G3_vs_BB} }
\end{figure}

\begin{figure}[t]
\centerline{
\epsfig{file=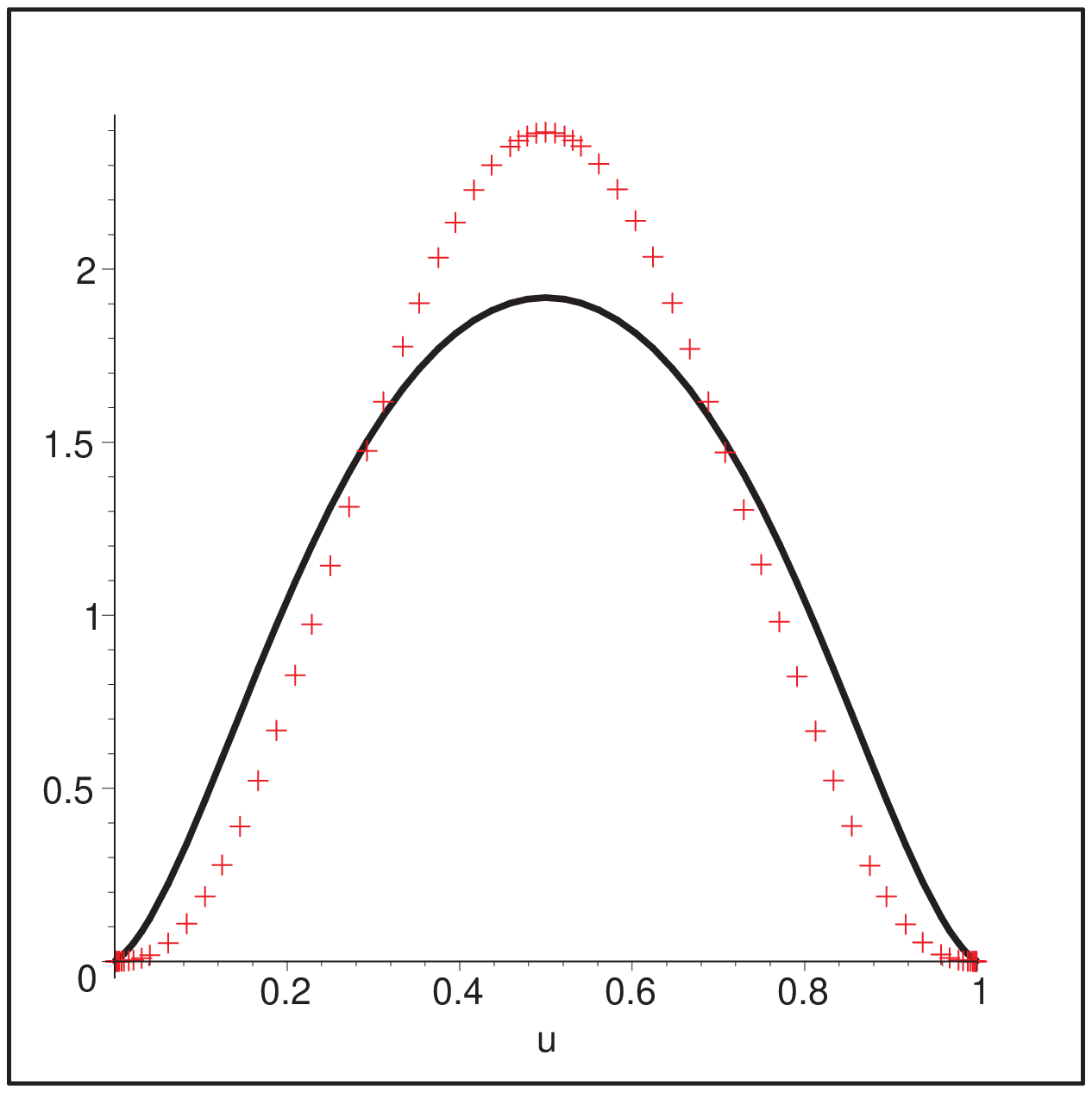,angle=-90,width=6.4cm}\hspace*{20pt}
\epsfig{file=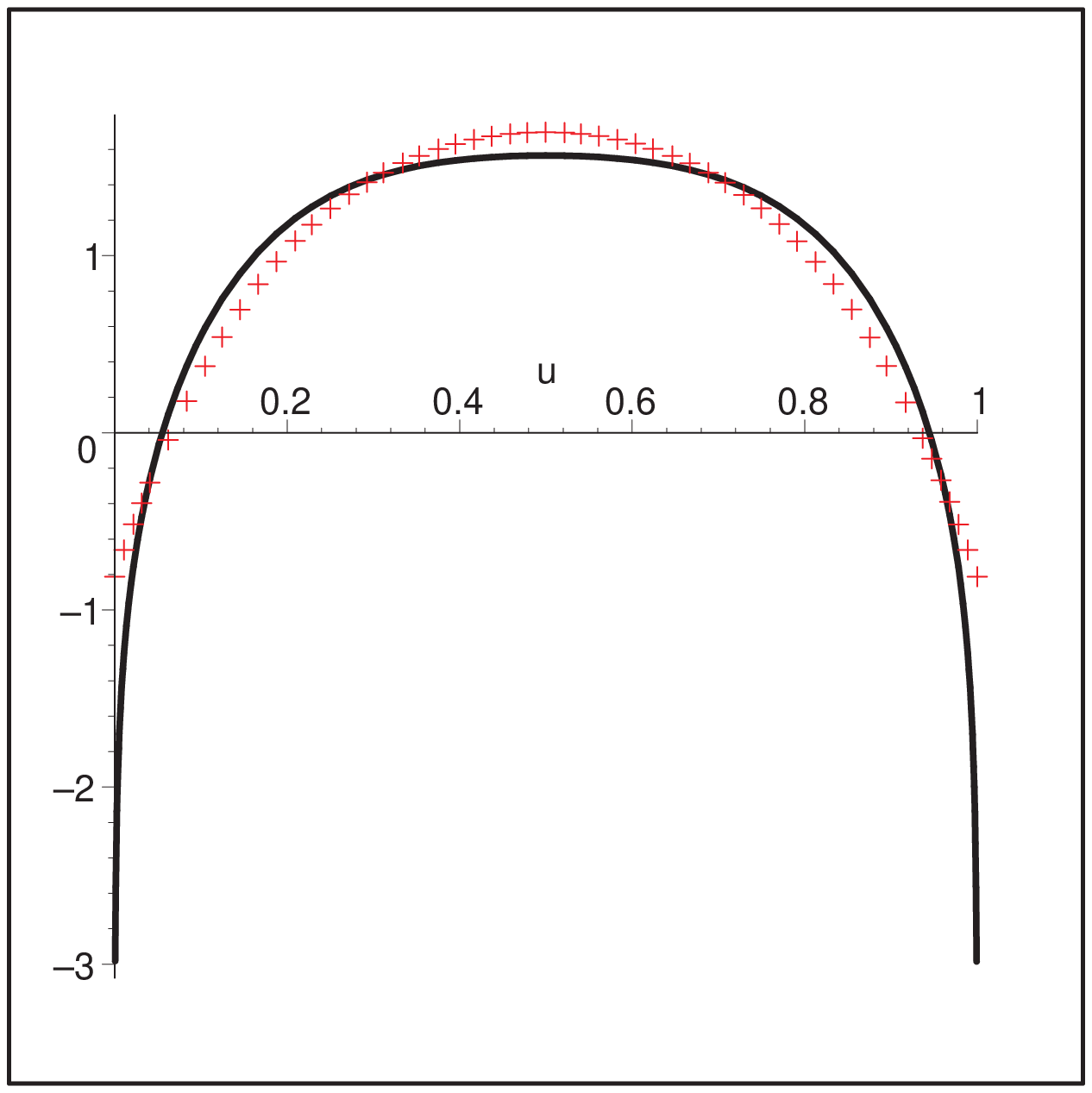,angle=-90,width=6.4cm}}
 \vspace*{20pt}
\caption{\small The same as Fig.~\ref{A_and_G3_vs_BB}, but including (in both models) 
the Wandzura--Wilczek contributions as given by Eqs. (4.22) and (4.23) in Ref. \cite{BB98} 
with the parameters of Table 2 there, except that we consistently use the asymptotic leading--twist 
DA, so $a_2^{\parallel}=0$.\label{A_and_G3_vs_BB_inc_WW}}
\end{figure}

Figure \ref{A_and_G3_vs_BB} compares the renormalon model with that of Ref. \cite{BB98}.
For $g_3(u)$ the main difference is in the asymptotic end--point behavior: it is logarithmic in the 
renormalon model and constant in the model of Ref. \cite{BB98}. For $\mathbb{A}(u)$ the difference
is more pronounced --- the asymptotic behavior is linear and quadratic in the two models, 
respectively --- and, moreover, it extends to the central region because of the very different $J=4$ 
contributions\footnote{Note that there is no $J=4$ contribution 
for $g_3(u)$.} which are determined  by \eq{omega_ren} and the central value of \eq{omega_SR}, respectively.
Figure~\ref{A_and_G3_vs_BB_inc_WW} makes a similar comparison but this time adding 
the Wandzura--Wilczek terms in both models.  
  
%
%

\subsection{Chiral--odd distribution amplitudes}
\subsubsection{Definitions}
For the chiral-odd operator involving the $\sigma_{\mu\nu}$--matrix  the
light--cone expansion to twist--four accuracy reads \cite{BB98}:
\begin{eqnarray}
\langle 0|\Big[\bar d(-x) \sigma_{\mu \nu}
u(x)\Big]_{\mu_F^2}\!|\rho^+\!(P,\lambda)\rangle & = &
 i f_{\rho}^{T} \!\left\{ (e^{(\lambda)}_{\mu}\!P_\nu -
e^{(\lambda)}_{\nu}\!P_\mu )
\!\!\int_{0}^{1}\!\!\! \!du\, e^{-i \xi P x}
\bigg[\phi_{\perp}(u) +\frac{m_\rho^2x^2}{4} {\mathbb A}_T(u)\bigg] \right.
\nonumber \\
& &{}+ (P_\mu x_\nu - P_\nu x_\mu )
\frac{e^{(\lambda)} x}{(P x)^{2}}
m_{\rho}^{2}
\int_{0}^{1} \!du\, e^{-i \xi P x} {\mathbb B}_T (u)
\nonumber \\
& & \left.{}+ \frac{1}{2}
(e^{(\lambda)}_{ \mu} x_\nu -e^{(\lambda)}_{ \nu} x_\mu)
\frac{m_{\rho}^{2}}{P  x}
\int_{0}^{1} \!du\, e^{-i \xi P x} {\mathbb C}_T(u) \right\},
\label{eq:OPE2}
\end{eqnarray}
where the tensor coupling $f_\rho^T$ is given by
\begin{equation}
\langle 0|\bar u(0) \sigma_{\mu \nu} 
d(0)|\rho^-(P,\lambda)\rangle = i f_{\rho}^{T}
(e_{\mu}^{(\lambda)}P_{\nu} - e_{\nu}^{(\lambda)}P_{\mu}).
\label{eq:frp}
\end{equation}
In comparison, the corresponding light--cone DA are defined as
\begin{eqnarray}
\langle 0|\Big[\bar d(-z) \sigma_{\mu \nu}
u(z)\Big]_{\mu_F^2}\!|\rho^+\!(P,\lambda)\rangle
& = & i f_{\rho}^{T} \bigg\{ ( e^{(\lambda)}_{\perp \mu}p_\nu -
e^{(\lambda)}_{\perp \nu}p_\mu )
\int_{0}^{1} \!du\, e^{-i \xi p  z} \phi_{\perp}(u, \mu_F^{2}) 
\nonumber \\
& &\hspace*{-20pt}+ (p_\mu z_\nu - p_\nu z_\mu )
\frac{e^{(\lambda)}  z}{(p z)^{2}}
m_{\rho}^{2}
\int_{0}^{1} \!du\, e^{-i \xi p  z} h_\parallel^{(t)} (u, \mu_F^{2})
\nonumber \\
& &\hspace*{-20pt} + \frac{1}{2}
(e^{(\lambda)}_{\perp \mu} z_\nu -e^{(\lambda)}_{\perp \nu} z_\mu)
\frac{m_{\rho}^{2}}{p  z}\!
\int_{0}^{1}\! \!du\, e^{-i \xi p  z} h_{3}(u, \mu_F^{2}) \bigg\},
\label{eq:tda} 
\end{eqnarray}
where $\phi_\perp(u)$ is the leading--twist DA of the transversely polarized
$\rho$--meson,  
$h_\parallel^{(t)}(u)$ is of twist--three and of no interest for our present purposes,  
and $h_3(u)$ is of twist--four. All three
functions $\phi = \{\phi_\perp,h_\parallel^{(t)},h_3\}$ are
normalized as 
$
\int_0^1 du\, \phi(u) = 1.
$
Matching (\ref{eq:tda}) with the light-cone expansion in Eq.~(\ref{eq:OPE2}) one obtains
\begin{eqnarray}
\label{BC_T_relations}
   {\mathbb B}_T(u) &=& h_\parallel^{(t)}(u) -\frac{1}{2}\phi_\perp(u)-
              \frac{1}{2} h_3(u),
\nonumber\\
   {\mathbb C}_T(u) &=& h_3(u)-\phi_\perp(u).
\end{eqnarray}
The remaining invariant amplitude ${\mathbb A}_T(u)$ describes the 
transverse momentum distribution in the leading--twist component of the 
wave function. We end up with two two--particle twist--four 
DA of the transversely polarized $\rho$--meson, ${\mathbb A}_T(u)$ and $h_3(u)$ which are counterparts 
of the distributions ${\mathbb A}(u)$ and $g_3(u)$ for the longitudinally polarized $\rho$--meson and
 of the pion DA $\phi_1^{(4)}(u)$ and $\phi_2^{(4)}(u)$ defined in \eq{off}. The precise correspondence will be given below.   

The three--particle DA are even more numerous than 
in the chiral-even case and can be defined as \cite{BB98}:
\begin{eqnarray}
\lefteqn{\langle 0|\bar d(-z) \sigma_{\alpha\beta}
         gG_{\mu\nu}(vz)
         u(z)|\rho^+(P,\lambda)\rangle \ =}\hspace*{1.6cm}\nonumber\\
&=& f_{\rho}^T m_{\rho}^2 \frac{e^{(\lambda)} z }{2 (p  z)}
    \bigg[ p_\alpha p_\mu g^\perp_{\beta\nu}
     -p_\beta p_\mu g^\perp_{\alpha\nu}
     -p_\alpha p_\nu g^\perp_{\beta\mu}
     +p_\beta p_\nu g^\perp_{\alpha\mu} \bigg]
      {\cal T}(v,pz)
\nonumber\\
&+& f_{\rho}^T m_{\rho}^2
    \bigg[ p_\alpha e^{(\lambda)}_{\perp\mu}g^\perp_{\beta\nu}
     -p_\beta e^{(\lambda)}_{\perp\mu}g^\perp_{\alpha\nu}
     -p_\alpha e^{(\lambda)}_{\perp\nu}g^\perp_{\beta\mu}
     +p_\beta e^{(\lambda)}_{\perp\nu}g^\perp_{\alpha\mu} \bigg]
      T_1(v,pz)
\nonumber\\
&+& f_{\rho}^T m_{\rho}^2
    \bigg[ p_\mu e^{(\lambda)}_{\perp\alpha}g^\perp_{\beta\nu}
     -p_\mu e^{(\lambda)}_{\perp\beta}g^\perp_{\alpha\nu}
     -p_\nu e^{(\lambda)}_{\perp\alpha}g^\perp_{\beta\mu}
     +p_\nu e^{(\lambda)}_{\perp\beta}g^\perp_{\alpha\mu} \bigg]
      T_2(v,pz)
\nonumber\\
&+& \frac{f_{\rho}^T m_{\rho}^2}{pz}
    \bigg[ p_\alpha p_\mu e^{(\lambda)}_{\perp\beta}z_\nu
     -p_\beta p_\mu e^{(\lambda)}_{\perp\alpha}z_\nu
     -p_\alpha p_\nu e^{(\lambda)}_{\perp\beta}z_\mu
     +p_\beta p_\nu e^{(\lambda)}_{\perp\alpha}z_\mu \bigg]
      T_3(v,pz)
\nonumber\\
&+& \frac{f_{\rho}^T m_{\rho}^2}{pz}
    \bigg[ p_\alpha p_\mu e^{(\lambda)}_{\perp\nu}z_\beta
     -p_\beta p_\mu e^{(\lambda)}_{\perp\nu}z_\alpha
     -p_\alpha p_\nu e^{(\lambda)}_{\perp\mu}z_\beta
     +p_\beta p_\nu e^{(\lambda)}_{\perp\mu}z_\alpha \bigg]
      T_4(v,pz),\hspace*{1cm}
\label{eq:T3}\\
\lefteqn{\langle 0|\bar d(-z)
         gG_{\mu\nu}(vz)
         u(z)|\rho^+(P,\lambda)\rangle
\ =\ i f_{\rho}^T m_{\rho}^2
 \bigg[e^{(\lambda)}_{\perp\mu}p_\nu-e^{(\lambda)}_{\perp\nu}p_\mu\bigg] S(v,pz),}
\hspace*{1.6cm}\nonumber\\
\lefteqn{\langle 0|\bar d(-z)
         ig\widetilde G_{\mu\nu}(vz)\gamma_5
         u(z)|\rho^+(P,\lambda)\rangle
\ =\ i f_{\rho}^T m_{\rho}^2
 \bigg[e^{(\lambda)}_{\perp\mu}p_\nu-e^{(\lambda)}_{\perp\nu}p_\mu\bigg]
  \widetilde S(v,pz).}\hspace*{1.6cm}\label{eq:2.21}
\end{eqnarray}
Of these seven amplitudes ${\cal T}$ is of twist three  and the other six
of twist four; higher--twist terms are suppressed.

We also introduce one more twist--four DA as follows
\begin{equation}
\label{Xi_T_rho}
\langle 0| \bar{d}(-z) \sigma_{\mu\nu}
 gD^{\alpha}G_{\alpha\beta}(vz) u(z) | \rho^+(P,\lambda)\rangle
= if_{\rho}m_{\rho}^2\,\bigg[e^{(\lambda)}_{\perp\mu}p_{\nu}-e^{(\lambda)}_{\perp\nu}p_{\mu}\bigg]
p_{\beta}\,\Xi^T_\rho(v,pz)\,.
\end{equation}

As in the other cases the twist--four two--particle DA do not present 
independent degrees of freedom and can be expressed in terms of three--particle 
DA using EOM~\cite{BB98}, see \eq{equ_motion_odd} below.

\subsubsection{Renormalon model and comparison with Ref. \cite{BB98}\label{ren_model_odd}}

Computing the first UV--renormalon contribution to the operator
 $\bar{d}(-z) \sigma_{\alpha\beta}G_{\mu\nu}(vz)u(z)$,
and taking the relevant Lorentz projections we find
\begin{eqnarray}
\label{UV_ren_rho_odd}
&&\hspace*{-5pt} \delta_{\UV}\Big\{\,\bar{d}(-z)
e_{\perp\nu}^{(\lambda)}z^\beta\,\sigma_{\mu\beta}\,g{G}^{\mu\nu} (vz) u(z) \Big\}
                    \nonumber       \\
&&\hspace*{50pt}=-ic\Lambda^2\, e_{\perp\rho}^{(\lambda)}z_{\varepsilon}
 \int_0^1 da (1+2a)\,\bigg[\bar d(-y)\sigma^{\rho\varepsilon} u(z)-
  \bar d(-z)\sigma^{\rho\varepsilon} u(\widetilde{y})
 \bigg],            \nonumber       \\
 &&\hspace*{-5pt}  \delta_{\UV}\Big\{\,\bar{d}(-z)
e^{(\lambda)}_{\perp\beta}z^\nu\,\sigma^{\mu\beta}\,g{G}_{\mu\nu} (vz) u(z) \Big\}
  \nonumber
  \\&&\hspace*{50pt}=ic\Lambda^2\, e^{(\lambda)}_{\perp\rho}z_{\varepsilon}\,\int_0^1 da \,(1-2a)\,
 \bigg[\bar d(-y)\sigma^{\rho\varepsilon} u(z)-
  \bar d(-z)\sigma^{\rho\varepsilon} u(\widetilde{y})
 \bigg],     \nonumber  \\
 &&\hspace*{-5pt} \delta_{\UV}\Big\{\,\bar{d}(-z)
e^{(\lambda)}_{\perp\beta} z_\alpha z_\mu\,\sigma^{\alpha\beta}\,gG^{\mu\nu} (vz) u(z) \Big\}
  \nonumber
  \\&&\hspace*{50pt}=-2ic\Lambda^2\,z^{\nu}z_{\rho}e^{(\lambda)}_{\perp\varepsilon} \int_0^1 da \,a\,
 \bigg[\bar d(-y)\sigma^{\rho\varepsilon} u(z)-
  \bar d(-z)\sigma^{\rho\varepsilon} u(\widetilde{y})
 \bigg],            \nonumber       \\
 &&\hspace*{-5pt}  \delta_{\UV}\Big\{\,\bar{d}(-z)
e^{(\lambda)}_{\perp\nu}z_\alpha z_\mu\,\sigma^{\alpha\beta}\,gG^{\mu\nu} (vz) u(z) \Big\}
  \nonumber
  \\&&\hspace*{50pt}=-ic\Lambda^2\,z^{\beta}z_{\rho}e^{(\lambda)}_{\perp\varepsilon}\int_0^1 da \,
 \bigg[\bar d(-y)\sigma^{\rho\varepsilon} u(z)-
  \bar d(-z)\sigma^{\rho\varepsilon} u(\widetilde{y})
 \bigg],            
\end{eqnarray}
where $y$ and $\widetilde{y}$ are defined below \eq{UV_div}.
In a similar manner, for the operators $\bar{d}(-z) G_{\alpha\beta}(vz)u(z)$ and $\bar{d}(-z)
 \gamma_5i\widetilde{G}_{\alpha\beta}(vz)u(z)$ we obtain
\begin{eqnarray}
\label{UV_ren_rho_odd_more}
&&\hspace*{-5pt} \delta_{\UV}\Big\{\bar{d}(-z)e^{(\lambda)}_{\perp\nu}z^{\mu}\,
gG_{\mu\nu} (vz) u(z) \Big\}
 =\,c\Lambda^2\, z^{\alpha} e^{(\lambda)}_{\perp\beta}
\int_0^1 da \bigg[\bar d(-y)\sigma_{\alpha\beta} u(z)+
  \bar d(-z)\sigma_{\alpha\beta} u(\widetilde{y})
 \bigg],
\nonumber \\                \nonumber 
 &&\hspace*{-5pt}  
\delta_{\UV}\Big\{\bar{d}(-z)e^{(\lambda)}_{\perp\nu}z^{\mu}  \gamma_{5}
ig\widetilde{G}_{\mu\nu} (vz) u(z) \Big\}
 =\,-c\Lambda^2\, z^{\alpha} e^{(\lambda)}_{\perp\beta}\int_0^1\!\! da \,
\bigg[\bar d(-y)\sigma_{\alpha\beta} u(z)+
  \bar d(-z)\sigma_{\alpha\beta} u(\widetilde{y})
 \bigg],\\
\end{eqnarray}
and finally for the operator $\bar{d}(-z) \sigma_{\mu\nu}
 gD^{\alpha}G_{\alpha\beta}(vz) u(z)$ we get
\begin{eqnarray}
\label{UV_ren_DG_odd}
&&\hspace*{-5pt} \delta_{\UV}\Big\{ z^{\beta}\bar d(-z)
gD^{\alpha}G_{\alpha\beta}(vz)\sigma_{\mu\nu} u(z)\Big\}
\nonumber \\&&\hspace*{50pt}=
 -4ic\Lambda^2\int_0^1da\,a\frac{d}{da}\bigg[\frac{1}{1+v}\,\bar d(-y)\sigma_{\mu\nu}
u(z)
+\,\frac{1}{1-v}\bar d(-z)\sigma_{\mu\nu} u(\widetilde{y}) \bigg].
\end{eqnarray}

Taking the matrix elements of the operators in
\eq{UV_ren_rho_odd} through \eq{UV_ren_DG_odd} between the vacuum and the $\rho$--meson state we extract
 the renormalon ambiguity for these twist--four DA in terms of $\phi_{\perp}(u)$.
Going over from the renormalon ambiguity to a model, one has to take into account that 
in the present case there exist {\em two} independent local operators of the lowest dimension
that have proper quantum numbers: 
\begin{eqnarray}
\label{eq:defzetaT}
\langle 0|\bar d(0) gG_{\mu \nu}(0)u(0)|\rho^+(P,\lambda)\rangle &=&
  if_\rho^T m_\rho^2 \zeta^T_4(
e^{(\lambda)}_{\mu}P_\nu - e^{(\lambda)}_{\nu}P_\mu),
\nonumber\\
\langle 0|\bar d(0) g\widetilde G_{\mu \nu}(0)i\gamma_5
   u(0)|\rho^+(P,\lambda)\rangle &=&
  if_\rho^T m_\rho^2 \widetilde \zeta^T_4(
e^{(\lambda)}_{\mu}P_\nu - e^{(\lambda)}_{\nu}P_\mu).
\end{eqnarray}
The parameters $\zeta_4^T \pm \widetilde\zeta_4^T$
renormalize multiplicatively with different anomalous dimensions \cite{BBK89}
and from the QCD sum rules one finds \cite{BBK89,BB98}
\begin{eqnarray}
\label{zeta_T_val}
      \zeta_4^T - \widetilde{\zeta}_4^T &=& 0.2\pm 0.1\,,
\nonumber\\
      \zeta_4^T + \widetilde{\zeta}_4^T &=& 0\,.
\end{eqnarray}
The vanishing (or smallness) of the second number in \eq{zeta_T_val} appears 
as a consequence of vanishing of the leading 
contribution to the corresponding correlation function, see Appendix C in \cite{BB98}. 

By comparison of the expressions in \eq{UV_ren_rho_odd_more} 
we observe that the leading UV--renormalon contribution to the operator  
$\bar d(-z)[gG_{\mu \nu}+g\widetilde G_{\mu \nu}i\gamma_5](vz)u(z)$\ and thus to 
$\zeta_4^T + \widetilde{\zeta}_4^T$ vanishes
as well. Therefore, within the renormalon model $\widetilde\zeta_4^T = - \zeta_4^T$ and similarly to 
the pion and the chiral--even $\rho$--meson cases the model has only one parameter. 

Collecting everything and making the substitution
\begin{equation}
    c\Lambda^2 \to \frac12 \zeta_4^T m_\rho^2
\label{zeta_4_T}
\end{equation}
we obtain
\begin{eqnarray}
\label{three_rho_odd}
&&\hspace*{-10pt}
  T_1(\alpha_i) =-T_3(\alpha_i)=
\zeta_4^T\left[
\frac{\alpha_2\phi_{\perp}(\alpha_1)}{(1-\alpha_1)^2}
-\frac{\alpha_1\phi_{\perp}
(\alpha_2)}{(1-\alpha_2)^2}
\right],
\nonumber\\
&&\hspace*{-10pt}
 T_2(\alpha_i) =\phantom{-}T_4(\alpha_i) =
-\frac12 \zeta_4^T\left[
\frac{\phi_{\perp}(\alpha_1)}{(1-\alpha_1)}
-\frac{\phi_{\perp}(\alpha_2)}{(1-\alpha_2)}
\right],
\nonumber\\
&&\hspace*{-7pt}
 S(\alpha_i)=  -\,\widetilde{S}(\alpha_i)
\,=\,
\frac12 \zeta_4^T\left[
\frac{\phi_{\perp}(\alpha_1)}{1-\alpha_1}+\frac{\phi_{\perp}(\alpha_2)}{1-\alpha_2}
\right],
\nonumber\\
&&\hspace*{-7pt}\Xi^T_{\rho}(\alpha_i) = -
 2\zeta_4^T
\left[\frac{\alpha_2}{1-\alpha_1}\,\phi_{\perp}(\alpha_1)
-\frac{\alpha_1}{1-\alpha_2}\phi_{\perp}(\alpha_2)
\right],
\end{eqnarray}
where we used the symmetry $\phi_{\perp}(u)=\phi_{\perp}(1-u)$.
The correspondence with the pion DA is as follows: upon replacing $\delta^2$ by $3\zeta_4^T$,
$\Phi_{\parallel}^{\pi} \longrightarrow  T_1^{\rho}$, $\Phi_{\perp}^{\pi} \longrightarrow  -T_2^{\rho}$,
$\Psi_{\perp}^{\pi} \longrightarrow  S^{\rho}$ and $\Xi_\pi\longrightarrow \Xi^T_{\rho}$.
There is no analog for $\Psi_{\parallel}^{\pi}$.

Finally, the two--particle twist--four DA are restored using EOM (see Appendix~\ref{EOM_UV}):
\begin{eqnarray}
\label{two_rho_odd}
  \mathbb{A}_T(u)&=&8\zeta_4^T\int_0^1
dv\,\phi_\perp(v)\Bigg[\theta(u>v)\;\frac{1}{\bar v^2}\left({\bar u +(u-v)
  \ln\frac{u-v}{\bar v}}\right)\nonumber\\
  &&\hspace*{80pt}+\theta(u<v)\;\frac{1}{v^2}\left({u+(v-u)\ln\frac{v-u}{v}}\right)\Bigg],
\nonumber\\
h_3(u) &=& 0\,.
\end{eqnarray}
Note that there is no renormalon ambiguity  in $h_3(u)$ at the level of a single dressed gluon,
and therefore in our model this DA vanishes. 
Also note that apart from the different overall normalization ($\delta^2 \longrightarrow 3\zeta_4^T$) 
$\mathbb{A}_T(u)$ is the same as  $16\phi^{(4)}_1(u)$ in the pion 
case.

Choosing the asymptotic leading--twist DA $\phi_\perp(u) = 6u(1-u)$ we obtain simple expressions
\begin{eqnarray}
\label{three_rho_odd_as}
&&\hspace*{-10pt}
  T_1(\alpha_i) =-T_3(\alpha_i)=6\zeta_4^T
\alpha_1\alpha_2\left[\frac{1}{1-\alpha_1}-\frac{1}{1-\alpha_2}\right]\,,
\nonumber\\
&&\hspace*{-10pt}
 T_2(\alpha_i) =\phantom{-}T_4(\alpha_i) =3\zeta_4^T [\alpha_2-\alpha_1]\,,
\nonumber\\
&&\hspace*{-7pt}
 S(\alpha_i)=  -\,\widetilde{S}(\alpha_i) 
\,=\, 3\zeta_4^T \left[\alpha_1+\alpha_2\right],
\nonumber\\
&&\hspace*{-7pt}
\Xi^T_{\rho}(\alpha_i) =0
\end{eqnarray} 
and 
\begin{eqnarray}
\label{two_rho_odd_as}
  \mathbb{A}_T(u)&=&48\zeta_4^T\bigg\{\bar u \bigg[\ln(\bar u) - \mbox{\rm Li}_2 (\bar u)\bigg]
+u\bigg[\ln (u)- \mbox{\rm Li}_2( u )\bigg]
-u\bar u+\frac{\pi^2}6\bigg\},
\nonumber\\
h_3(u) &=& 0\,.
\end{eqnarray}

These results can be compared with the model of \cite{BB98}.
The structure of the conformal expansion to spin $J=4$ accuracy is more complicated in this case
as it involves three parameters. Following \cite{BB98} we write 
\begin{eqnarray}
\label{odd_conf_exp}
T_1({\alpha_i}) & = & \phantom{-}120 t_{10} (\alpha_2-\alpha_1)
\alpha_2\alpha_1\alpha_3,\nonumber\\
T_2({\alpha_i}) & = & -30\alpha_3^2 (\alpha_2-\alpha_1)
\left[ \widetilde
  s_{00} + \frac{1}{2}\,\widetilde s_{10}\, (5\alpha_3-3) + \widetilde
  s_{01} \alpha_3\right],\nonumber\\
T_3({\alpha_i}) & = & -120 \widetilde t_{10} (\alpha_2-\alpha_1)
\alpha_2\alpha_1\alpha_3,\nonumber\\
T_4({\alpha_i}) & = & \phantom{-}
30\alpha_3^2 (\alpha_2-\alpha_1) \left[
  s_{00} + \frac{1}{2}\, s_{10}\, (5\alpha_3-3) +
  s_{01} \alpha_3\right], \\
{ S}({\alpha_i}) & = & 30\alpha_3^2\! \left
  [ s_{00}\,(1\!-\!\alpha_3) +
  s_{10} \left\{\!\alpha_3 (1\!-\!\alpha_3) - \frac{3}{2}\, (\alpha_2^2 +
  \alpha_1^2) \right\} + s_{01} \left\{ \alpha_3 (1\!-\!\alpha_3) -
  6\alpha_2\alpha_1\right\} \right],\nonumber\\
\widetilde{ S}({\alpha_i}) & = & 30\alpha_3^2\! \left[ \widetilde
  s_{00}\,(1\!-\!\alpha_3)  +
  \widetilde s_{10} \left\{\!\alpha_3 (1\!-\!\alpha_3) - \frac{3}{2}\, (\alpha_2^2 +
  \alpha_1^2) \right\} + \widetilde s_{01} \left\{ \alpha_3 (1\!-\!\alpha_3) -
  6\alpha_2\alpha_1\right\} \right],\nonumber 
\end{eqnarray}
where, assuming that $\widetilde\zeta_4^T = -   \zeta_4^T$, one obtains
\begin{equation}
s_{00}  =  -\widetilde{s}_{00}=\zeta_4^T,
\end{equation}
and the remaining six coefficients involve three parameters
$\langle\!\langle Q^{(1)}\rangle\!\rangle$,
$\langle\!\langle Q^{(3)}\rangle\!\rangle$
and $\langle\!\langle Q^{(5)}\rangle\!\rangle$ defined by reduced matrix elements of local 
operators specified in Eq. (5.20) in \cite{BB98}:    
\begin{eqnarray}
\label{eq:CO}
s_{10} & = & 
 \frac{28}{55}\, \langle\!\langle Q^{(1)}\rangle\!\rangle +
\frac{7}{11}\, \langle\!\langle Q^{(3)}\rangle\!\rangle +
\frac{14}{3}\, \langle\!\langle Q^{(5)}\rangle\!\rangle,\nonumber\\
\widetilde{s}_{10} & = & 
 - \frac{28}{55}\, \langle\!\langle Q^{(1)}\rangle\!\rangle -
\frac{7}{11}\, \langle\!\langle Q^{(3)}\rangle\!\rangle
+ \frac{14}{3}\, \langle\!\langle Q^{(5)}\rangle\!\rangle,\nonumber\\
s_{01} & = & 
 + \frac{49}{110}\, \langle\!\langle Q^{(1)}\rangle\!\rangle -
\frac{7}{22}\, \langle\!\langle Q^{(3)}\rangle\!\rangle
+ \frac{7}{3}\, \langle\!\langle Q^{(5)}\rangle\!\rangle,\nonumber\\
\widetilde s_{01} & = & 
 - \frac{49}{110}\, \langle\!\langle Q^{(1)}\rangle\!\rangle +
\frac{7}{22}\, \langle\!\langle Q^{(3)}\rangle\!\rangle
 + \frac{7}{3}\, \langle\!\langle Q^{(5)}\rangle\!\rangle,\nonumber\\
t_{10} & = & 
 - \frac{63}{220}\, \langle\!\langle Q^{(1)}\rangle\!\rangle +
\frac{119}{44}\, \langle\!\langle Q^{(3)}\rangle\!\rangle,\nonumber\\
\widetilde t_{10} & = & \phantom{-} 
 \frac{63}{220}\, \langle\!\langle Q^{(1)}\rangle\!\rangle +
\frac{35}{44}\, \langle\!\langle Q^{(3)}\rangle\!\rangle.
\end{eqnarray}
Here we only show genuine twist--four contributions and suppress the Wandzura-Wilczek terms.
To the same accuracy 
\begin{eqnarray}
\label{eq:ATexp}
{\mathbb A}_T^{\rm BB}(u) & = & 120\,  \zeta_4^T \, u^2 \bar u^2 
- \Big(\frac{126}{55}\, \langle\!\langle Q^{(1)}\rangle\!\rangle
       + \frac{70}{11}\, \langle\!\langle Q^{(3)}\rangle\!\rangle\Big)
 \Big[ u\bar u (2+13 u\bar u)
\nonumber\\&&{}\hspace*{10pt}
 + \, 2u^3 (10-15u+6u^2) \ln u 
 + 2\bar u^3 (10-15\bar u + 6\bar u^2) \ln \bar u\Big],
\nonumber\\
h_3^{\rm BB} (u) & = & 0\,.
\end{eqnarray}
The particular model suggested in \cite{BB98} makes use of the QCD sum--rule estimate
\begin{equation}
\label{odd_sum-rules}
\langle\langle Q^{(1)}\rangle\rangle\big\vert_{\rm SR} = -0.15 \pm 0.15~\mbox{\cite{BBK89}}\,,
\qquad
\langle\langle Q^{(3)} \, \rangle \rangle\big\vert_{\rm SR} = 
\langle\langle Q^{(5)} \, \rangle \rangle\big\vert_{\rm SR}
=0\,.
\end{equation}

On the other hand, starting with the renormalon model in \eq{three_rho_odd_as} and \eq{two_rho_odd_as}
and isolating the $J=4$ contribution\footnote{Note that $T_2=T_4$ implies that
$\langle\langle Q^{(5)}\rangle\rangle=0$ while $T_1=-T_3$ leads to
$\langle\langle Q^{(3)}\rangle\rangle=\frac{3}{10}\langle\langle Q^{(1)}\rangle\rangle$.}
we obtain, using \eq{zeta_T_val}, 
\begin{equation}
\label{odd_renormalon}
\langle\langle Q^{(1)}\rangle\rangle\big\vert_{\rm Ren} 
= \frac{10}{3} \langle\langle Q^{(3)} \, \rangle \rangle\big\vert_{\rm Ren},  
\quad
\langle\langle Q^{(3)} \, \rangle \rangle\big\vert_{\rm Ren} =-\zeta_4^T=-0.10\pm0.05 
  , \quad 
\langle\langle Q^{(5)} \, \rangle \rangle\big\vert_{\rm Ren}
=0\,.
\end{equation} 
Note that the renormalon model prediction for $\langle\langle Q^{(1)}\rangle\rangle$ is consistent with  
the sum--rule estimate within errors and the main difference is that $\langle\langle Q^{(3)}\rangle\rangle$
is non-vanishing. With the numbers from \eq{odd_renormalon} the 
$J=4$ contribution to ${\mathbb A}_T(u)$ is enhanced by roughly factor four compared to the 
sum--rule estimate, but is still smaller than the leading $J=3$ term. 

We conclude by a numerical comparison of the two--particle DA $\mathbb{A}_T(u)$ in the renormalon model, 
\eq{two_rho_odd_as}, and the model of \cite{BB98} in \eq{eq:ATexp} with QCD sum--rule estimates of 
the parameters. As in previous cases, the difference is most pronounced in the end--point regions 
where the two expressions have different asymptotic behavior.  
\begin{figure}[t]
\centerline{
\epsfig{file=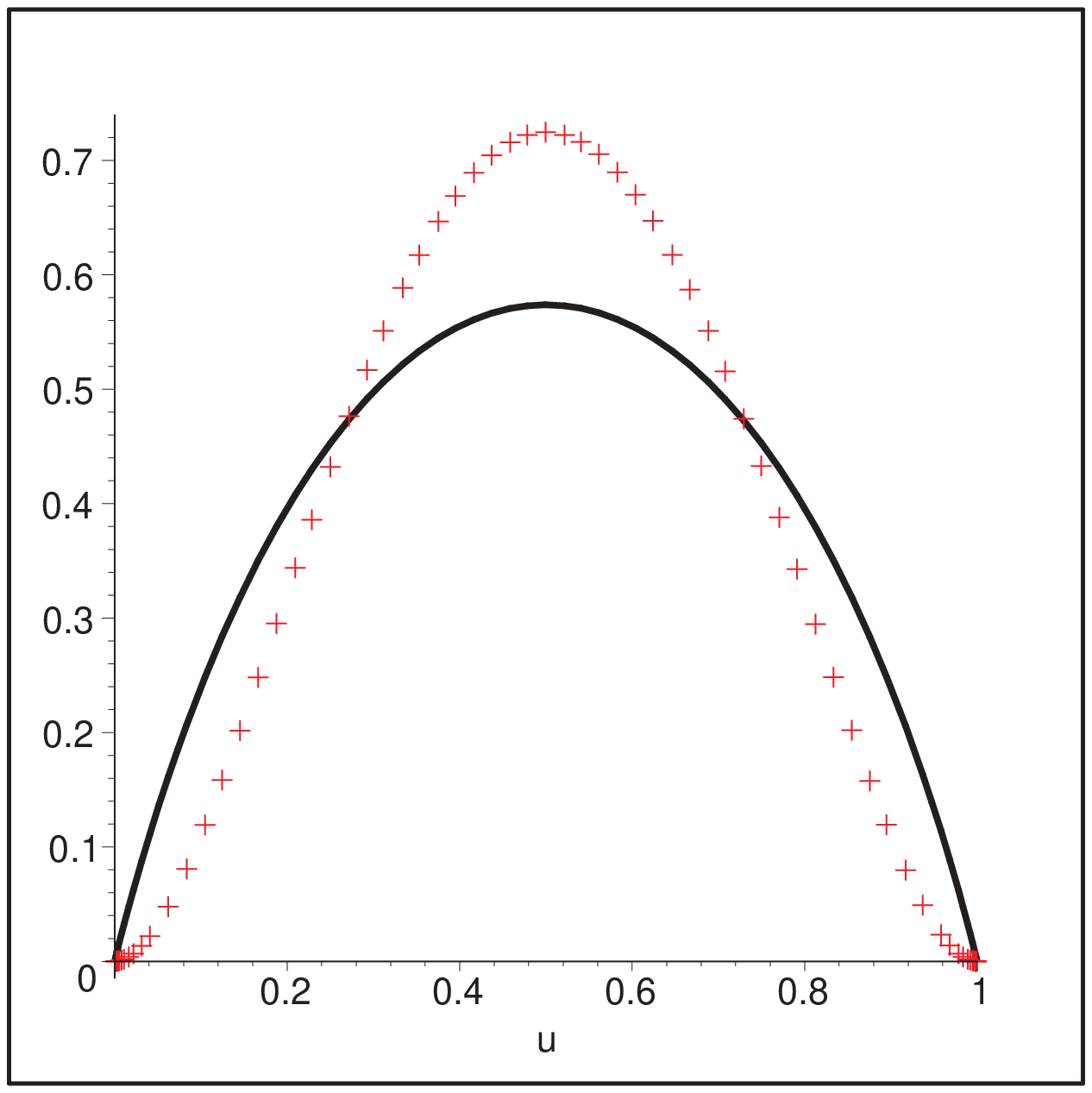,angle=-90,width=6.4cm}\hspace*{20pt}
\epsfig{file=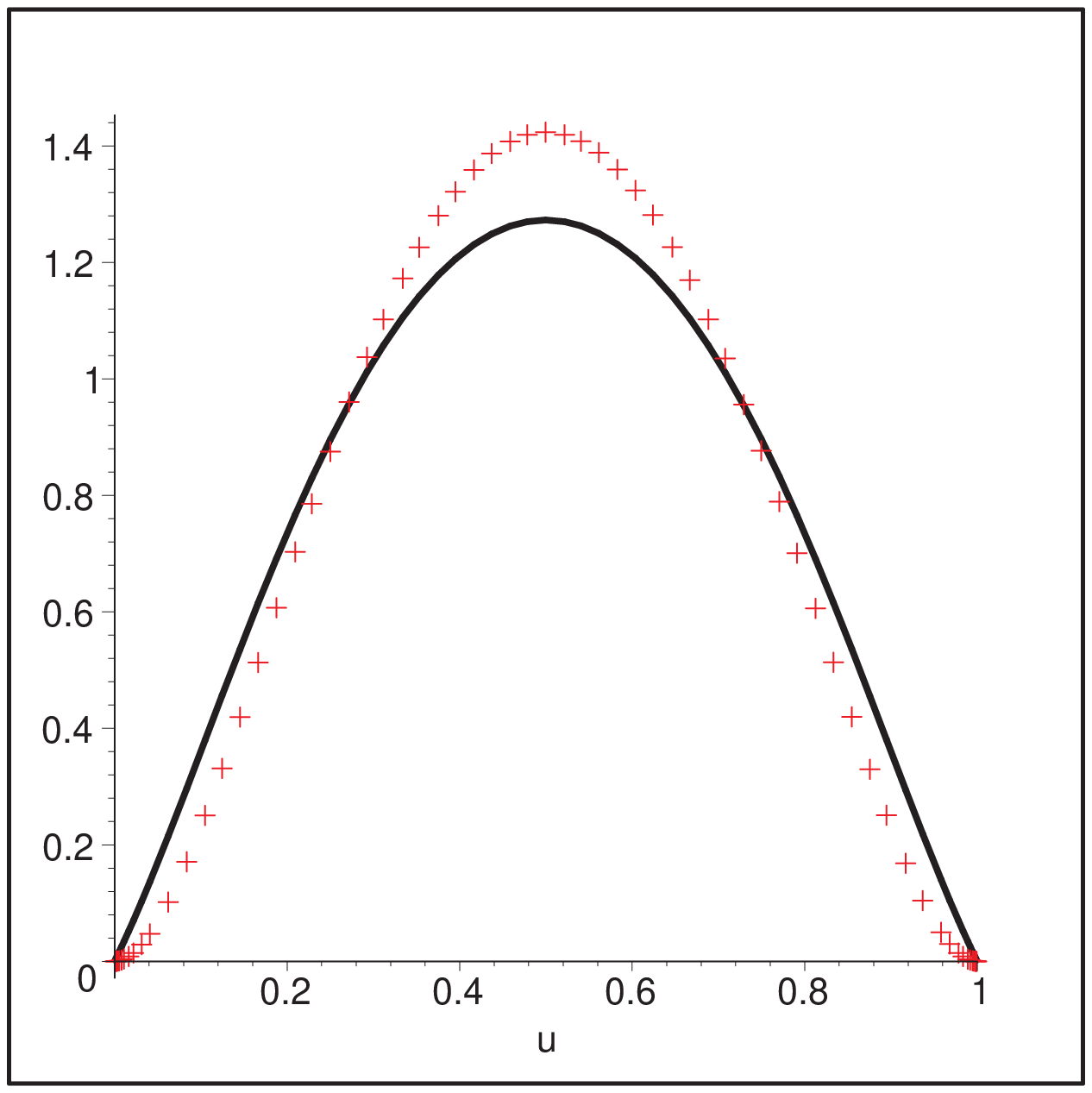,angle=-90,width=6.4cm}}
 \vspace*{20pt}
\caption{The two--particle twist--four DA $\mathbb{A}_T(u)$ without (left)
and with (right) the Wandzura--Wilczek terms. Each figure compares between the renormalon model (full line)
and model of Ref. \cite{BB98} (crosses) which uses the first two orders in the conformal expansion
with the sum--rule estimate of \eq{odd_sum-rules}. Note the different scales on the vertical axis.
\label{rho_off_plot} }
\end{figure}

\section{Conclusions\label{conc}}
\setcounter{equation}{0}

In this paper we have presented for the first time a systematic analysis of twist--four 
meson distribution amplitudes that goes beyond the first few orders in the conformal expansion.   
Our analysis is based on the study of the high--order behavior of perturbation theory 
in the single--dressed--gluon approximation which is equivalent to the study of one--loop power divergences 
of the contributing twist--four operators. In general, this calculation supports the conjecture that 
the shape of higher--twist DA is not far from the asymptotic form which is associated with  
operators with the lowest conformal spin. 
However, we find that the conformal expansion of the renormalon model
does not converge uniformly at the end points.
Consequently, the end--point behavior corresponding to the situation where one valence quark is soft, 
is qualitatively different between the sum over all spins  and any fixed order in the conformal expansion.
As the principal result, we obtain that the two--particle twist--four DA that describes the 
$k_\perp^2$ distribution of valence quarks in the meson --- $\phi^{(4)}_1(u)$ for pion, $\mathbb{A}(u)$  and 
$\mathbb{A}^T(u)$
for $\rho$--meson --- has the same linear falloff $\sim u(1-u)$ at $u\to 0,1$ as the leading--twist DA,
compared to the quadratic behavior $\sim u^2(1-u)^2$ of the asymptotic DA (lowest conformal spin).
Taking into account the limitations of our analysis this has to be understood as an upper bound.
The existence of such a bound is important for proofs of factorization theorems.

An attractive feature  of the renormalon approach is that it allows one to construct simple models of 
higher--twist DA with minimum number of non-perturbative parameters. 
In this paper we constructed such models for the pion
and for the $\rho$-meson with both the longitudinal and the transverse polarizations. The corresponding expressions 
are given in Sect.~\ref{pion_model}, \ref{ren_model_even} and \ref{ren_model_odd}, respectively. 
In each of these cases the {\em entire set} of twist--four DA is determined 
in terms of the leading--twist DA  with just one free parameter, the overall normalization, 
corresponding to the matrix element of a certain local operator.  
This approach presents a viable alternative to the models of Refs.~\cite{BF90,BB98} based on the two 
first orders in the conformal expansion. The spread between the predictions of these two approaches is a fair  
measure of uncertainty in our present understanding of higher--twist effects.    

In addition to giving, for the first time, an upper bound for the possible contribution of the operators with 
large conformal spins $J$, the renormalon model can also be used to estimate the next--to--lowest spin $J=4$ 
contributions. The corresponding estimates are given in \eq{eps_ren}, \eq{omega_ren} and \eq{odd_renormalon}
for the pion, longitudinal $\rho$--meson and transverse $\rho$--meson, respectively. These estimates 
are probably more reliable than the corresponding QCD sum--rule results, 
as it is known that the sum--rule approach does not work well for operators containing derivatives. 
In particular for the longitudinal $\rho$--meson there is a significant difference, compare \eq{omega_ren}   
and \eq{omega_SR}.

The renormalon approach can be applied in a straightforward manner to study yet higher power corrections,
of twist six and above, which are otherwise inaccessible. 
Higher infrared--renormalon ambiguities in the coefficient functions of
\eq{B} at $w_0=2,3$, etc. corresponding
to twist six, eight, etc. can be extracted from \eq{B1} and \eq{B2}
in full analogy with the calculation of the leading renormalon ambiguity  in Sect.~\ref{Borel_sing}. 
Assuming for simplicity the asymptotic leading--twist
DA, the ambiguities in the leading--twist part of $G_1$ read:
\begin{eqnarray}
\label{ambi_higher_ren}
 \delta_{\IR}^{w=2}\left\{C_1\otimes\phi_\pi\right\} &=&
-\frac{\pi C_F}{2\beta_0}\frac{{\rm e}^{\frac{10}{3}}}{4^2}
\left(-\Delta^2\Lambda^2\right)^2\, 2\left[u^3\ln(u)+\bar{u}^3\ln(\bar{u})+u\bar{u}\right],
\\ \nonumber
\delta_{\IR}^{w=3}\left\{C_1\otimes\phi_\pi\right\} &=&-
\frac{\pi C_F}{2\beta_0}\frac{{\rm e}^{\frac{15}{3}}}{4^3}
\left(-\Delta^2\Lambda^2\right)^3\, \frac{8}{3}u\bar{u}
\left[1+7u\bar{u} \right],
\\ \nonumber
\delta_{\IR}^{w=4}\left\{C_1\otimes\phi_\pi\right\} &=&-
\frac{\pi C_F}{2\beta_0}\frac{{\rm e}^{\frac{20}{3}}}{4^4}
\left(-\Delta^2\Lambda^2\right)^4\, \frac{1}{75} u\bar{u}
\left[31+199u\bar{u}-2u^2\bar{u}^2 \right].
\end{eqnarray}
In general, we find in the renormalon approach
that the asymptotic behavior of the form $\sim u\bar{u}$ persists in
the case of $G_1$ to all twists\footnote{An exception is twist six (the first line in
\eq{ambi_higher_ren}) where the behavior is $\sim u^2\bar{u}^2$ owing to a complete
cancellation between diagrams 1 and 3 in figure \ref{diagrams}.}.
For $G_2$, on the other hand, we find at any twist higher than four an
asymptotic behavior of the form $\sim u^2\bar{u}^2$. This result suggests that the DA of all twists
probably have the same universal power behavior in the end--point regions. This strong conjecture implies, 
in particular, that the twist expansion breaks down owing to the increasing singularity of 
higher--twist coefficient functions. 

The present study can be extended in several respects. From the theoretical point of view the renormalon
calculation that uses the modified gluon propagator in \eq{propa} can be understood as changing the scaling
dimension of the gluon field. It is, therefore, tempting to try to reconcile the renormalon model 
with conformal symmetry of QCD with modified conformal spin assignment for the fields. 
Alternatively, since the renormalon model predictions effectively reduce to the analysis of quadratic 
divergences of twist--four operators, they may have the symmetry of QCD  in two dimensions. 
Finally, there is a general problem of going beyond the large--$\beta_0$ approximation and taking anomalous 
dimensions into account. 

To summarize, we believe that the renormalon approach is useful for understanding the structure of 
higher--twist contributions in hard exclusive processes and allows one to obtain quantitative estimates. 
Phenomenological applications are numerous but go beyond the tasks of this work.    

\section*{Acknowledgements}
The work of E.G. and S.G. was supported by the DFG.

\appendix  
\renewcommand{\theequation}{\Alph{section}.\arabic{equation}}  

\section*{Appendices}

\section{\label{IRR}Single--dressed--gluon  calculation of the twist--two coefficient function}
\setcounter{equation}{0}

In this Appendix we present the detailed calculation of the
diagrams in Fig.~\ref{diagrams} to arrive at the
Borel--regularized expression for the leading--twist coefficient
function of the non--local operator in \eq{NL}. The calculation is
done in the Feynman gauge and for brevity we do not write
explicitly the Wilson line connecting operators at different
points.

We begin with  Diagram 1 in Fig.~\ref{diagrams} where the
gluon is exchanged between the quark and the Wilson line. The
latter is defined by~\eq{def_wilson_line}. The calculation of this
diagram plus its symmetric counterpart, where the gluon line is
attached to the other quark, yields:
\begin{eqnarray}
  \label{calc_vertex}
  \left\langle p_2 \left| {\rm T}\{\bar{d}(x_2) \gamma_{\nu} \gamma_{5} u(x_1)\}\right|p_1\right\rangle_1
  &=&\frac{4\pi^2 C_F}{\beta_0}\int_0^\infty \! dw\,
  {\rm e}^{\frac53 w}\,(-\Lambda^2)^w\int_0^1\! da \, e^{i(p_2 x_2-p_1 x_1)}
  \nonumber\\
  &&\hspace*{-110pt}\times\,\left[\bar{d}_{p_2}\gamma_\sigma \gamma_\mu
  \gamma_\nu \gamma_5 u_{p_1}\Delta^\sigma {\rm I}_1^\mu(p_2,a\Delta)+
  \bar{d}_{p_2}  \gamma_\nu \gamma_5 \gamma_\mu \gamma_\sigma u_{p_1}
  \Delta^\sigma {\rm I}_1^\mu(p_1,a\Delta)\right],
\end{eqnarray}
where $\bar{d}_{p_2}$ and $u_{p_1}$ are quark spinors, $\Delta = x_2-x_1$ 
and the momentum integral ${\rm I}_1^\mu(p,\Delta)$ is given by
\begin{eqnarray}
  \label{int1}
  {\rm I}_1^\mu(p,z)&=& \int\!\frac{d^4k}{(2\pi)^4}{\rm e}^{-ik\Delta}
  \frac{(p+k)^\mu}{(k^2)^{1+w}(p+k)^2}
\nonumber \\
&=& \frac{1}{16\pi^2}\frac{\Gamma(-w)}{\Gamma(1+w)}
  \left(\frac{z^2}{4}\right)^w \int_0^1\!db\, b^w {\rm e}^{i(1-b)p\Delta}
  \left(ibp-2w\frac{\Delta}{\Delta^2}\right)^\mu\,.
\end{eqnarray}
Inserting the result of the momentum integration into~\eq{calc_vertex},
contracting the indices and using the Dirac equation for massless quarks gives
\begin{eqnarray}
  \label{calc_vertex_2}
  \left<p_2\left|{\rm T}\{\bar{d}(x_2) \gamma_{\nu} \gamma_{5} u(x_1)\}\right|p_1\right>_1
  &=&-\frac{C_F}{2\beta_0}\int_0^\infty \!\! dw\,{\rm e}^{\frac53 w}\,
  \frac{\Gamma(-w)}{\Gamma(1+w)}\left(\frac{-\Lambda^2\Delta^2}{4}\right)^w
  \!\int_0^1\!\! da\, a^{2w}\int_0^1\!\!db\, b^w\nonumber\\
  &&\hspace*{-100pt}\times\,\left[\left(-i b p_2\cdot\Delta+\frac{w}{a}\right)
  {\rm e}^{ia(1-b) p_2\cdot\Delta}+\left(-i b p_1\cdot\Delta+\frac{w}{a}\right)
  {\rm e}^{ia(1-b) p_1\cdot\Delta}\right]\\
  &&\hspace*{-100pt}\times\,e^{i(p_2 x_2-p_1 x_1)}\bar{d}_{p_2}  \gamma_\nu
  \gamma_5 u_{p_1}\nonumber.
\end{eqnarray}
The terms proportional to $p_{1(2)}\cdot\Delta$ can be removed
using $-ip\cdot\Delta {\rm
e}^{ia(1-b)p\cdot\Delta}=\frac{1}{a}\frac{d}{db}{\rm
e}^{ia(1-b)p\cdot\Delta}$ and then integrating by parts. Now the
dependence on the external momenta is only in the spinors and the
phase and can be absorbed in external quark states, so that the
result takes the form:
\begin{eqnarray}
  \label{calc_vertex_3}
  \left<p_2\left|{\rm T}\{\bar{d}(x_2) \gamma_{\nu}  \gamma_{5}u(x_1)\}\right|p_1\right>_1
  &=&-\frac{C_F}{2\beta_0}\int_0^\infty \! dw\,{\rm e}^{\frac53 w}\,
  \frac{\Gamma(-w)}{\Gamma(1+w)}\left(\frac{-\Lambda^2\Delta^2}{4}\right)^w
  \int_0^1\!d\alpha\int_0^{1-\alpha}\!\!d\beta \nonumber \\
  &&\hspace*{-138pt}\times\int_0^1\!\!da\, a^{2w-1}\Big\{-2\delta(\alpha)
  \delta(\beta)+\int_0^1\!\!db\, b^w \left[\delta(\beta)\delta
  \left(\alpha-a(1-b)\right)+\delta(\alpha)\delta\left(\beta-a(1-b)\right)
  \right]\Big\}\nonumber  \\
  &&\hspace*{-138pt}\times\left<p_2\left|\bar{d}(x_2+\alpha\Delta)
  \gamma_{\nu}  \gamma_{5} u(x_1-\beta\Delta)\right|p_1\right>.
\end{eqnarray}
Here we introduced for later convenience two new variables
$\alpha$ and $\beta$. The integration over $a$ can be taken
and after rearranging the terms the result can be represented as an OPE
\begin{eqnarray}
  \label{calc_vertex_res}
  {\rm T}\{\bar{d}(x_2) \gamma_{\nu} \gamma_{5} u(x_1)\}_1
  &=& \frac{C_F}{2\beta_0}\,\int_0^{\infty}\!
  dw\,{\rm e}^{\frac53 w}\,\frac{\Gamma(-w)}{\Gamma(1+w)}
  \left(\frac{-\Delta^2\Lambda^2}{4}\right)^{w} \,
  \int_0^1d\alpha\int_0^{1-\alpha} d\beta \nonumber\\
  &&\hspace*{-90pt}\times\,\bigg[(1+w) \bigg(f_+^{(w)}(\beta)\delta(\alpha)
  +f_+^{(w)}(\alpha)\delta(\beta)\bigg)
  -w\bigg(f^{(w)}(\beta)\delta(\alpha)+f^{(w)}(\alpha)\delta(\beta)\bigg)
  \bigg]\nonumber\\
  &&\hspace*{-90pt}\times\,\bar{d}(x_2+\alpha\Delta)
  \gamma^{\mu} \gamma_{5} u(x_1-\beta\Delta)\,,
\end{eqnarray}
where it is understood that only leading--twist operators are retained on the 
right--hand side,  
\begin{eqnarray}
  \label{fw_app}
  f^{(w)}(\beta)&\equiv& \beta^{2w-1}\int_0^{1-\beta}db\,(1-b)^{-2w}b^w
  \nonumber \\
  &=&
\frac{\beta^{2w-1} (1-\beta)^{1+w}}{1+w} \,_2\!
  F_1([2w,w+1],[2+w];1-\beta),
\end{eqnarray}
and the ``$+$" prescription is defined, as usual, by \eq{def_plus}.

Next we consider Diagram 2 where the gluon is
exchanged between the quarks. This contribution reads:
\begin{eqnarray}
  \label{calc_box}
  \left<p_2\left|{\rm T}\{\bar{d}(x_2) \gamma_{\nu} \gamma_{5} u(x_1)\}\right|p_1\right>_2
  &=&-\frac{i8\pi^2 C_F}{\beta_0}\int_0^\infty \! dw\,
  {\rm e}^{\frac53 w}\,(-\Lambda^2)^w \,e^{i(p_2 x_2-p_1 x_1)} \nonumber\\
  &&\,\times \, 
\bar{d}_{p_2}\gamma_\mu \gamma_\nu \gamma_5\gamma_\sigma  u_{p_1}
{\rm I}_2^{\mu\sigma}(p_1,p_2,\Delta)\,,
\end{eqnarray}
where the integral ${\rm I}_2^{\mu\sigma}(p_1,p_2,\Delta)$ is given by the expression:
\begin{eqnarray}
\label{int2}
  {\rm I}_2^{\mu\sigma}(p_1,p_2,\Delta)&=&\int\!\frac{d^4k}{(2\pi)^4}
  {\rm e}^{-ik\Delta}\frac{(p_1+k)^\mu(p_2+k)^\sigma}{(k^2)^{1+w}(p_1+k)^2
  (p_2+k)^2}\nonumber\\
  &=& \frac{i}{32\pi^2}\frac{\Gamma(-w)}{\Gamma(1+w)}\left(\frac{\Delta^2}{4}
  \right)^w\int_0^1\!d\alpha\int_0^{1-\alpha}\!d\beta\,(1-\alpha-\beta)^w \\
  &&\,\times\,{\rm e}^{i(\alpha p_2\Delta+\beta p_1\Delta)}\left\{\left[
  g^{\mu\sigma}+2w\frac{\Delta^\mu \Delta^\sigma}{\Delta^2}\right]+
  \ldots\right\}\nonumber.
\end{eqnarray}
Here the dots represent terms proportional to at least one external momentum
$p_{1(2)}^{\mu(\sigma)}$,  which do not contribute to \eq{calc_box} by
virtue of the Dirac equation. Inserting~\eq{int2} in~\eq{calc_box},
contracting the indices and absorbing the dependence on the momenta $p_1$ and
$p_2$ in external quark states, one immediately obtains
\begin{eqnarray}
  \label{calc_box_res}
{\rm T}\{\bar{d}(x_2) \gamma_{\nu} \gamma_{5} u(x_1)\}_2
  &=&\frac{C_F}{2\beta_0}\int_0^\infty \! dw\,{\rm e}^{\frac53 w}\,
  \frac{\Gamma(-w)}{\Gamma(1+w)}\left(\frac{-\Lambda^2\Delta^2}{4}\right)^w
  \int_0^1\!d\alpha\int_0^{1-\alpha}\!d\beta\,\nonumber\\
  &&\times\,(1-\alpha-\beta)^w\left[g_{\mu\nu}(1+w)-2w\frac{\Delta_\mu
  \Delta_\nu}{\Delta^2}\right]\nonumber\\
  &&\times\,
\bar{d}(x_2+\alpha\Delta) \gamma^{\mu}  \gamma_{5}  u(x_1-\beta\Delta)\,.
\end{eqnarray}
For the last step it was important to consider a non-forward matrix element
($p_1\neq p_2$), because otherwise one could not identify how the
quark--field operators get shifted.

The last contribution comes from Diagram 3 since
Diagram 4, describing the self energy of the incoming quark, has no scale and
thus vanishes. We obtain:
\begin{equation}
  \label{calc_wilson}
{\rm T}\{\bar{d}(x_2) \gamma_{\nu} \gamma_{5} u(x_1)\}_3
  =\frac{i4\pi^2 C_F}{\beta_0}\Delta^2\!\!\int_0^\infty \!\!\!\! dw\,
  {\rm e}^{\frac53 w}\,(-\Lambda^2)^w {\rm I}_3(\Delta)
\bar{d}(x_2) \gamma_{\nu}
 \gamma_{5} u(x_1)\,,
\end{equation}
where 
\begin{equation}
  \label{int3}
  {\rm I}_3(\Delta^2)=\int_0^1\!da\int_0^a\!db\,\int\!\frac{d^4k}{(2\pi)^4}
  {\rm e}^{-i(a-b)k\Delta}\frac{1}{(k^2)^{1+w}}
  =\frac{i}{32\pi^2}\frac{\Gamma(-w)}{(1-2w)\Gamma(1+w)}
  \left(\frac{\Delta^2}{4}\right)^{w-1}.
\end{equation}
Collecting the contributions in
\eq{calc_vertex_res}, \eq{calc_box_res} and \eq{calc_wilson} we
get a gauge--invariant result for the OPE of the T--product of quark 
fields to all orders in the strong coupling in in the large-$\beta_0$ limit:
\begin{eqnarray}
  \label{calc}
  {\rm T}\{\bar{d}(x_2) \gamma_{\nu}  \gamma_{5} [x_2,x_1] u(x_1)\}
   &=& \frac{C_F}{2\beta_0}\,\int_0^{\infty}\!
  dw\,{\rm e}^{\frac53 w}\,\frac{\Gamma(-w)}{\Gamma(1+w)}
  \left(\frac{-\Delta^2\Lambda^2}{4}\right)^{w} \,
  \int_0^1d\alpha\int_0^{1-\alpha} d\beta    \nonumber
  \\ \nonumber
  && \hspace*{-110pt} \times\bigg\{
  g_{\mu \nu}\,\bigg[(1+w) \bigg(f_+^{(w)}(\beta)\delta(\alpha)
  +f_+^{(w)}(\alpha)\delta(\beta)\bigg)
  -w\bigg(f^{(w)}(\beta)\delta(\alpha)+f^{(w)}(\alpha)\delta(\beta)\bigg)\bigg]
  \\
  &&\hspace*{-90pt} + \,\left(g_{\mu \nu} (1+w) \, -w\frac{2\Delta_{\mu}
  \Delta_{\nu}}{\Delta^2}\right)\,(1-\alpha-\beta)^w  \\   \nonumber
  &&\hspace*{-90pt} -\, g_{\mu \nu}\frac{1}{1-2w} \delta(\alpha)\delta(\beta)
  \bigg\} \,\times \,\bar{d}(x_2+\alpha\Delta) \gamma^{\mu}  \gamma_{5}
  [x_2+\alpha\Delta,x_1-\beta\Delta] u(x_1-\beta\Delta),
\end{eqnarray}
where the terms in the curly brackets are grouped as they appear from
individual diagrams in Feynman gauge: the first line corresponds to
Diagram 1 (vertex correction) in Fig.~\ref{diagrams} and its symmetric
counterpart, the second line to Diagram 2 (box diagram),
and the third line to Diagram 3 (self--energy like correction to the Wilson
line). Disentangling the two Lorentz structures we end up with the answers for the 
unrenormalized coefficient functions given in \eq{B1} and \eq{B2}.

\section{The operator $\bar d \gamma_\mu\gamma_5 gD_\alpha G^{\alpha\beta} u $: UV divergence and EOM
relations\label{DG_eq_of_motion}}
\setcounter{equation}{0}

The calculation of the UV--renormalon ambiguity of the operator in \eq{Xi} goes along the same 
lines as in Sect.~2.3 and is fully analogous to the similar calculation 
(with no $\gamma_5$) in \cite{ren_review,GKRT}.  
{}For the matrix element between off--shell quark states (Fig.~\ref{diagrams_tw4}) we get 
\begin{eqnarray}
\label{UV_div_O7}
\left< q_2\left\vert \bar{d}(-z)  gD^{\alpha}G_{\alpha\beta}(vz)\Gamma u(z)
\right\vert q_1 \right>
 \,&=&\,
 i\frac{4\pi^2 C_F}{\beta_0} {\rm e}^{-i(q_1+q_2)z}\,\left(g^{\sigma\rho}g_{\beta \nu}
 -g^{\rho}_{\beta}g^{\sigma}_{\nu} \right)\\ \nonumber
&&\hspace*{-185pt}\times\int_0^{\infty}\!\!\! dw\,{\rm e}^{\frac53 w} (-\Lambda^2)^{w}\,
 \Big[\bar{d}_{q_2} \gamma^{\nu}\gamma^{\lambda}\Gamma u_{q_1}\,
 \bar{I}_{\lambda\sigma\rho}(q_2,(1+v)z)
 +\bar{d}_{q_2} \Gamma\gamma^{\lambda}\gamma^{\nu} u_{q_1}\,
 \bar{I}_{\lambda\sigma\rho}(q_1,(1-v)z)\Big],
\end{eqnarray}
where the momentum integral is
\begin{equation}
\bar{I}_{\lambda\sigma\rho}(q,z)=\int\frac{d^4k}{(2\pi)^4}
\frac{(q+k)_{\lambda}k_{\sigma}k_{\rho}}{\left(k^2
\right)^{(1+w)}(q+k)^2}\,{\rm e}^{-ikz}.
\end{equation}
Computing the integral, extracting the $w=1$ residue, specifying
$\Gamma=\gamma_{\mu}\gamma_5$ and projecting with
$x^{\mu}x^{\beta}$ we obtain\footnote{We performed integration by parts over $a$ to eliminate $q\cdot z$ factors
in order to convert to operator notation.} the result in \eq{DG}.
Taking the matrix element of \eq{DG} between the vacuum and a pion state we end up with
the renormalon ambiguity of the DA $\Xi_{\pi}(\alpha_i)$ given in the last line of \eq{three_part_dist_amb}.

Going from an ambiguity to a model involves a replacement of the large--$\beta_0$
renormalon residue by a physical non-perturbative parameter, a
certain local matrix element. From \cite{BF90} it is known that the normalization of all 
four DA $\Phi_{\perp,\parallel}$ and $\Psi_{\perp,\parallel}$ is controlled by a single non-perturbative 
parameter $\delta^2$ for $J=3$. In the renormalon model, contributions of higher spins $J=4,5,\ldots$ 
to $\Phi_{\perp,\parallel}$ and $\Psi_{\perp,\parallel}$ are fixed uniquely in terms of $\delta^2$ so that 
no new parameters appear. We are going to argue that within this construction the normalization of $\Xi_{\pi}$ is also fixed 
uniquely by EOM that relate it to the $J=5$ contributions to the other DA. Thus, all five DA are given in terms  
of  $\delta^2$ and the leading--twist pion DA. 

The necessary constraint can be derived from the operator identity for the second derivative 
\begin{eqnarray}
\lefteqn{
\frac{\partial^2}{\partial x_\alpha \partial x^\alpha} \bar d(-x)\Gamma\, u(x)
 = -\partial^2 \bar d(-x)\Gamma\, u(x)
+g\bar d(-x)\Big[\sigma G(-x)\,\Gamma + \Gamma \sigma
G(x)\Big]u(x)}
\nonumber\\
&+&{}2igx^\nu \frac{\partial}{\partial
x_\mu}\!\int_{-1}^1\!\!dv\,v\,\bar d(-x) \Gamma G_{\nu\mu}(vx)u(x)
+2igx^\nu \partial_\mu\! \int_{-1}^1\!\!dv\,\bar d(-x) \Gamma
G_{\nu\mu}(vx)u(x)
\nonumber\\
&&{}+2g\int_{-1}^1\!dv\int_{-1}^v\!dt\,(1+vt)\bar d(-x)\Gamma x^\mu x^\nu
G_{\mu\rho}(-vx)G^{\rho}_{\phantom{\nu}\nu}(-tx)u(x)
\nonumber\\
&&{}-igx^\nu \int_{-1}^1\!\!dv\,(1+v^2)\,\bar d(-x)\Gamma
[D_\mu,G^{\mu}_{\phantom{\nu}\nu}](vx) u(x)
\label{twoderiv}
\end{eqnarray}
given in Eq.~(A.9) in \cite{BB98} (see also \cite{BB89}). Here $\sigma G \equiv \sigma_{\alpha\beta}G^{\alpha\beta}$
and $\partial_\mu$ stands for the derivative with respect to the total translation. 

Taking the matrix element of (\ref{twoderiv}) between the vacuum and a pion state
and using the definitions of the DA we obtain
\begin{eqnarray}
\label{DG_EOM_rel}
 &&2\,i\,\int_0^1 du {\rm e}^{ -
i\,\mathit{pz}\,(2\,u - 1)}\,\bigg[(4 - 2\,i\,\mathit{pz}\,(2\,u -
1))\,\phi_1(u) + ( - 3 + i\mathit{pz}\,(2
\,u - 1))\,\phi_2(u)\bigg]\nonumber \\
&&= \int D\alpha_i\,\,
{\rm e}^{ - i\,\mathit{pz}\,(\alpha_1 - \alpha_2)}
\bigg[ 2\left(\left({ \frac {1}{\mathit{pz}^{2}\,\alpha_3^{2
}}}  - 1\right)\,\mathrm{sin}(\mathit{pz}\,\alpha_3) - {
\frac {\mathrm{cos}(\mathit{pz}\,\alpha_3)}{\mathit{pz}\,
\alpha_3}} \right)\,{\Phi_{\perp}(\alpha_i)}  \nonumber\\
&&\!\!+ \left(\left({ \frac {1}{\alpha_3}}  + 2 -
{ \frac {i\,\alpha_1}{\mathit{pz}\,\alpha_3^{2}}}  +
{ \frac {i\,\alpha_2}{\mathit{pz}\,\alpha_3^{2}} }
\right)\,\mathrm{sin}(\mathit{pz}\,\alpha_3) +
\frac{i}{\alpha_3}\,( {\alpha_1}  - {\alpha_2}
)\,\mathrm{cos}(\mathit{pz}\,\alpha_3)\right)\,{
\Phi_{\parallel}(\alpha_i)}
\\ \nonumber
&&\!\! + \left(\left({ \frac {1}{\alpha_3}}  -
{ \frac {1}{\mathit{pz}^{2}\,\alpha_3^{3}}} \right)\,
\mathrm{sin}(\mathit{pz}\,\alpha_3) + { \frac {
\mathrm{cos}(\mathit{pz}\,\alpha_3)}{\mathit{pz}\,\alpha_3^{2}}}
\right)\Xi_{\pi}(\alpha_i) + i\left(2{\Psi_{\perp}(\alpha_i)} -
{\Psi_{\parallel}(\alpha_i)}\right)\, \mathrm{cos}(\mathit{pz}\,
\alpha_3)\bigg],
\end{eqnarray}
where we omitted  the contributions of the
two--gluon operator from the right--hand side. These can be systematically put to zero 
to our accuracy as they start contributing at higher order in the flavor expansion.

Expansion of \eq{DG_EOM_rel} in powers of 
$pz$ yields simple relations between integrals that involve all five three--particle DA. 
The odd powers are
trivial: both the right--hand side  and the left--hand side  vanish by symmetry.
The first two non-trivial relations are:
\begin{eqnarray}
&&\int_0^1 du   \Big[-8\,\phi_1(u) + 6\,\phi_2(u)\Big] = \int
D\alpha_i\,\, \Big[ -2\,{{ \Psi_{\perp}(\alpha_i)}} +
\,{{\Psi_{\parallel}(\alpha_i)}} \Big],
\\ \nonumber
&& \int_0^1 du\,(u - \bar{u})^{2}\,\Big[-16\,\phi_1(u) + 10\,\phi_2(u) \Big]=\int D\alpha_i\,\,
\bigg[-{ \frac {4}{3}} \,(\alpha_1 - \alpha_2)\,\Xi_{\pi}(\alpha_i) \\
\nonumber &&\hspace{30pt}+ { \frac {2}{3}}
\,(\alpha_1-\alpha_2)\left(7\,(\alpha_1+\alpha 2\right) -10)
\,{\Phi_{\parallel}(\alpha_i)} - { \frac {8}{3}} \,(
\alpha_1-\alpha_2)(\alpha_1
 +\alpha _2 - 1)\,\Phi_{\perp}(\alpha_i) \\
\nonumber &&\hspace{30pt} \mbox{} + \left(1 - 2\,\alpha_2 -
2\,\alpha_1 + 2\,\alpha_2^{2} + 2\,
\alpha_1^{2}\right)\,{{\left(\Psi_{\parallel}(\alpha_i)-2\Psi_{\perp}(\alpha_i)\right)}}
\bigg].
\end{eqnarray}
The first relation does not involve $\Xi_{\pi}(\alpha_i)$ and is satisfied identically both by the 
model of \cite{BF90} and the renormalon model.
The second relation gives the required constraint for the normalization 
integral $\int {\cal D}\alpha_i\, (\alpha_1-\alpha_2)\,\Xi_{\pi}(\alpha_i)$ 
in terms of the other four DA. It is easy to verify that in order to satisfy 
this constraint one  must assume in the last line of \eq{three_part_dist_amb} the {\em same} replacement
$c\Lambda^2\longrightarrow \delta^2/6$ as in the other DA. 


\section{Cancellation of renormalons for the $\rho$--meson amplitudes\label{Amb_Cancel_rho}}
\setcounter{equation}{0}

Here we want to demonstrate cancellation of IR renormalon ambiguities in the
leading--twist coefficient functions with the UV--renormalon ambiguities in the
matrix elements of twist--four DA for the case of exclusive amplitudes involving
a vector $\rho$--meson. Similarly to the pion case, we consider the simplest example:
a gauge--invariant T--product of quark fields sandwiched between the vacuum and the meson state.


\subsection{UV renormalons in two--particle DA\label{EOM_UV}}


To begin with, we calculate the UV--renormalon ambiguities in the two--particle DA of twist four.
These can be obtained from the three--particle ones using EOM.

The specific EOM relations we need in the chiral--even sector are~\cite{BB98}:
\begin{eqnarray}
  \label{equ_motion_even}
  g_3(u)&=&\phi_\parallel - 2\frac{d}{du}\int_0^u\!d\alpha_1\int_0^{1-u}\!
  d\alpha_2\frac{1}{\alpha_3}\left[2\Phi(\alpha_i)+\Psi(\alpha_i)\right]
  \nonumber\\
  \mathbb{A}(u)&=&32\int_0^u\!dv\int_0^v\!dw\left(g^v_\perp(w)-\phi_\parallel(w)
  \right)+32\int_0^u\!dv\int_0^v\!d\alpha_1\int_0^{1-v}\!d\alpha_2
  \frac{1}{\alpha_3}\left[2\Phi(\alpha_i)+\Psi(\alpha_i)\right],\nonumber \\
  &&+8\int_0^u\!d\alpha_1\int_0^{1-u}\!d\alpha_2
  \frac{u \alpha_2-(1-u)\alpha_1}{\alpha_3^2}\left[2\Phi(\alpha_i)+
  \Psi(\alpha_i)\right].
\end{eqnarray}
This implies that the ambiguities of $g_3(u)$ and $\mathbb{A}(u)$ due
to UV renormalons at $w=1$ are given by
\begin{eqnarray}
\delta_{\UV}\left\{g_3(u)\right\}&=&- 2\frac{d}{du}\int_0^u\!d\alpha_1\int_0^{1-u}\!
  d\alpha_2\frac{1}{\alpha_3}\delta_{\UV}\left\{2\Phi(\alpha_i)+\Psi(\alpha_i)\right\},
  \nonumber\\
 \delta_{\UV}\left\{\mathbb{A}(u)\right\} &=&32\int_0^u\!dv\int_0^v\!d\alpha_1\int_0^{1-v}\!
d\alpha_2
  \frac{1}{\alpha_3}\delta_{\UV}\left\{2\Phi(\alpha_i)+\Psi(\alpha_i)\right\},
  \nonumber \\
  &&+8\int_0^u\!d\alpha_1\int_0^{1-u}\!d\alpha_2
  \frac{u \alpha_2-(1-u)\alpha_1}{\alpha_3^2}\delta_{\UV}\left\{2\Phi(\alpha_i)+
  \Psi(\alpha_i)\right\},
\end{eqnarray}
where the ambiguities of the
three--particle  DA can be read from \eq{three_rho_even} using the inverse 
substitution in \eq{zeta_4}.
{}For $g_3(u)$ we get
\begin{eqnarray}
 \delta_{\UV}
\left\{g_3(u)\right\}&=&4c\frac{\Lambda^2}{m_{\rho}^2}\frac{d}{du}
\int_0^u\!d\alpha_1\int_0^{1-u}\!
  d\alpha_2\,\left[\frac{\phi_{\parallel}(\alpha_1)}{(1-\alpha_1)^2}
-\frac{\phi_{\parallel}(\alpha_2)}{(1-\alpha_2)^2}\right]\\ \nonumber
&=&4c\frac{\Lambda^2}{m_{\rho}^2}\frac{d}{du}\left[(1-u)\int_0^u d\alpha\,
\frac{\phi_{\parallel}(\alpha)}{(1-\alpha)^2}-u\int_0^{1-u}
d\alpha\,\frac{\phi_{\parallel}(\alpha)}{(1-\alpha)^2}\right],
\end{eqnarray}
where we evaluated one of the $\alpha$ integrals. Taking the
derivative in respect to $u$ and using the symmetry
$\phi_{\parallel}(u)=\phi_{\parallel}(1-u)$ yields the desired result:
\begin{eqnarray}
\label{g3_UV}
  \delta_{\UV}\left\{g_3(u)\right\}
&=&-4c\frac{\Lambda^2}{m_\rho^2}\left\{\int_0^1\!dv\,
\phi_\parallel(v)\left[\theta(u>v)\;\frac{1}{\bar v ^2}+\theta(u<v)\;
\frac{1}{v^2}\right]-\frac{\phi_\parallel(u)}{u\bar u}\right\}.
\end{eqnarray}

In turn, for the DA $\mathbb{A}(u)$ we get the following expression:
\begin{eqnarray}
 \delta_{\UV}\left\{\mathbb{A}(u)\right\}&=&-16c\frac{\Lambda^2}{m_{\rho}^2}
\Bigg\{4\int_0^u\!dv\int_0^v\!d\alpha_1\int_0^{1-v}\!
  d\alpha_2\,\left[\frac{\phi_{\parallel}(\alpha_1)}{(1-\alpha_1)^2}
-\frac{\phi_{\parallel}(\alpha_2)}{(1-\alpha_2)^2}\right]\\ \nonumber
&&+\int_0^u\!d\alpha_1\int_0^{1-u}\!
  d\alpha_2\,\frac{u\alpha_2-(1-u)\alpha_1}{1-\alpha_1-\alpha_2}
\left[\frac{\phi_{\parallel}(\alpha_1)}{(1-\alpha_1)^2}
-\frac{\phi_{\parallel}(\alpha_2)}{(1-\alpha_2)^2}\right]\Bigg\}.
\end{eqnarray}
Taking one of the $\alpha$ integrals in each line yields:
\begin{eqnarray}
\delta_{\UV}\left\{\mathbb{A}(u)\right\}&=&-16c\frac{\Lambda^2}{m_{\rho}^2}
\Bigg\{4\int_0^u\!dv\,\left[(1-v)\int_0^v\!d\alpha\,\frac{\phi_{\parallel}(\alpha)}{(1-\alpha)^2}
-v\int_0^{1-v}\!
  d\alpha\,\frac{\phi_{\parallel}(\alpha)}{(1-\alpha)^2}\right]\nonumber\\
&&-\left[\int_0^u\!d\alpha\,\left((u-\alpha)\ln\left(\frac{u-\alpha}{1-\alpha}\right)
+u(1-u)\right)\frac{\phi_{\parallel}(\alpha)}{(1-\alpha)^2}\right.\\
&&-\left.\int_0^{1-u}\!d\alpha\,\left((1-u-\alpha)\ln\left(\frac{1-u-\alpha}{1-\alpha}\right)
+u(1-u)\right)\frac{\phi_{\parallel}(\alpha)}{(1-\alpha)^2}\right]\Bigg\}.\nonumber
\end{eqnarray}
Finally, integrating by parts over $v$ in the first line 
and rearranging the terms we obtain: 
\begin{eqnarray}
\label{A_UV}
 \delta_{\UV}\left\{\mathbb{A}(u)\right\}&=&16c\frac{\Lambda^2}{m_\rho^2}\int_0^1 dv\,
\phi_\parallel(v) \Bigg\{\theta(u>v)\;\frac{1}{\bar v^2}\left[
{\bar u+\bar u^2+(u-v)  \ln\frac{u-v}{\bar v}}\right]\nonumber\\
  &&\hspace*{20pt}+\theta(u<v)\;\frac{1}{v^2}\left[{u+u^2+(v-u)\ln\frac{v-u}{v}}\right]\Bigg\}.
\end{eqnarray}


In the chiral--odd sector the EOM relations between the
two--particle and the three--particle DA of twist four are given by~\cite{BB98}:
\begin{eqnarray}
  \label{equ_motion_odd}
  h_3(u)&=& 2 h_\parallel^{(s)}(u)-\Phi_\perp(u)\nonumber\\
&&-2\frac{d}{du}\int_0^u\!d\alpha_1\int_0^{1-u}
d\alpha_2\left[\frac{\alpha_1-\alpha_2-(2u-1)}{\alpha_3^2}S(\alpha_i)-\frac{1}{\alpha_3}
\left(T_2(\alpha_i)-T_3(\alpha_i)\right)\right],\nonumber\\
\mathbb{A}_T(u)&=&-2\int_0^u\!dv\,(2v-1)\left[\Phi_\perp(v)+h_3(v)\right]
+8\int_0^u\!dv\int_0^{1-u}\!dw\left[h_3(w)-\Phi_\perp(w)\right] \\
&&+4\int_0^u\!d\alpha_1\int_0^{1-u}\!d\alpha_2\left[\frac{1}{\alpha_3}S(\alpha_i)
-\frac{\alpha_1-\alpha_2-(2u-1)}{\alpha_3^2}\left(T_2(\alpha_i)
-T_3(\alpha_i)\right)\right].\nonumber
\end{eqnarray}
As a consequence, the ambiguity of $h_3(u)$ and $\mathbb{A}_T(u)$ due to UV 
renormalons is related to that of the
three--particle DA $S(\alpha_i)$,
$T_2(\alpha_i)$ and $T_3(\alpha_i)$. Using the results for the ambiguities of the three--particle DA in 
\eq{three_rho_odd} with the inverse substitution in \eq{zeta_4_T} we obtain
\begin{eqnarray}
  \delta_{\UV}\left\{h_3(u)\right\}&=&
4c\frac{\Lambda^2}{m_\rho^2}\frac{d}{du}\int_0^u\!d\alpha_1\int_0^{1-u}
d\alpha_2\left\{\left[\frac{u-\alpha_1}{(1-\alpha_1-\alpha_2)^2}
-\frac{1}{1-\alpha_1}\right]\frac{\phi_{\perp}(\alpha_1)}{(1-\alpha_1)}\right.\nonumber\\
&&-\left.\left[\frac{1-u-\alpha_2}{(1-\alpha_1-\alpha_2)^2}-\frac{1}{1-\alpha_2}\right]
\frac{\phi_{\perp}(\alpha_2)}{(1-\alpha_2)}\right\}.
\end{eqnarray}
The square bracket in the first line vanishes upon taking the
$\alpha_2$ integral and that in the second line 
upon performing the $\alpha_1$ integral. Thus there is no
UV renormalon ambiguity (at $w=1$):
\begin{eqnarray}
\label{delta_UV_h_0}
 \delta_{\UV}\left\{h_3(u)\right\}&=&0.
\end{eqnarray}

For $\mathbb{A}_T(u)$ on the other hand
\begin{eqnarray}
\delta_{\UV}\left\{\mathbb{A}_T(u)\right\}&=&8c\frac{\Lambda^2}{m_\rho^2}
\int_0^u\!\!d\alpha_1\!\int_0^{1-u}\!\!\!\!
d\alpha_2\left\{\left[1-\frac{2(u-\alpha_1)}{(1-\alpha_1-\alpha_2)}
+\frac{(u-\alpha_1)(1-\alpha_1)}{(1-\alpha_1-\alpha_2)^2}\right]
\frac{\phi_{\perp}(\alpha_1)}{(1-\alpha_1)^2}\right.\nonumber\\
&&+\left.\left[1-\frac{2(1-u-\alpha_2)}{(1-\alpha_1-\alpha_2)}
+\frac{(1-u-\alpha_2)(1-\alpha_2)}{(1-\alpha_1-\alpha_2)^2}\right]
\frac{\phi_{\perp}(\alpha_2)}{(1-\alpha_2)^2}\right\}.
\end{eqnarray}
Taking one $\alpha$ integral we obtain
\begin{eqnarray}
\label{A_T_amb}
  \delta_{\UV}\left\{\mathbb{A}_T(u)\right\}&=&16c\frac{\Lambda^2}{m_\rho^2}\int_0^1
dv\,\phi_\parallel(v)\Bigg[\theta(u>v)\;\frac{1}{(1-v)^2}\left({(1-u)+(u-v)
  \ln\frac{u-v}{1-v}}\right)\nonumber\\
  &&\hspace*{100pt}+\theta(u<v)\;\frac{1}{v^2}\left({u+(v-u)\ln\frac{v-u}{v}}\right)\Bigg].
\end{eqnarray}

\subsection{IR renormalons in coefficient functions\label{rho_IR}}

The IR  $w=1$ renormalon ambiguity in the all--order perturbative calculation of the
leading--twist coefficient functions can be obtained from a generalization
of the OPE relation in \eq{calc} for the case of an arbitrary Dirac matrix
$\Gamma$ between the quark fields:
\begin{eqnarray}
  \label{op_rel_tw2}
  \delta_{\IR}\left\{{\rm T}\{\bar{d}(x_2)\Gamma u(x_1)\}\right\}&=&-c
  \Lambda^2 \Delta^2\int_0^1\!d\alpha\int_0^{1-\alpha}\!d\beta\,
\nonumber\\&&\hspace*{-62pt}{}\times
\Big\{ \left(f^{(1)}(\beta)\delta(\alpha)+f^{(1)}(\alpha)\delta(\beta)
  \right)\left[\bar{d}(x_2+\alpha\Delta)\Gamma
u(x_1-\beta\Delta)\right]
  \\
   &&\hspace*{-48pt}+\frac{1}{4}(1-\alpha-\beta)\left(g_{\mu\nu}+
  2\frac{\Delta_{\mu}\Delta_{\nu}}{\Delta^2}\right)\,
  \left[\bar{d}(x_2+\alpha\Delta)\gamma_{\alpha}
  \gamma^{\mu}\Gamma\gamma^{\nu}\gamma^{\alpha}
u(x_1-\beta\Delta)\right]\Big\},
\nonumber
\end{eqnarray}
where we replaced the $w$--integral by $\pi$ times
the $w=1$ residue and suppressed the gauge--link; $c$ was defined in \eq{c}.

For vector mesons there are two relevant structures, chiral--even $\Gamma=\gamma_\mu$, and
chiral--odd $\Gamma=\sigma_{\mu\nu}$. We will discuss these two sectors
separately.

\subsubsection*{Chiral--even amplitudes}

The matrix element of the T--product of quark operators  between the $\rho^+$ and the
vacuum state contains three Lorentz structures 
which we parametrize using the structure function~$F_i$:
\begin{eqnarray}
  \label{def_rho_even}
  \langle 0 |{\rm T}\{\bar{d}(-x)\gamma_\mu u(x)\}|\rho^{+}(P,\lambda)\rangle_{\mu^2}&=&
    f_\rho m_\rho\int_0^1\!du\, e^{-ipx(u-\bar{u})}\, \\ \nonumber
  &&\hspace*{-162pt}
\times \Bigg\{\frac{e^{(\lambda)}x}{Px}P_\mu F_1(u,x^2;\mu^2)+
\left(e_\mu^{(\lambda)}-P_\mu\frac{e^{(\lambda)}x}{Px}\right)
  F_2(u,x^2;\mu^2)
-\frac{1}{2}x_\mu\frac{e^{(\lambda)}x}{(Px)^2}m_\rho^2 F_3(u,x^2;\mu^2) \Bigg\}.
\end{eqnarray}
The OPE to twist--four accuracy reads:
\begin{eqnarray}
\label{def_rho_even2}
F_1(u,x^2;\mu^2)&=&C_1^{(2)}\otimes \phi_{\parallel}(u)+\frac{m_\rho^2 x^2}{4}
  \mathbb{A}(u)\,,
\nonumber \\
F_2(u,x^2;\mu^2)&=&C_2^{(3)}\otimes g_\perp^{(v)}(u)+\ldots\,,
 \nonumber\\
F_3(u,x^2;\mu^2)&=&C_3^{(2)}\otimes \phi_{\parallel}(u)+\mathbb{C}(u).
\end{eqnarray}
Here $\phi_{\parallel}(u)$ is the leading--twist DA of the longitudinally polarized
$\rho$--meson, $g_\perp^{(v)}(u)$ is the DA of twist--three corresponding to the
contribution of the transversely polarized $\rho$--meson  and the
functions $ \mathbb{A}(u)$, $ \mathbb{C}(u)$ represent higher--twist contributions.
At leading order in $\alpha_s$ $C_1^{(2)}(u,v)= C_2^{(3)}(u,v) = \delta(u-v)$, alias
$F_1(u,x^2;\mu^2)=\phi_{\parallel}(u)$ and $F_2(u,x^2;\mu^2)=g_\perp^{(v)}(u)$,
while the remaining twist--two coefficient function $C_3^{(2)}$ vanishes.

In order to calculate the IR--renormalon ambiguity in the leading--twist
part of the amplitudes in \eq{def_rho_even} we take the appropriate matrix element of
\eq{op_rel_tw2} retaining the leading  terms in \eq{def_rho_even2}.
The result reads
\footnote{ The IR ambiguity of $C_2^{(3)}\otimes g_\perp^{(v)}$ is more difficult to obtain
 because the twist--three matrix element vanishes for on--shell massless quark states.
This  ambiguity must be compensated by contributions of twist--five operators and is
of no interest for our purposes.}:
\begin{eqnarray}
\label{A_IR}
 \delta_{\IR}
\left\{C_1^{(2)}\otimes \phi_{\parallel}(u)\right\}&\!=\!&\frac{-16c\Lambda^2}{m_\rho^2}
\int_0^1\!dv\int_0^1\!d\alpha\int_0^{1-\alpha}
\!d\beta\,\delta\big(v(1-\alpha-\beta)+\alpha-u\big)\,\phi_\parallel(v)
\nonumber \\
 &&\hspace*{90pt}\times \left[\left(f^{(1)}(\beta)
  \delta(\alpha)+f^{(1)}(\alpha)\delta(\beta)\right)+2(1-\alpha-\beta)\right]
\nonumber\\
 &=&\frac{-16c\Lambda^2}{m_\rho^2}\int_0^1dv\,
\phi_\parallel(v)\bigg[\theta(u<v)\;\frac{1}{v^2}\left({u+u^2+(v-u)\ln\frac{v-u}{v}}\right)
\nonumber \\&&
\hspace*{90pt}+
\theta(u>v)\;\frac{1}{\bar v^2}\left({\bar u +\bar u^2+(u-v)  \ln\frac{u-v}{\bar v}}\right)\bigg],
\nonumber\\
  \delta_{\IR}
\left\{C_3^{(2)}\otimes \phi_{\parallel}(u)\right\}&=&\frac{4c\Lambda^2}{m_\rho^2}\frac{d^2}{du^2}\int_0^1
\!dv\!\int_0^1\!\!d\alpha\!\int_0^{1-\alpha}\!\!\!\!\!\!\!d\beta(1\!-\!\alpha\!-\!\beta)
\,\delta\big(v(1\!-\!\alpha\!-\!\beta)+\alpha-u\big)\,\phi_\parallel(v)
\nonumber\\
&=&\frac{4c\Lambda^2}{m_\rho^2}\left\{\int_0^1\!dv\,\phi_\parallel(v)\left[\theta(u>v)\;
\frac{1}{\bar v^2}+\theta(u<v)\;\frac{1}{v^2}\right]-\frac{\phi_\parallel(u)}{u\bar u}\right\}.
\end{eqnarray}
Comparing these expressions with the UV--renormalon ambiguities in the twist--four
contributions 
for $ \mathbb{A}(u)$ and $ \mathbb{C}(u)$ in \eq{A_UV} and\footnote{Taking into account the
relation of \eq{BC_relations} between $ \mathbb{C}(u)$ and $g_3(u)$.}
 \eq{g3_UV}, respectively,
we observe that
the ambiguities in the structure functions $F_i$ cancel out, as expected.

\subsubsection*{Chiral--odd amplitudes}

The calculation for chiral-odd amplitudes goes along similar lines.
We parametrize the matrix element in terms of three more structure functions
\begin{eqnarray}
  \label{def_rho_odd}
  \langle 0 |\bar{d}(-x)\sigma_{\mu\nu} u(x)|\rho^{+}(P,\lambda)\rangle=
if_\rho^T \int_0^1\!du\,
  e^{-ipx(u-\bar{u})}\bigg\{\left(e^{(\lambda)}_\mu P_\nu-e^{(\lambda)}_\nu P_\mu\right)
F^{T}_1(u,x^2;\mu^2)\nonumber \\
  &&\hspace*{-420pt}+\left(P_\mu x_\nu -P_\nu x_\mu\right)\frac{e^{(\lambda)}x}{(Px)^2} m_\rho^2
F^{T}_2(u,x^2;\mu^2)+\frac{1}{2}\left(e^{(\lambda)}_\mu x_\nu-e^{(\lambda)}_\nu x_\mu\right)
  \frac{m_\rho^2}{Px} F^{T}_3(u,x^2;\mu^2)\bigg\}.
\end{eqnarray}
To twist--four accuracy
\begin{eqnarray}
  \label{def_rho_odd2}
F^{T}_1(u,x^2;\mu^2)&=& {C}_1^{T(2)}\otimes\phi_{\perp}(u)+\frac{m_\rho^2 x^2}{4}\mathbb{A}_T(u)\,,
\nonumber\\
F^{T}_2(u,x^2;\mu^2)&=& {C}_2^{T(3)}\otimes h_{\parallel}^{(t)}(u)+\ldots\,,
\nonumber\\
F^{T}_3(u,x^2;\mu^2)&=& {C}_3^{T(2)}\otimes\phi_{\perp}(u)+\mathbb{C}_T(u)\,,
\end{eqnarray}
where $\phi_{\perp}(u)$ parametrize the leading--twist part.
Repeating the procedure of the previous section we obtain the following ambiguity:
\begin{eqnarray}
\label{IR_AT}
  \delta_{\IR}
\left\{{C}_1^{T(2)}\otimes\phi_{\perp}(u)\right\}&=&-16c\frac{\Lambda^2}{m_\rho^2}\int_0^1
dv\int_0^1
d\alpha\int_0^{1-\alpha}\!d\beta\,\delta\left(v(1-\alpha-\beta)+\alpha-u\right)\phi_T(v)\nonumber\\
  &&\hspace*{90pt}\times\left(f^{(1)}(\beta)
  \delta(\alpha)+f^{(1)}(\alpha)\delta(\beta)\right) \nonumber\\ &=&-16c
  \frac{\Lambda^2}{m_\rho^2}\int_0^1\!dv\,\phi_T(v)
\Bigg[\theta(u<v)\;\frac{1}{v^2}\left({u+(v-u)\ln\frac{v-u}{v}}\right)
\nonumber\\
  && \hspace*{90pt}+
\theta(u>v)\;
\frac{1}{\bar v^2}\left({\bar u+(u-v)
  \ln\frac{u-v}{\bar v}}\right)
\Bigg],\nonumber\\
\delta_{\IR}
\left\{{C}_3^{T(2)}\otimes\phi_{\perp}(u)\right\}&=&0,
\end{eqnarray}
The absence of an ambiguity in ${C}_3^{T(2)}\otimes\phi_{\perp}(u)$,
which is to be associated with $\mathbb{C}_T(u)$, results from the fact that Diagram 2 in Fig.~\ref{diagrams} vanishes
owing to the identity $\gamma_\alpha \sigma_{\mu\nu}\gamma^\alpha=0$.
The absence of this ambiguity is expected based on the fact that $h_3(u)$ has 
no corresponding UV--renormalon ambiguity --- see \eq{delta_UV_h_0} --- and the relation
between $\mathbb{C}_T(u)$ and $h_3(u)$ in \eq{BC_T_relations}.
Comparing the result for $\delta_{\IR}\{{C}_1^{T(2)}\otimes\phi_{\perp}(u)\}$ with 
the UV--renormalon ambiguity in $\mathbb{A}^T(u)$ in \eq{A_T_amb} 
we observe that the ambiguities in the structure function $F_1^T$ cancel out. 
This completes the calculation.


\end{document}